\documentclass{jaa}
\usepackage{graphicx}
\usepackage[utf8]{inputenc}
\usepackage{natbib}
\usepackage{amsmath}
\usepackage{mathtools}
\usepackage{caption}
\usepackage{subcaption}
\usepackage{todonotes}
\usepackage{hyperref}
\hypersetup{backref=true,
    pagebackref=true,
    hyperindex=true,
    colorlinks=true,
    breaklinks=true,
    urlcolor= black,
    linkcolor= blue,
    bookmarks=true,
    bookmarksopen=false,
    filecolor=black,
    citecolor=blue,
    linkbordercolor=blue
}
\usepackage{xcolor}

\newcommand{\apjl}{The Astrophysical Journal Letter}
\newcommand{\apj}{The Astrophysical Journal}
\newcommand{\mnras}{MNRAS}
\newcommand{\procspie}{SPIE}

\definecolor{mygray}{gray}{0.45}
\begin{document}\sloppy

%%paper title
\title{Sub$-$MeV spectroscopy with AstroSat$-$CZT Imager for Gamma Ray Bursts}

%%author names are separated by comma (,) 
%%use \and before the last author name 
%%use a * along with the number separated by comma
%% for the  author for correspondence
%%\textsuperscript{number} is used for affiliation
%%\affilOne, \affilTwo etc., upto \affilTwentyfive is possible
%%Please note the first letter after \affil is capitalised in the command
%%

\author{Tanmoy Chattopadhyay\textsuperscript{1,*}, Soumya Gupta\textsuperscript{2}, Vidushi Sharma\textsuperscript{2, 3}, Shabnam Iyyani\textsuperscript{2}, Ajay Ratheesh\textsuperscript{4,5,6}, N. P. S. Mithun\textsuperscript{7}, E. Aarthy\textsuperscript{7}, 
Sourav Palit\textsuperscript{8},
Abhay Kumar\textsuperscript{7}, 
Santosh V Vadawale\textsuperscript{7}, A.R. Rao\textsuperscript{2,9}, Varun Bhalerao\textsuperscript{8} and Dipankar Bhattacharya\textsuperscript{2}}

\affilOne{\textsuperscript{1}Kavli Institute of Astrophysics and Cosmology, 452 Lomita Mall, Stanford, CA 94305, USA\\}
%\affilOne{\textsuperscript{1}Department of Physics, Stanford University, 382 Via Pueblo Mall, Stanford CA 94305, USA\\}
%\affilTwo{\textsuperscript{2}Kavli Institute of Astrophysics and Cosmology, 452 Lomita Mall, Stanford, CA 94305, USA\\}
\affilThree{\textsuperscript{2}The Inter-University Centre for Astronomy and Astrophysics, Pune, India\\}
\affilFour{\textsuperscript{3}Department of Physics, KTH Royal Institute of Technology, AlbaNova, 10691 Stockholm, Sweden\\}
\affilFive{\textsuperscript{4}Dipartimento di Fisica, Università di Roma Tor Vergata, Via della Ricerca Scientifica 1, I-00133 Roma, Italy\\}
\affilSix{\textsuperscript{5}INAF Istituto di Astrofisica e Planetologia Spaziali, Via del Fosso del Cavaliere 100,
00133 Roma (RM), Italy\\}
\affilSeven{\textsuperscript{6}Dipartimento di Fisica, Università La Sapienza, P. le A. Moro 2, 00185 Roma, Italy\\}
\affilEight{\textsuperscript{7}Physical Research Laboratory, 
Ahmedabad, Gujarat, India\\}
\affilNine{\textsuperscript{8}Indian Institute of Technology Bombay, Mumbai, India\\}
\affilTen{\textsuperscript{9}Tata Institute of Fundamental Research,Mumbai, India\\}

%%escape two column mode for title, affiliation and abstract
%%by giving \twocolumn command as shown

\twocolumn[{

\maketitle

%%include \corres to print the corresponding author Email id
\corres{tanmoyc@stanford.edu, tanmoyrng1@gmail.com}

%%include \msinfo for
%%manuscript information such as
%%received, revised and accepted dates
%%
\msinfo{---}{---}

%%abstract
\begin{abstract}
Cadmium Zinc Telluride Imager (CZTI) onboard {\em AstroSat} has been a prolific Gamma-Ray Burst (GRB) monitor.
%\textcolor{mygray}{and a sensitive GRB polarimeter}. 
While the $2-$pixel Compton scattered events ($100-300$ keV) are used to extract sensitive %\textcolor{mygray}{polarization and}
spectroscopic information, the inclusion of the low-gain pixels ($\sim 20 \%$ of the detector plane) after careful calibration extends the energy range of Compton energy spectra 
%\textcolor{mygray}{and polarization} 
to 600 keV. The new feature also allows single-pixel spectroscopy of the GRBs to the sub-MeV range which is otherwise limited to 150 keV. We also introduced a new noise rejection algorithm in the analysis (`Compton noise'). These new additions not only enhances the spectroscopic %\textcolor{mygray}{-polarimetry} 
sensitivity of CZTI, but the sub-MeV spectroscopy will also allow proper characterization of the GRBs not detected by {\em Fermi}. This article describes the methodology of single, Compton event and veto spectroscopy 
%\textcolor{mygray}{along with the updated polarization results} 
in $100-600$ keV for the GRBs detected in the first year of operation. CZTI in last five years has detected $\sim$20 bright GRBs. The new methodologies, when applied on the spectral 
%\textcolor{mygray}{-polarimetric} 
analysis for this large sample of GRBs, has the potential to improve the results significantly and help in better understanding the prompt emission mechanism.      

\end{abstract}

%%insert keywords separated by 3 hyphens using \keywords{words}

}]
%%close the twocolumn escape here

%%include \doinum{number}for the DOI number in the header
%%include \volnum{number} for the volume number in the header
%%include \year{yyyy} for  year of publication in the header
%%include \pgrange{num--num} page range of article in the header
%%include \artcitid{num} for the article citation id
%%include \lp to print last page of the article
%%include \setcounter{page}{pagenum} for the exact starting page of the article

\doinum{12.3456/s78910-011-012-3}
\artcitid{\#\#\#\#}
\volnum{000}
\myyear{2021}
\pgrange{1--}
\setcounter{page}{1}
\lp{1}

\section{Introduction}
Cadmium Zinc Telluride Imager (hereafter CZTI) on board {\em AstroSat} \citep{singh14,paul13}, India's first dedicated Astronomical satellite, has been demonstrated as a prolific Gamma Ray Burst (GRB) monitor, 
%and polarimeter 
since the launch of {\em AstroSat} \citep{rao16,chattopadhyay19}.  
CZTI is one of the two hard X-ray detectors sensitive in 20 $-$ 150 keV. The instrument employs an array of CZT detectors, each 40 mm $\times$ 40 mm $\times$ 5 mm in size, totalling to a collecting of 924 cm$^2$. Each detector is further segmented spatially to 256 pixels with a pitch of $\sim$2.5 mm. Use of collimator made of 0.07 mm Tantalum and 1 mm Aluminum sheets restricts the field of view of the instrument to $\sim$4$^\circ$. Details of the payload design and function are given in \citet{bhalerao16} and \citet{rao16}. At energies beyond 100 keV, the increasing transparency of the collimators and the supporting structure enables CZTI to work as an all-sky monitor.      
Because of this all-sky sensitivity, CZTI instrument since the launch of {\em AstroSat} has been working as an efficient GRB monitor with around $\sim 83$ GRB detections per year\footnote{http://astrosat.iucaa.in/czti/?q=grb}. %A 5 mm thick CZT has a significant Compton scattering efficiency and therefore, works as a sensitive Compton polarimeter in 100 $-$ 300 keV where the Compton scattered events between the adjacent CZTI pixels have a preferential distribution that depends on the polarization angle and fraction of the incident radiation from the GRBs. Polarization sensitivity of CZTI has been demonstrated both before the launch of {\em AstroSat} during ground calibration \citep{chattopadhyay14,vadawale15}, as well as in space for Crab pulsar and nebula \citep{vadawale17}, and for a sample of GRBs detected in the first year of operation of CZTI \citep{chattopadhyay19}.   

In last one year, we have explored a number of new techniques in the spectral analysis for bright ON-axis sources like Crab and Cygnus X-1 (Chattopadhyay {\em et al.} (2021), under preparation). We also identified a number of possible improvements in the {\em AstroSat} mass model for better spectroscopic and polarimetry analysis for these sources. Implementation of these new techniques (listed below) will yield a significant improvement in the overall spectro-polarimetric 
sensitivity of GRBs detected by the CZTI. 

\begin{itemize}

\item { After the launch of {\em AstroSat}, $\sim$20 \% of the CZTI pixels were found to have electronic gains significantly lower than the ground calibrated gain values. Majority of these pixels now possesses gain around 2 $-$ 4 times lower than expected. However, the gain for these pixels is now stable since
launch i.e. the gain change was a one-time phenomenon whose origin remains
unknown. As a result of the lower gain, these pixels have a higher energy
threshold of $\sim$60 keV for X-ray photon detection but are also sensitive to
photons of much higher energies up to $\sim$800 keV. We refer to these pixels as
low-gain pixels, which were originally excluded from any scientific
analysis. However, after a careful and detailed analysis of the events
from these pixels, here we attempt to include these pixels to increase the
spectroscopic energy ranges for GRBs.}

\item From the detector plane histogram (DPH) images of the valid Compton scattered events, we further identify the noisy pixels giving rise to 2-pixel events. Filtering out this `Compton noise' which is otherwise not removed from the standard noise rejection algorithm. 
%further improves the signal to noise ratio in polarization data.
\end{itemize}

%The above new techniques in the polarization analysis are expected to improve the source polarization signature and in particular, the inclusion of low gain pixels not only enhances the polarimetric sensitivity of the instrument with the utilization of the full available collection area of CZTI but also allows to explore the polarization properties of the GRBs in the sub-MeV region extending up to 600 keV. This is particularly interesting because for some of the {\em AstroSat} detected GRBs e.g. GRB 171010A \citep{chand18b}, a hint of variation in polarization parameters was seen with respect to energy. Some of the emission models like subphotospheric dissipation model \citep{lundman14} predict such a change in polarization below and near the peak energy of the bursts ($E_{peak}$). The extended energy range, thus, will now allow us to explore such scenarios by attempting energy resolved polarimetry analysis for the bright {\em AstroSat} GRBs. 

The new techniques allow us to explore the capability of CZTI as a sub-MeV GRB spectrometer. 
In the standard CZTI analysis pipeline, the prompt emission spectroscopy of the bursts are limited only in 100 $-$ 200 keV whereas even with the 2-pixel Compton scattering events, the spectroscopy can only be extended up to $\sim$350 keV.
With the utilization of the low-gain pixels, the spectroscopy of the GRBs are now extended all the way up to $\sim$1 MeV. There are three different ways the CZTI instrument provides spectroscopic information for the GRBs $-$ (1) 1-pixel or single pixel events from CZT detectors in 100 $-$ 900 keV, (2) 2-pixel or Compton scattering events from CZT detectors in 100 $-$ 700 keV which are used to extract polarization information and (3) four CsI-Veto detectors below the CZTI sensitive in 100 $-$ 500 keV. We use the {\em AstroSat} mass model to generate the effective area as a function of energy and response matrix for each of these spectroscopic techniques and perform broadband spectral analysis along with {\em Fermi} and {\em Swift}$-$BAT data. 
Proper spectral fits and constraining the spectral parameters critically depend on the correct estimation of response matrix elements which are different for different GRB direction with respect to the satellite orientation. Although the mass model has been validated and tested in detail using imaging method (Mate {\em et al.,} 2020, this issue), spectroscopic analysis of the eleven GRBs which cover the full sky with respect to the {\em AstroSat} satellite indirectly tests the mass model further. This also helps in identifying the shortcomings in some parts of the mass model and quantifying those from the spectral fits. CZTI sub-MeV spectroscopy is particularly valuable for those GRBs which are detected by {\em AstroSat} and Niel Gehrels {\it Swift} BAT but not by {\it Fermi}, as it allows us to constrain the spectral parameters including the peak energy in the energy range (15 $-$ 900 keV), which otherwise generally is  not possible because of the narrow energy range of BAT. 
%There are multiple such GRBs for which simultaneous spectro-polarimetry analysis might now be feasible with CZTI. Even though for some of the GRBs polarization study might not be possible because of either low number of Compton events or viewing angle greater than 60$^\circ$ (for $\theta = >60^\circ$, the polarization analyzing power of CZTI is low), the spectroscopic analysis can still be availed from single, Compton and veto detected events.      
In this article, we explore CZTI as a sub-MeV spectroscopy and 
report the spectroscopic 
%and polarization 
measurements for the eleven bright GRBs detected in the first year of CZTI operation with the implementation of these new developments 
for the entire burst time interval which is obtained using the Bayesian block technique on the GRB single event data. 
The new techniques and the burst selection methods are described in section \ref{method} In section \ref{spec}, we describe the spectroscopy methods followed by broadband spectral analysis in section \ref{result_spec}. %In section \ref{pol} and \ref{result_pol}, the methodology for sub-MeV polarimetry and polarization results are described respectively. 
While this article primarily outlines the methodologies of sub-MeV spectroscopy for GRBs, we plan to apply the new techniques to a sample of $\sim$20 bright GRBs detected in last five years of operation of {\em AstroSat}. 
%The new techniques can be the key to perform sensitive spectro-polarimetry study of this large sample and will help in developing a better understanding of the GRB prompt emission.     
\section{New Techniques in the Spectrum Analysis}
\label{method}

In this section, we describe the new techniques implemented in the spectral
%and polarimetry 
analysis compared to that discussed in \citep{chattopadhyay19}. 
In the previous polarimetry reports on {\em AstroSat} GRB by \citet{chattopadhyay19,chand18a,chand18b,vidushi19}, we utilized only 75$-$80 \% of the CZTI collecting area consisting only the `spectroscopically good' pixels. A fraction of CZTI pixels are found to have lower gains (gain value 3 $-$ 4 times lower than that of the normal or good pixels) and therefore are sensitive at higher energies. We performed a detailed characterization of these pixels and utilize them in the spectrum analysis extending the overall spectroscopic energy range to $\sim$1 MeV. 
%and Compton polarimetry to 600 keV. We further improved the signal to noise ratio in the polarization sensitive Compton events by introducing a new algorithm to filter out the noisy 2-pixel events (within the CZTI coincidence window). 
We discuss these new developments below along with our new strategy of selecting the burst interval for 
%polarization and 
spectroscopic analysis of the GRBs. 
%This is particularly important given the fact that polarization of the GRBs is extremely sensitive to the selection of the region as demonstrated by \citet{vidushi19} where polarization angle was shown to change within a single broad pulse of GRB 160821A and is suspected to be a generic feature for the GRBs. Full burst polarization analysis is therefore expected to result in null polarization as seen in case of POLAR GRBs \citep{zhang19} whereas for the same bursts, a high polarization can be achieved from careful selection of time and energy range as seen in case of {\em AstroSat} GRBs \citep{chattopadhyay19}.   

\subsection{Characterization of low-gain pixels}
\begin{figure}
\centering
\includegraphics[width=1.1\hsize]{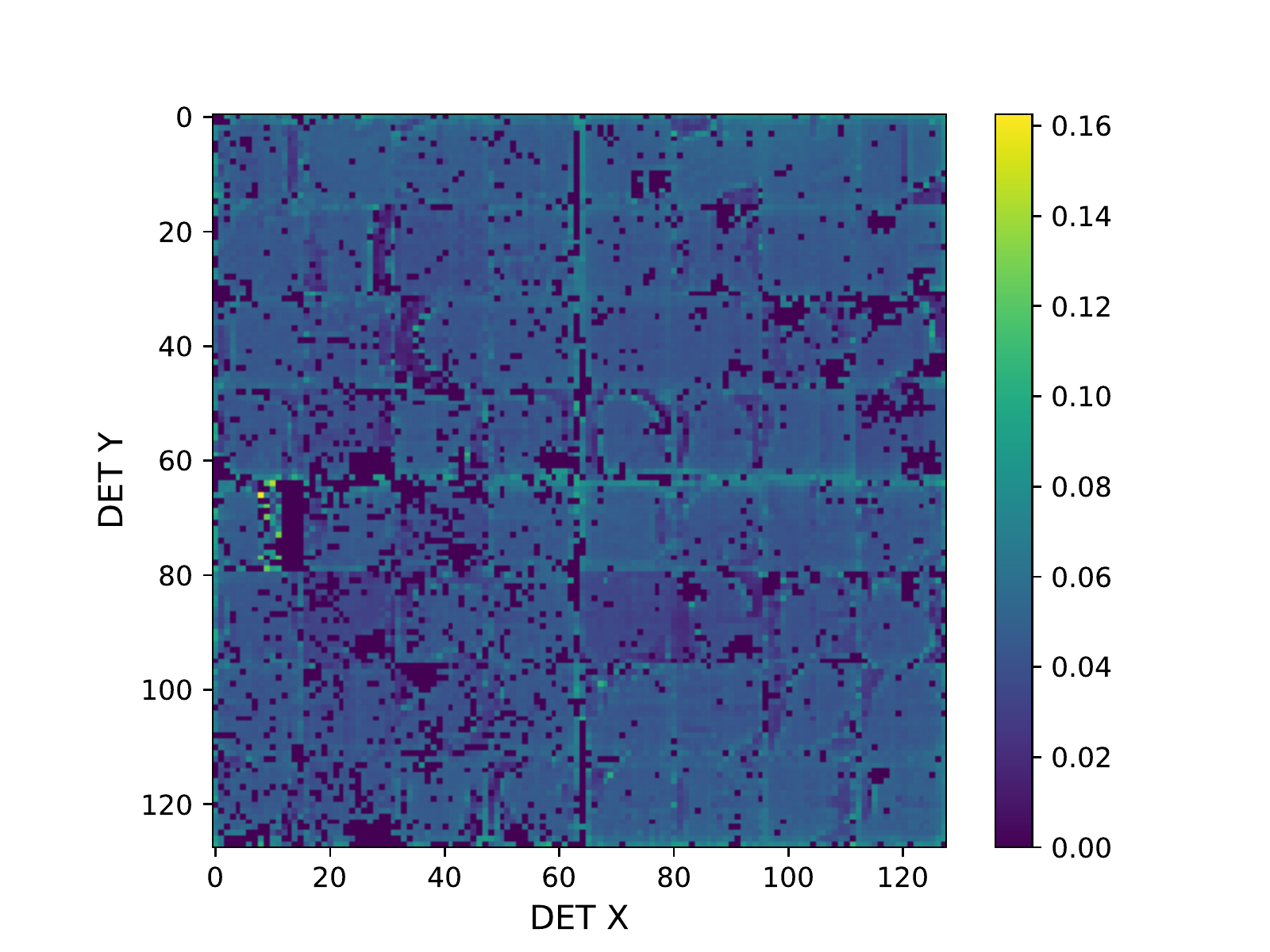}
\caption{The detector plane histogram (DPH) of all the CZT detector quadrants for the obsID:~9000000618 (data from 2016-August). The lower count rates detected in a fraction of the pixels are seen as patches in the DPH which is because of the relatively higher gain values of the pixels. The color bars indicate the count rate.}
\label{DPH_banana}
\end{figure}

From the detector plane histogram (DPH) of the on-board data, it was seen that the count rate in some spatially clustered pixels were significantly lower compared to the mean count rate (see Figure \ref{DPH_banana}). 
Even though most of the pixels are found in clusters, there are instances of isolated pixels as well. These pixels also did not show the alpha tagged line at 60 keV from the on board calibration source $^{241}$Am, indicating that the gain has shifted at least by a factor of two or three (hereafter we refer to these pixels as low-gain pixels). The reason for the shift is unknown, however since no shift was seen in the laboratory measurements during calibration and appeared right after the launch, mechanical stress during the launch is thought to be one of the possibilities. 

From the light curve analysis from the low-gain pixels with different time bins, we found that the count rates detected by these pixels are of Poissonian nature and therefore the detected events are not spurious pixel noise and could be real X-ray events.
\begin{figure}[h]
    \centering
    \includegraphics[width=1.05\linewidth]{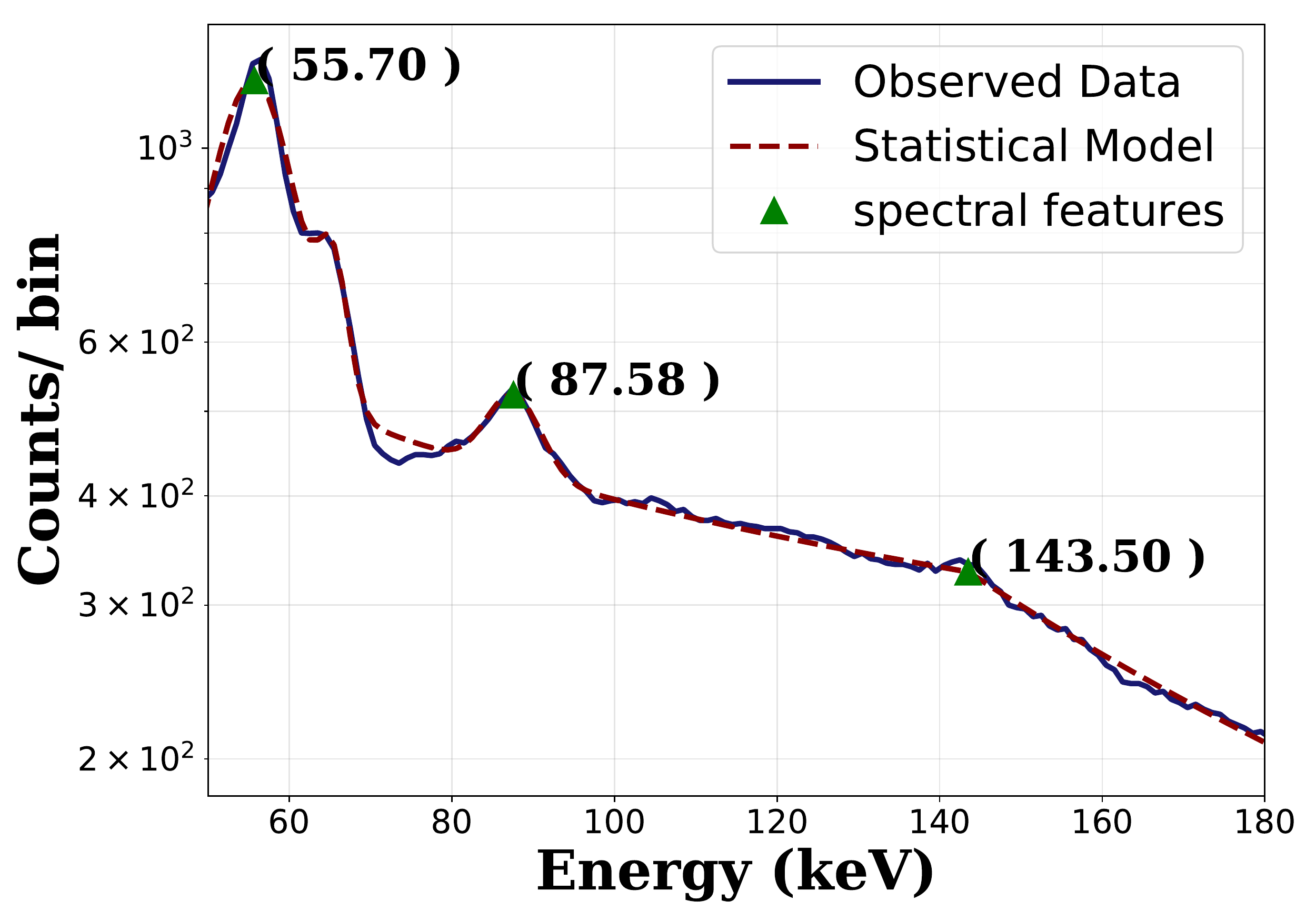}
    \caption{
    Continuum spectra from the spectroscopically good pixels (in blue) for one of the detector modules (data taken from July 2016). The spectrum is fitted with an empirical model (red) consisting of three Gaussian (1. Tantalum line at 54 keV, 2. a bump structure near the Tantalum line and 3. an arbitrary line around 90 keV which is most likely a proton induced background feature \citep{odaka18}) and a broken power law (break energy around 140 keV which denotes the onset of falling detection efficiency for the 5 mm thick CZT detectors). This template has been used to compare the spectra of the low-gain pixels to estimate their gains (see text for more details).}
    \label{model_good_pixel}
\end{figure} 
\begin{figure}[h]
    \centering
    \includegraphics[width=1\linewidth]{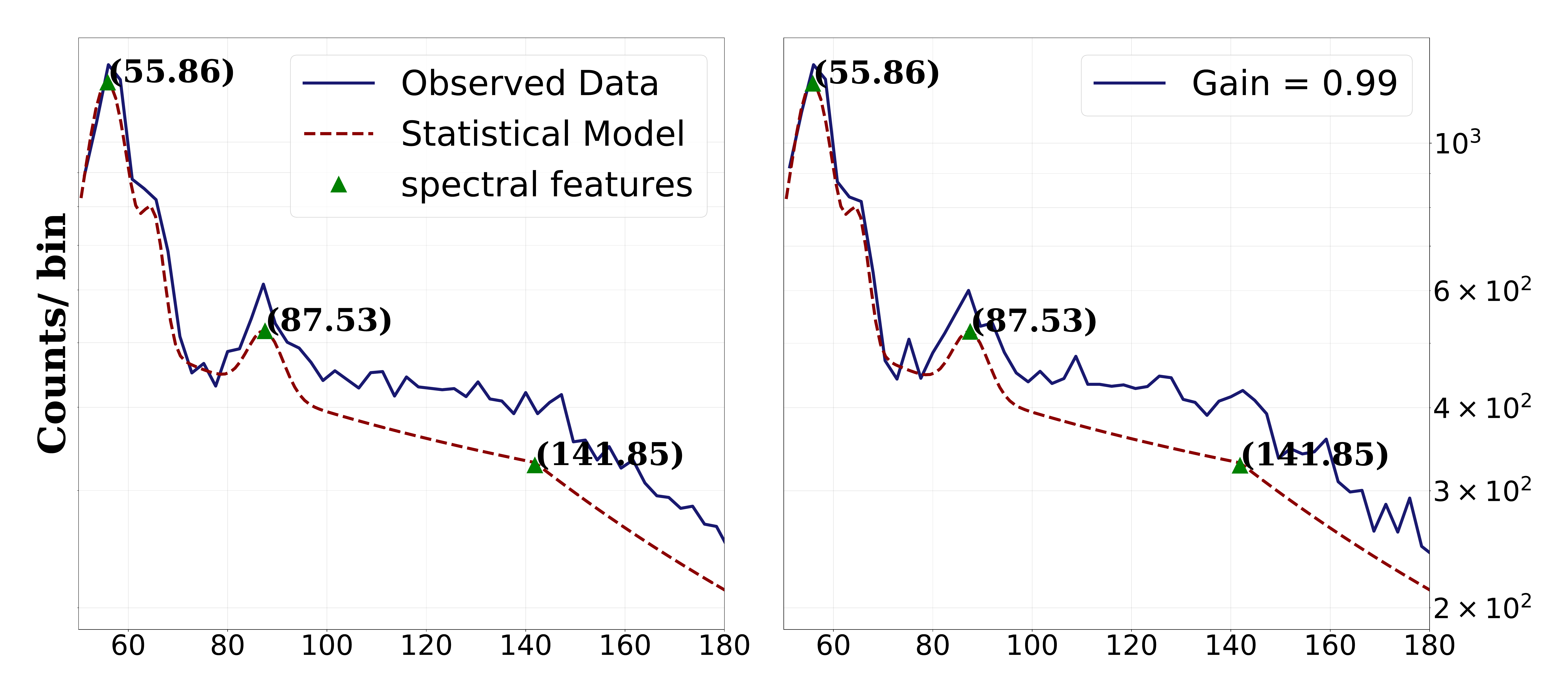}
    \includegraphics[width=1\linewidth]{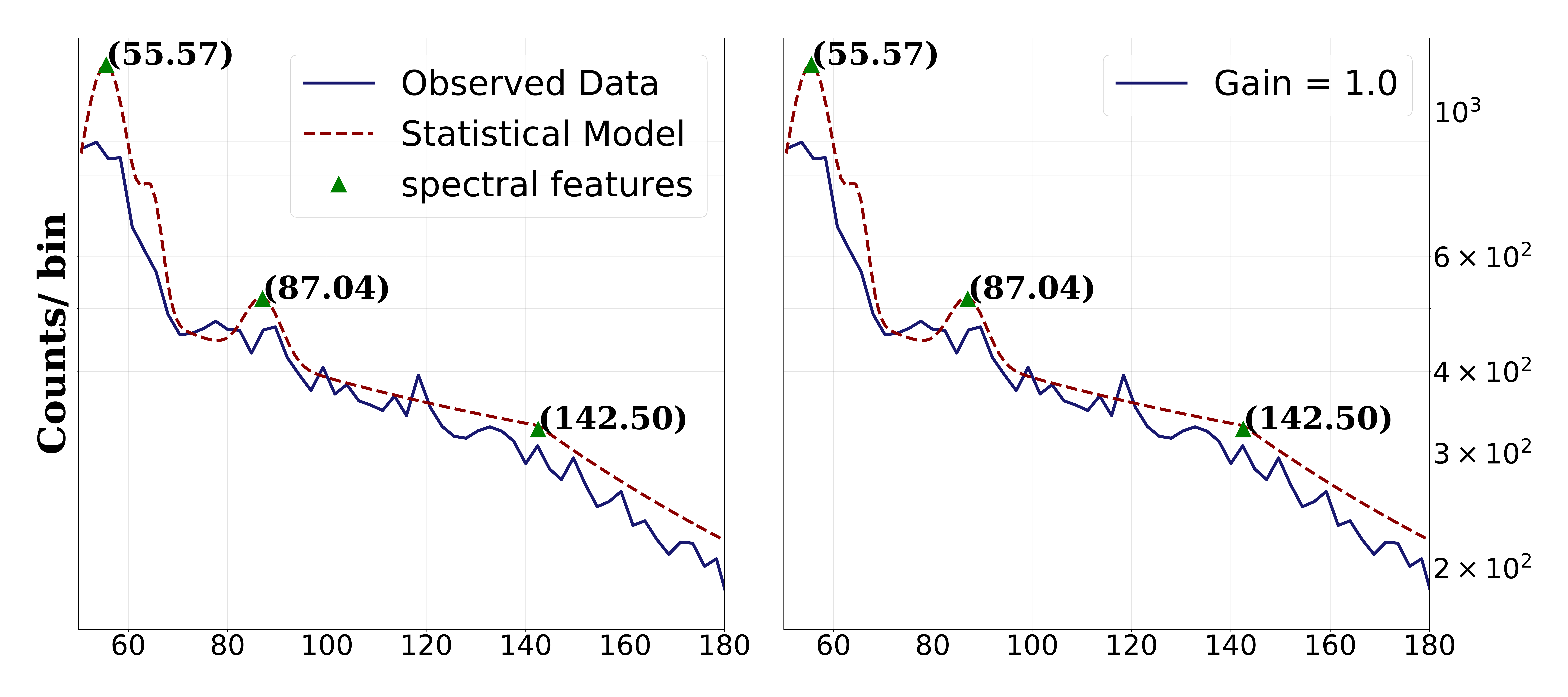}
    \includegraphics[width=1\linewidth]{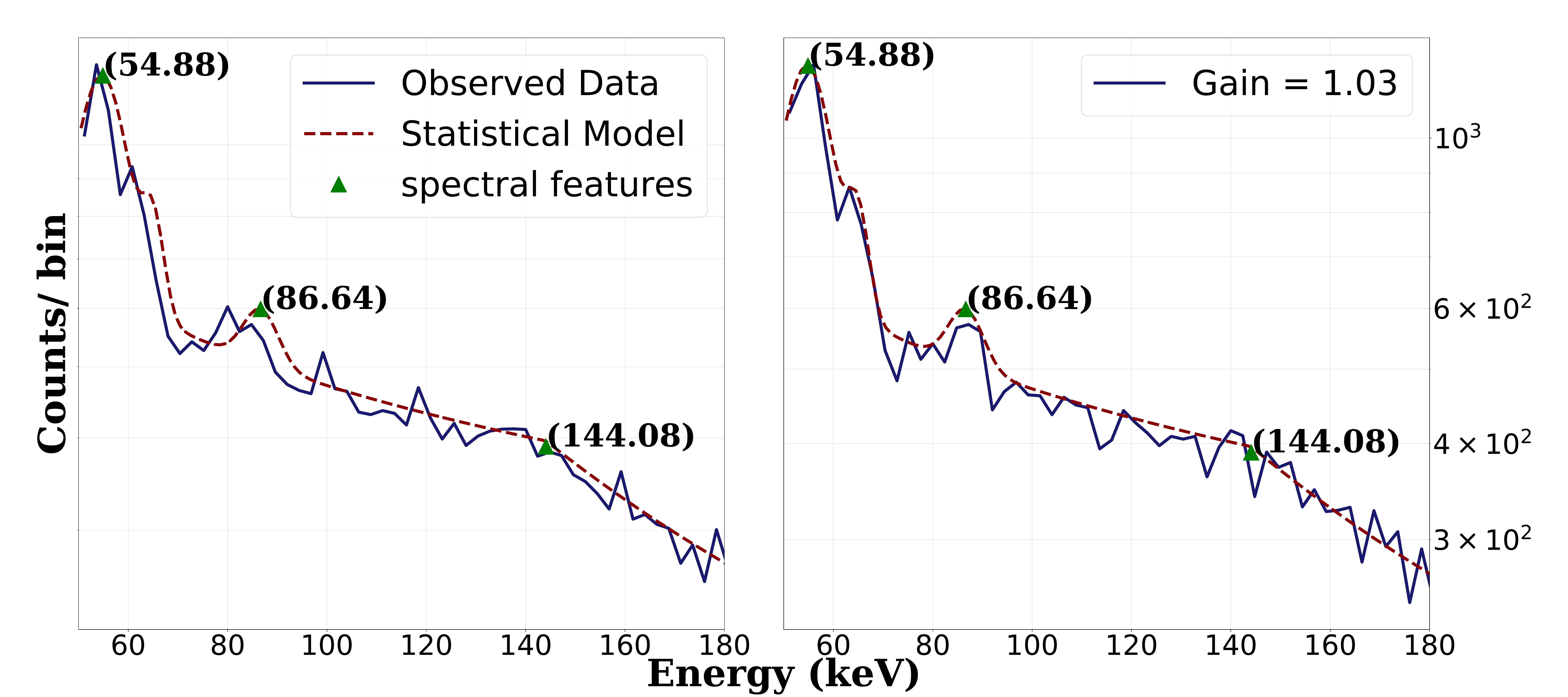}
    \caption{Spectra of three Low-Gain pixels of type I before and after applying gain correction given in left and right panel respectively. top- pixel number 161 from module 4, middle- pixel number 248 from module 5, and bottom- pixel number 45 from module 13. The red lines are the empirical models used to compare the overlapping region of 45 $-$ 180 keV of the low gain pixels. After comparison, the fitted gain shift factors (a multiplication factor to the ground calibrated gain) are found to be between 0.8 and 1.5 for the type I low gain pixels.}
    \label{lg_type1}
\end{figure}
Because these pixels consist almost 20 \% of the CZTI active area, we explored the possibility of characterizing the pixels in detail. In absence of any mono-energetic line at higher energies to calculate the correct gain for the low-gain pixels, we compare the overlapping region of the continuum spectrum of these pixels and the spectroscopically good pixels.
For that purpose we first fitted the good pixel spectra for each of the 64 detector modules with an empirical model in 45 $-$ 180 keV range using three Gaussian (Tantalum k$_\alpha$ line at 54 keV, a bump structure around 65 keV source of which is unknown and a Tellurium activation line at around 88 keV) and a broken power law with a break energy around 140 keV as shown in Figure \ref{model_good_pixel}. 
The break energy denotes the onset of falling detection efficiency for a 5 mm thick CZT detector and therefore can also be used to calibrate the low-gain pixels along with the continuum comparison.  
 The strong line around 88 keV seen in the spectra is supposed to originate from high energy particle induced Tellurium activation ($^{127m}Te$ with half life of 9.17$\times$10$^6$ seconds \citep{odaka18}). We also see a hint of a line feature around 145 keV which could also be from activated Tellurium ($^{125m}Te$ with half life of 4.96$\times$10$^6$ seconds). Because of the large half life of the isotopes, we see the lines even far from the SAA region where the activation is supposed to take place \citep{odaka18}.
Since the number of good pixels vary in each module, the count rate was normalised by the total number of good pixels in that module. 

\begin{figure}[h]
    \centering
    \includegraphics[width=1\linewidth]{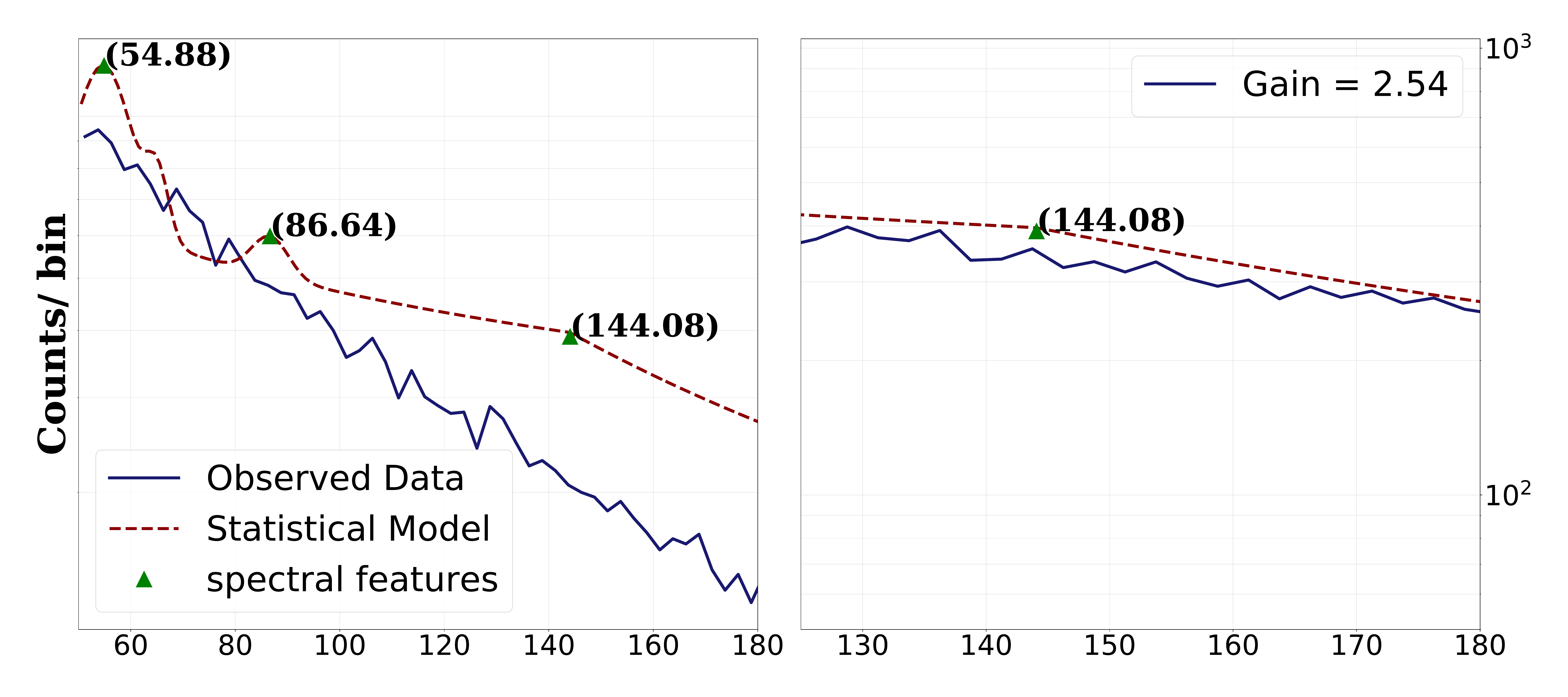}
    \includegraphics[width=1\linewidth]{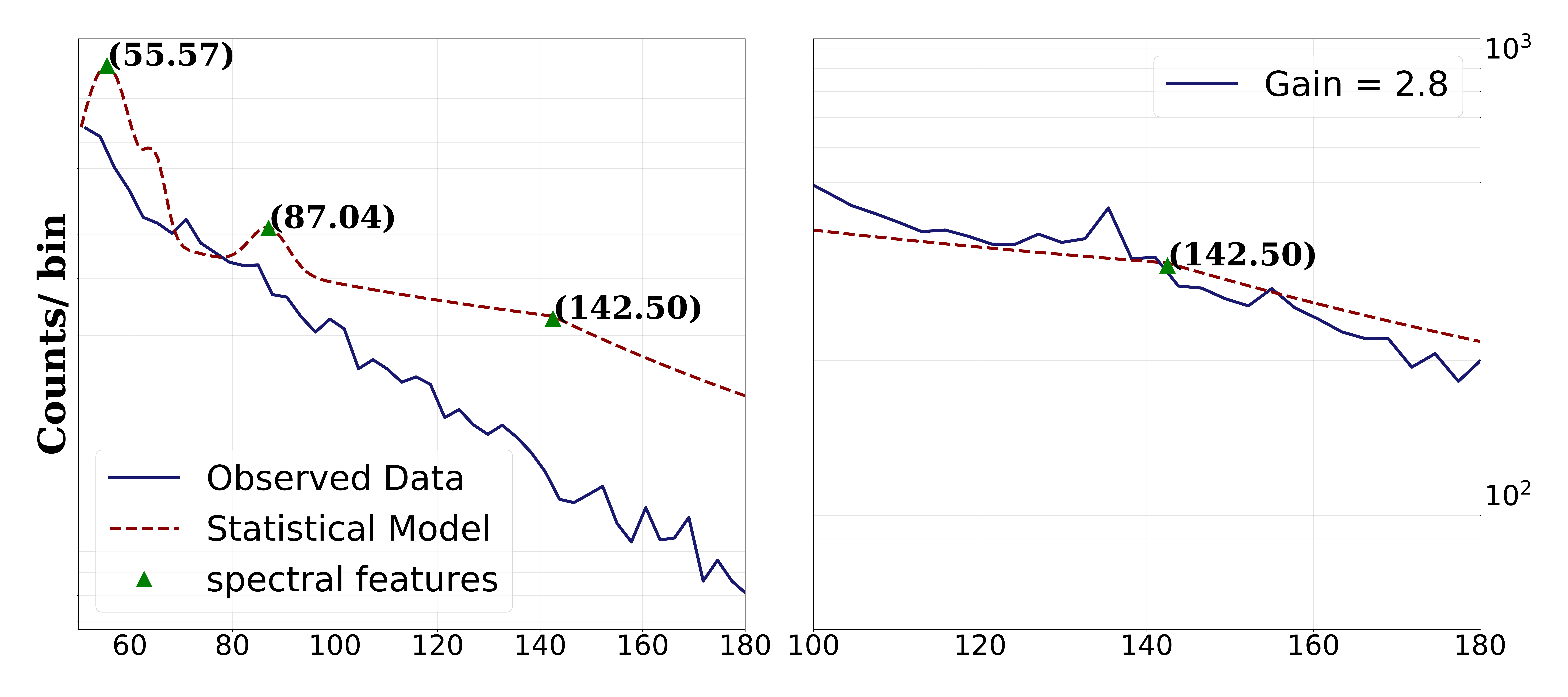}
    \includegraphics[width=1\linewidth]{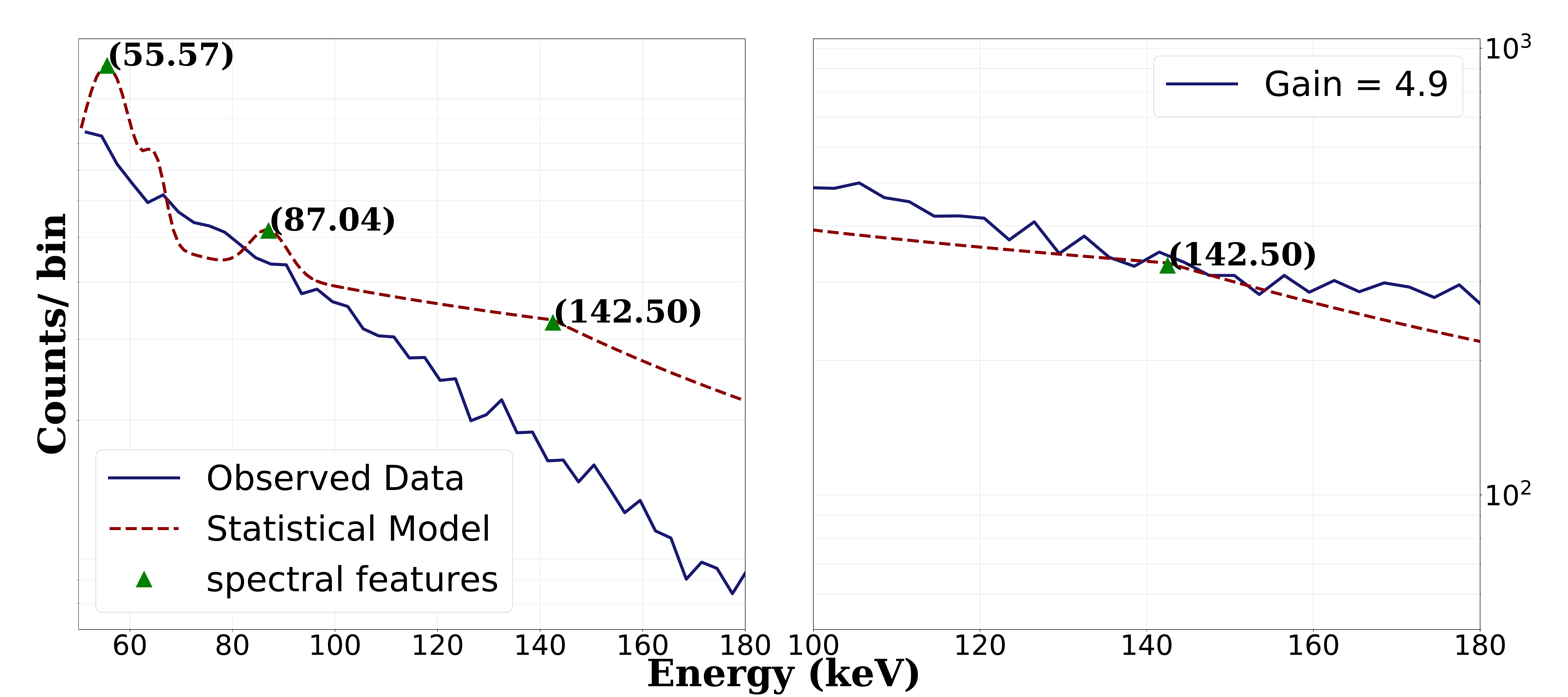}
    \caption{ Same as Figure \ref{lg_type1} but for three type II Low-Gain pixels (top- pixel number 223 from module 13, middle- pixel number 0 from module 5, and bottom- pixel number 1 from module 5). For type II low-gain pixels, we found the fitted gain shift factors between 1.5 and 4.}
    \label{lg_type2}
\end{figure}

In order to have sufficient statistics in the spectra of both good and low-gain pixels, we took a long one month data ($\sim$1 million seconds of exposure). The South Atlantic Anamoly (SAA) regions are normally excluded in the raw data itself based on the count rates from an on board charge particle monitor. In addition to that, a time interval of 500 seconds was ignored before and after the already excluded SAA region in order to filter out the high particle background regions. The fitted module wise good pixel models were then used to compare the spectra of each of the low-gain pixels in 100 $-$ 180 keV range and reduced $\chi^2$ values were calculated by varying a multiplication factor to the ground calibrated gain of the low-gain pixels in the range of 0.8 $-$ 5.0 at an interval of 0.01. We call this multiplication factor to the gain as `gain shift factor'. Based on the fitting results, we classified the low-gain pixels into three subcategories $-$ 
 \begin{enumerate}
        \item Low-Gain pixels type I: These pixels were found to have gain shift factor between 0.8 and 1.5 and are seen to have spectral features like the Tantalum line, the 90 keV background and spectral break at around 140 keV as also seen in the good pixel spectra. Left column of Figure \ref{lg_type1} shows the comparison of the spectra (shown in blue) with the good pixel model (shown in red) for three pixels whereas the right panel shows the same comparison after correcting for the gains of the pixels. These pixels were previously identified as low-gain
        pixels in CALDB. Since we find them to have gains very close their ground calibration values, we plan to calibrate them with the on board calibration source for further validation, details of which will be presented elsewhere (Mithun {\em et al.,} in prep.). 
        \item Low-Gain pixels type II: These pixels were found to have relatively higher gain shift values between 1.5 and 5.0. Comparison of the continuum spectra in 100 to 180 keV before and after gain correction is shown in Figure \ref{lg_type2}. 
        \item Low-Gain pixels type III: For a fraction of pixels we could not get satisfactory fit in the common 100 $-$ 180 keV range even for the maximum gain shift values. These pixels are ignored from any further analysis.
        
\end{enumerate}

We carried out the analysis for each of the five years of CZTI data (normally June/July of each year depending on the data availability) to check for repeatability or any possible time evolution in the obtained gain values for the type I and type II low-gain pixels. 
We use the gain list from the year of detection of a given GRB (note for this paper we use gain list of the year 2016 as all the GRBs analyzed here are detected in 2016). In future, we plan to characterize the low-gain pixels (particularly the type II pixels) using various particle-induced radioactivation background lines \citep{odaka18} for further verification. 
It might be possible to further verify the gain values by looking at the Crab pulse profile from these pixels and calculate the ratio of pulsed fractions in two pulses as they are known to be energy dependent. 
We also plan to validate the gain values of the type I pixels by investigating the alpha tagged spectra from these pixels.  
    
\begin{figure}[h]
\centering
\includegraphics[width=1.1\hsize]{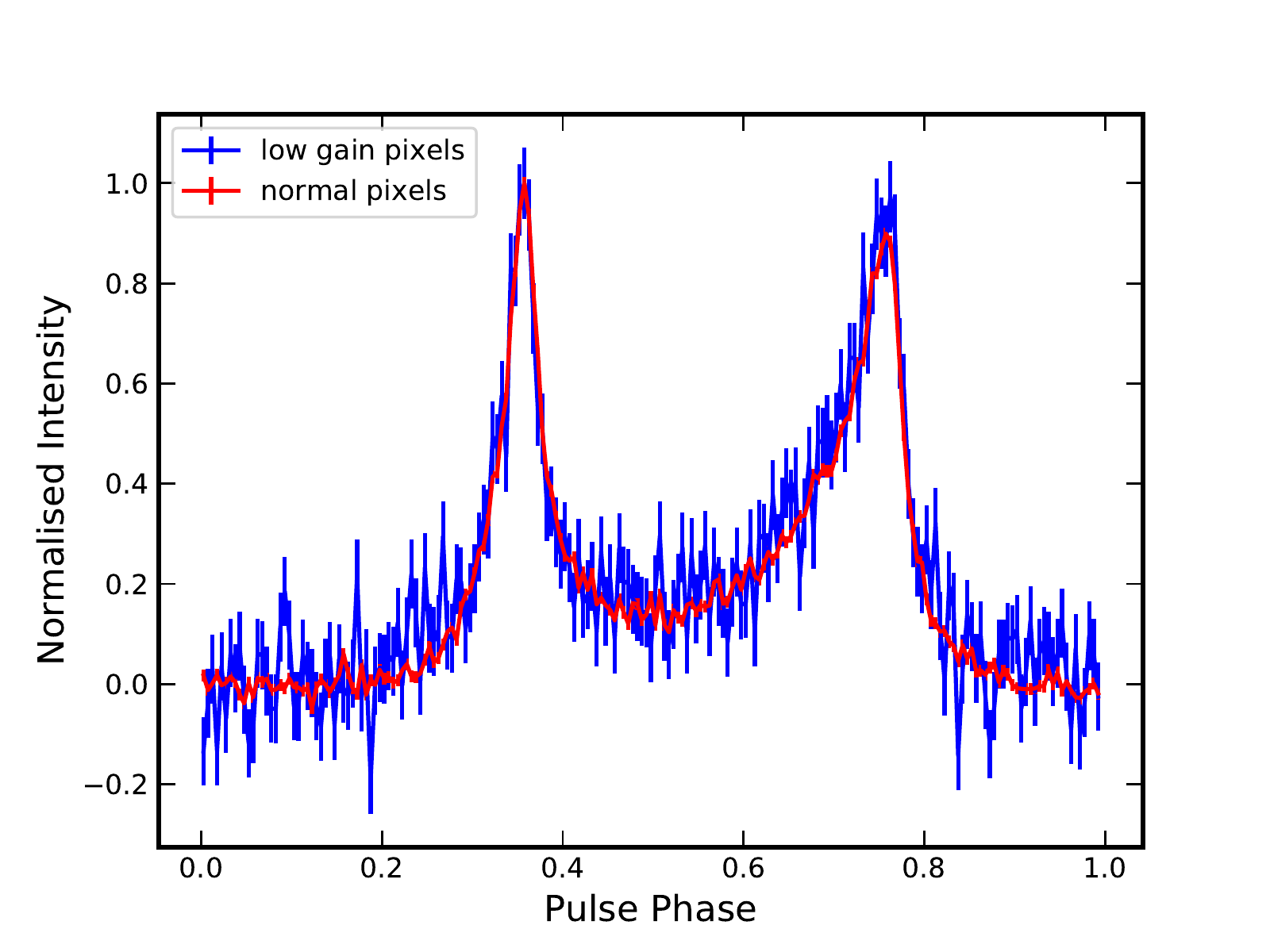}
\caption{The pulse profile of Crab pulsar in low-gain pixels (blue) of all the CZT quadrants after gain correction. For comparison the pulse profile in the spectroscopically good pixels are plotted against it (red). }
\label{crab_banana}
\end{figure}

\begin{figure}
\centering
\includegraphics[width=1.1\hsize]{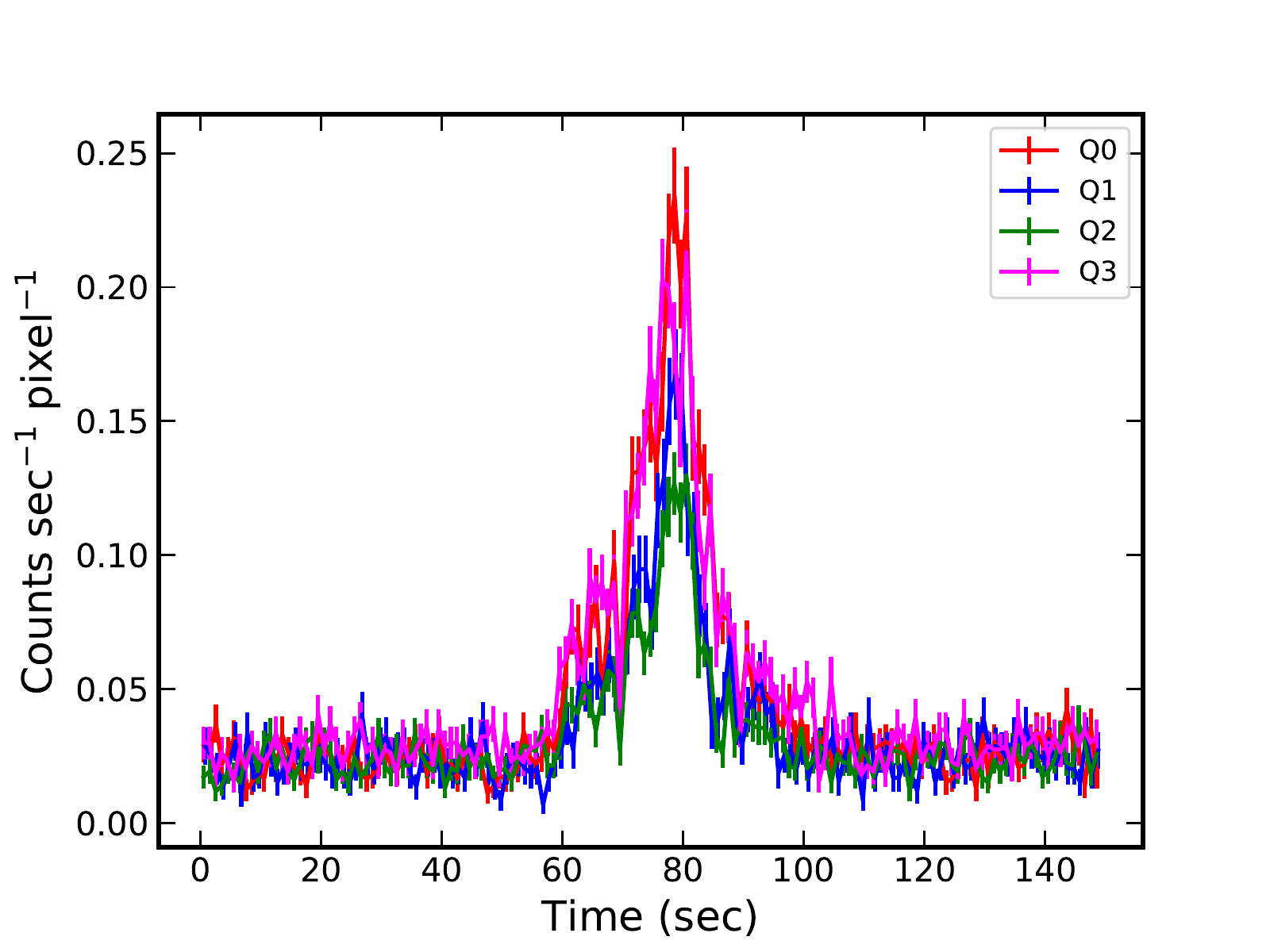}
\caption{Light curve of GRB 160821A in the low-gain pixels with corrected gains. Different colors represent the four different CZTI quadrants as indicated inside the plot. The time axis is plotted from {\em AstroSat} time 209507728 seconds (marked as zero). Each CZTI quadrant is shadowed by different degree for each GRB according to its location with respect to the spacecraft giving rise to unequal flux levels in different CZTI quadrants.}
\label{grb160821A_banana}
\end{figure}

In order to boost the confidence in the use of low-gain pixels, we attempted to reconstruct the Crab pulse profile using these pixels after gain correction.
Figure \ref{crab_banana} shows the pulse profile of Crab pulsar in low-gain pixels from all the four CZTI quadrants during a $\sim$78 ks observation on 14$^{th}$ Jan 2017. { This further verifies that the events from these pixels are genuine X-ray events and not random noise.} 
We used the crab ephemeris at MJD 57769.0 from \citet{crab_Lyne1999JodrellBC}. The events are folded from {\em AstroSat} time of 222220803.426 seconds. The background is subtracted by the counts in the off pulse region and the pulse profile is normalised by the maximum peak counts for the purpose of visualization. We could also detect the GRBs in these pixels as shown in  Figure \ref{grb160821A_banana} for GRB 160821A.
Since the number of low-gain pixels vary in different quadrants, the count rate is normalized by the total number of low-gain pixels in a quadrant. Detection of astrophysical sources in low-gain pixels therefore presents a strong case in using them for future spectral analysis. 

\subsection{Compton Noise}
\begin{figure}[h]
\centering
\includegraphics[width=1.1\hsize]{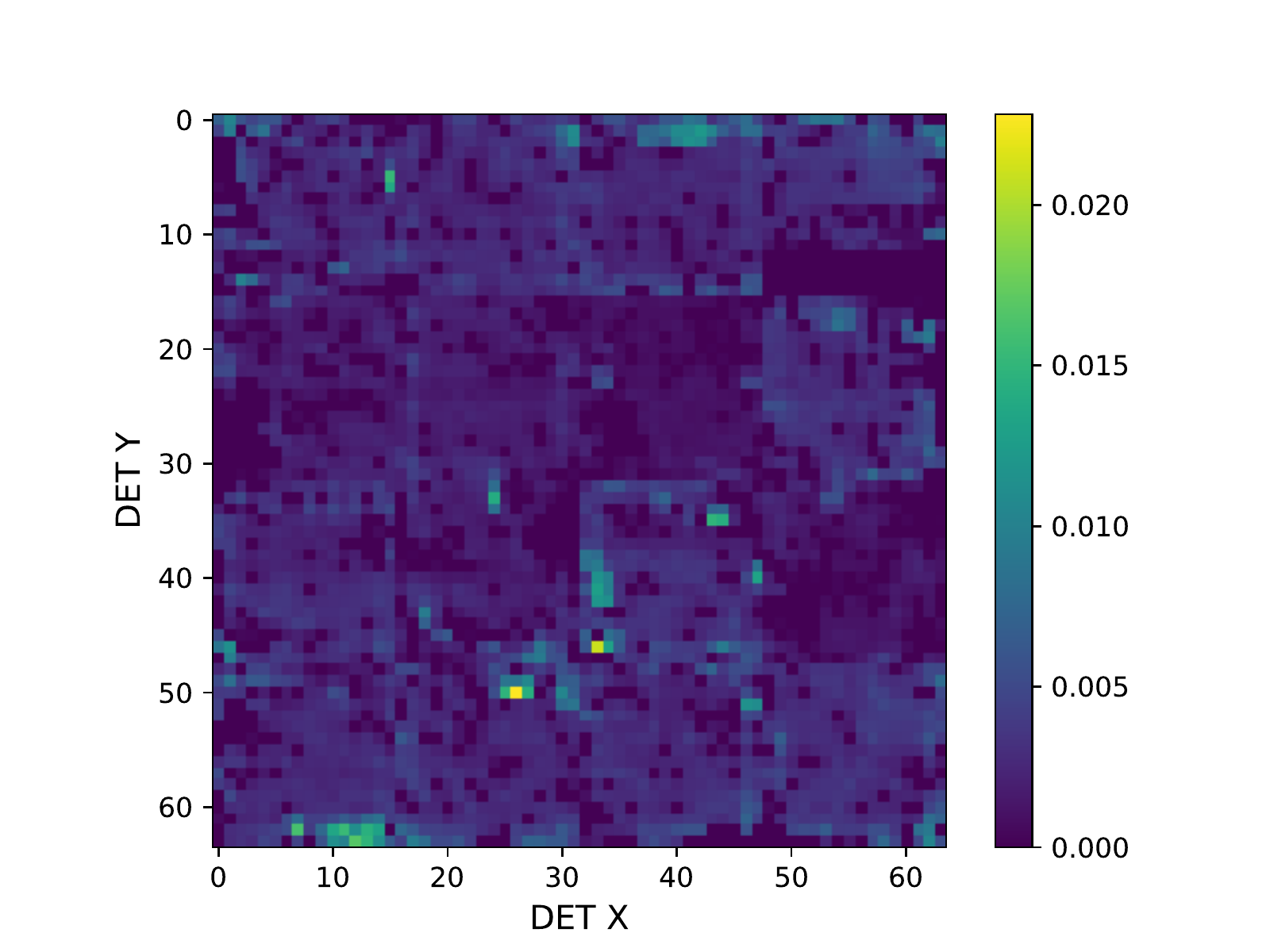}
\caption{Detector plane histogram of the neighbouring 2-pixel Compton events for the 3$^{rd}$ CZTI quadrant. The plotted data belongs to  obsID:~9000000618 (data from 2016-August). The colorbar indicates count rate. { The brighter spots in the image correspond to the Compton noisy events arising from noisy neighboring pixels. These events are removed from further analysis.}}
\label{comp_noise_DPH}
\end{figure}

%Hard X-ray polarimetry using CZTI is sensitive to pixel noises. 
{ CZTI has already been demonstrated as a sensitive ON-axis and GRB polarimeter in 100$-$350 keV in \citet{vadawale17} and \citet{chattopadhyay19} respectively, where the Compton scattered events are used to generate the azimuthal angle histogram. The same Compton events can be used in spectroscopy of the GRBs.  
These events are selected through strict Compton kinematics criteria $-$
\begin{itemize}
    \item identify the adjacent 2-pixel events from 20 $\mu$s coincidence window both from the spectroscopically good pixels and low-gain pixels,
   \item impose criteria of ratio of the energies deposited in two pixels between 1 and 6 in order to filter out the noisy chance events. This is motivated by the fact that in a true Compton scattering event, the electron recoil energy deposited in one of the pixels, is much lower than the scattered photon energy deposited in the other pixel.
\end{itemize}
In spite of the strict selection criteria, there is still a significant amount of overlapping noise events.} Neighbouring pixels can flicker at time scales lower than the coincidence time window of 20 $\mu$s causing some of these events to permeate into the Compton event selection and thus causing instrumental artifacts in the modulation curve. A DPH showing outliers in 2-pixel events is shown in Figure \ref{comp_noise_DPH}. 
%These noise events can become crucial if they exist in the corner pixels (corresponding to azimuthal scattering angles of 45, 135, 225, 315$^\circ$, see section \ref{pol}) as the noise gets amplified while correcting for the pixel geometry.
These events can be identified as outliers from the DPH of neighbouring 2-pixel events and can be removed from further analysis. Threshold for an outlier is kept at four sigma and three sigma from the mean for normal and low-gain pixels respectively. Due to the difference in count rate between the side and corner pixel double events, we identify the noisy pixels in the side and corner pixels separately. When a pixel is identified as noisy, no events from that pixel is considered for further analysis.
%and polarization. 
Further details on the Compton noise analysis can be found in Ratheesh {\em et al.,} 2020, this issue.

\subsection{Selection of burst interval}
\label{interval}

\begin{figure}
\centering
\includegraphics[width=.49\linewidth]{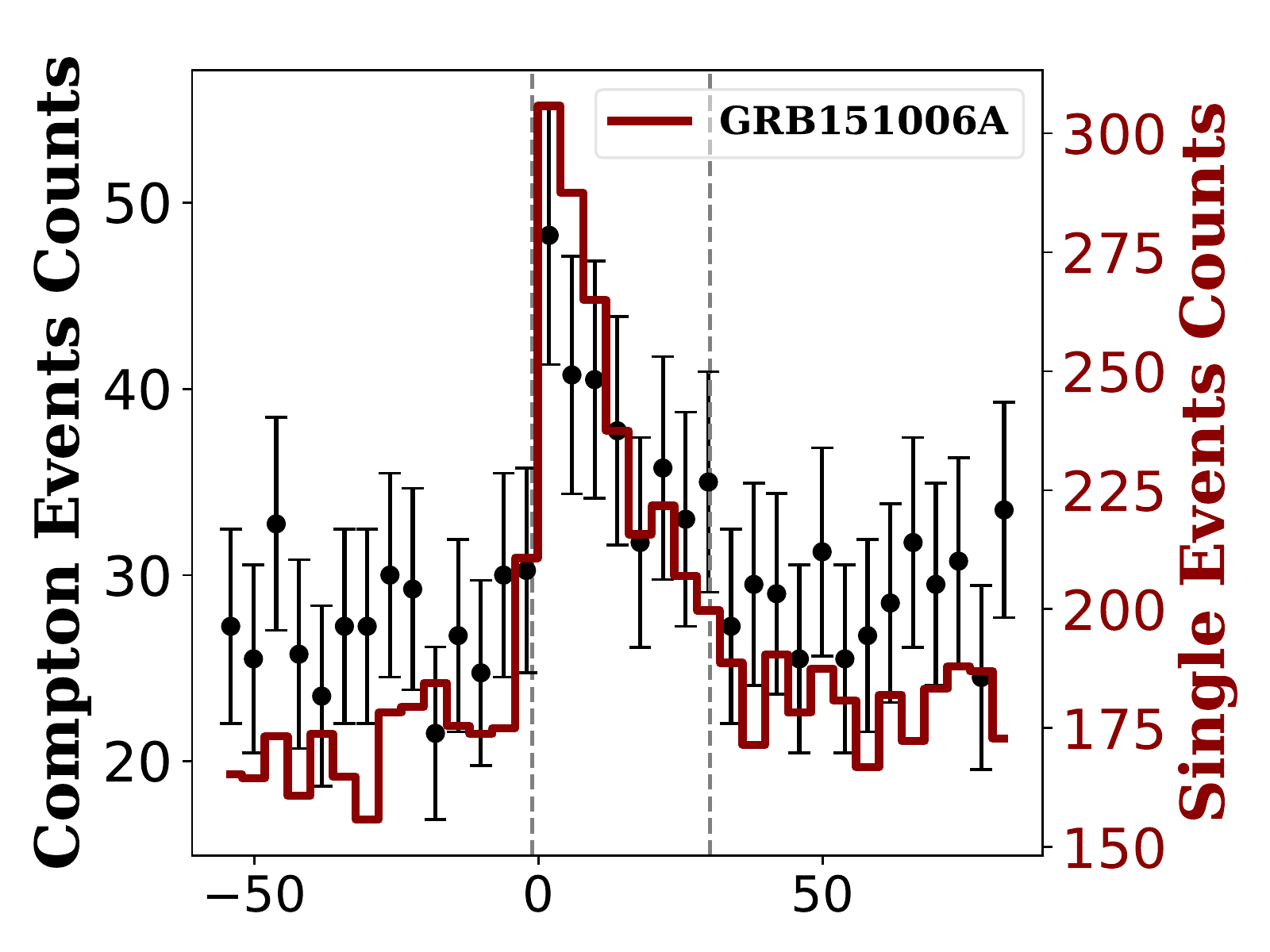}
\includegraphics[width=.49\linewidth]{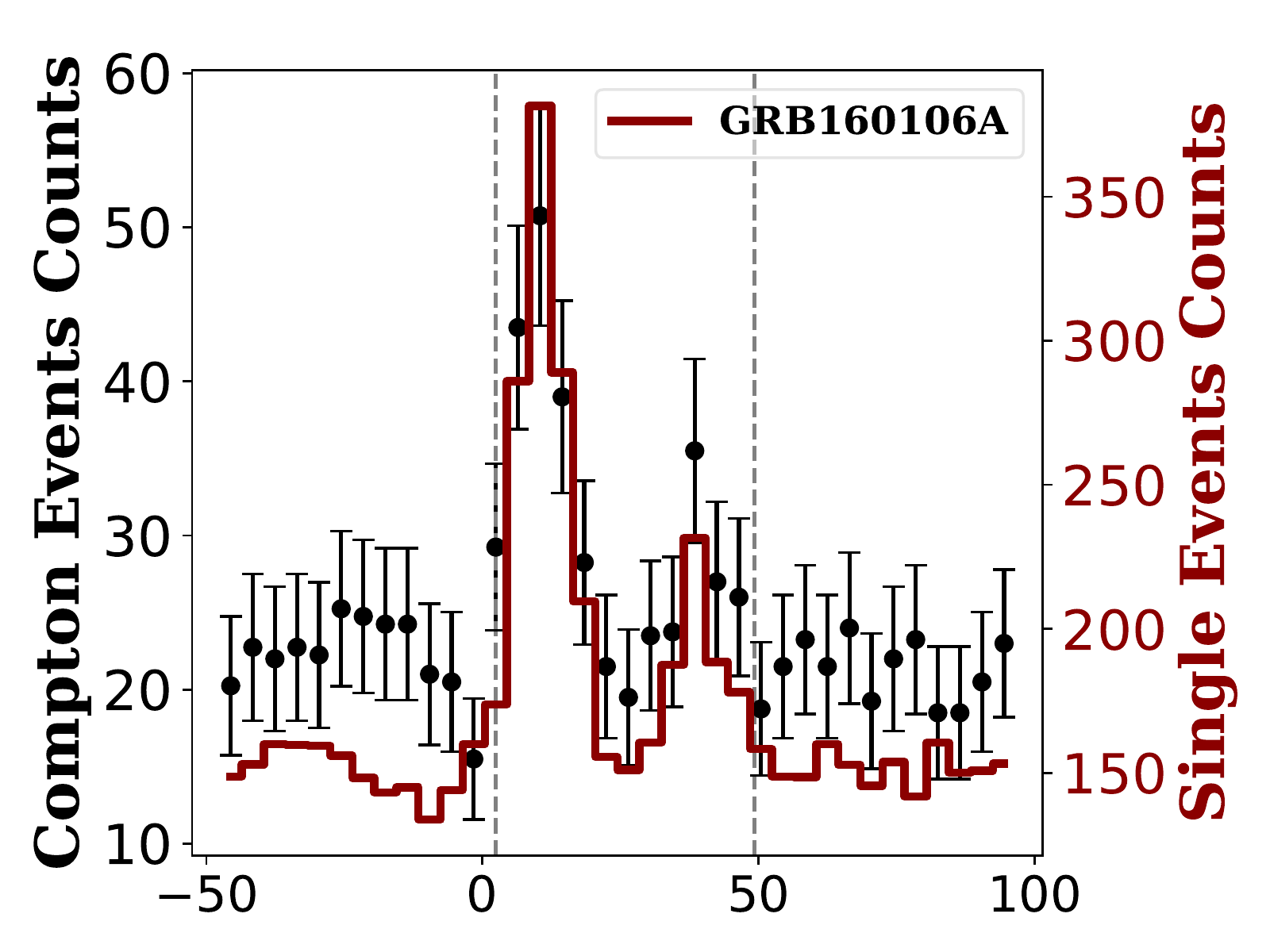}
\includegraphics[width=.49\linewidth]{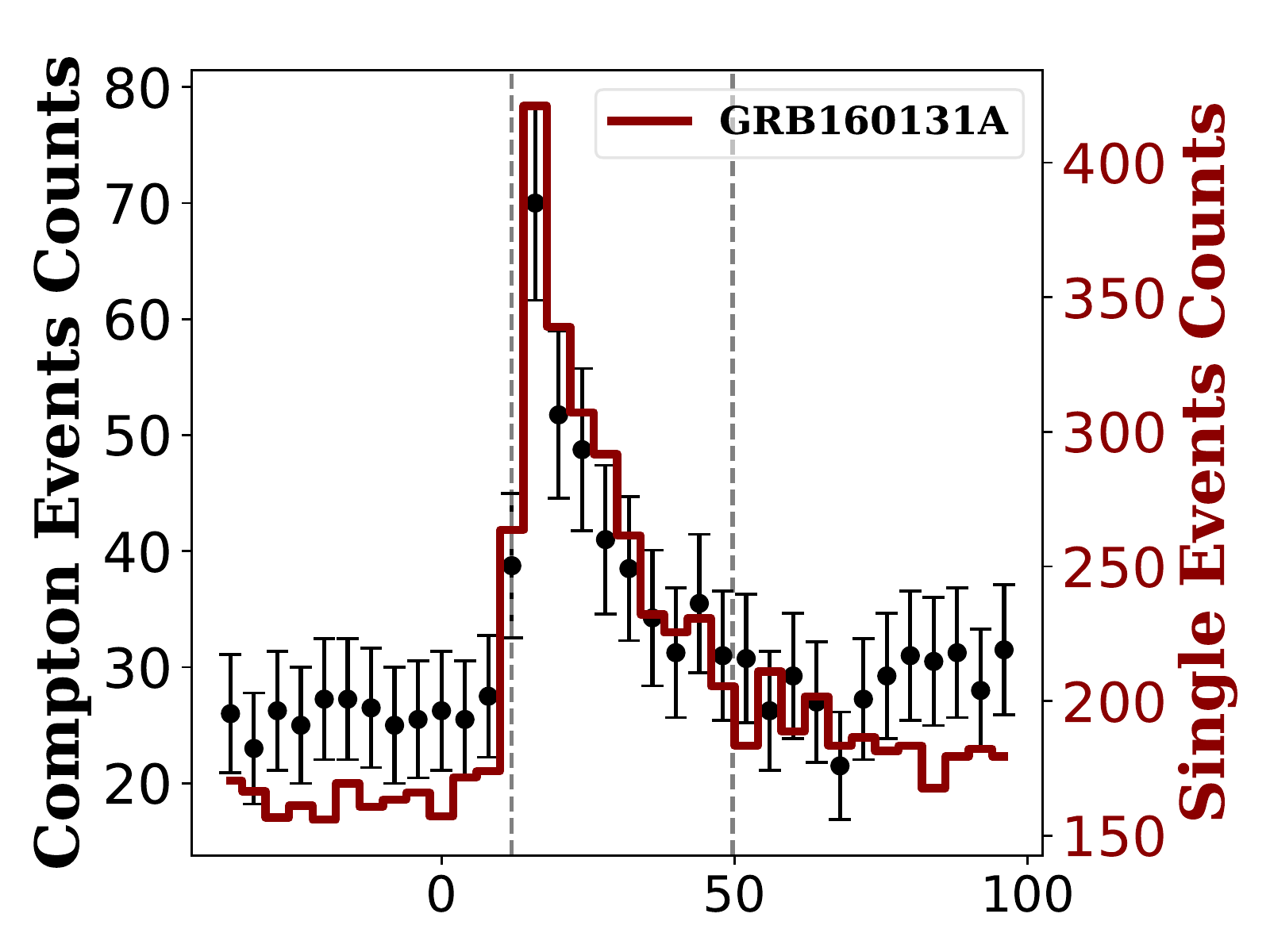}
\includegraphics[width=.49\linewidth]{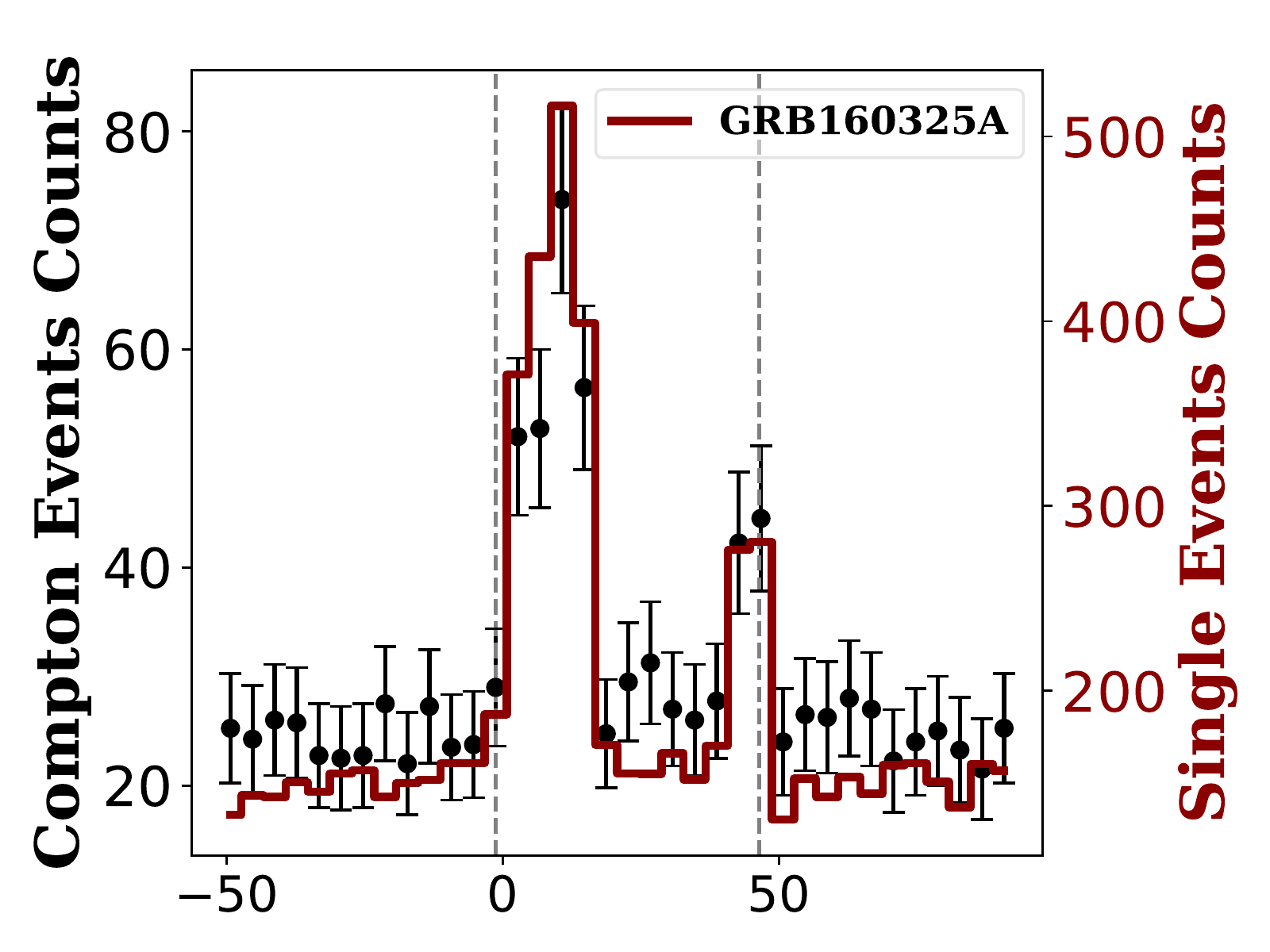}
\includegraphics[width=.49\linewidth]{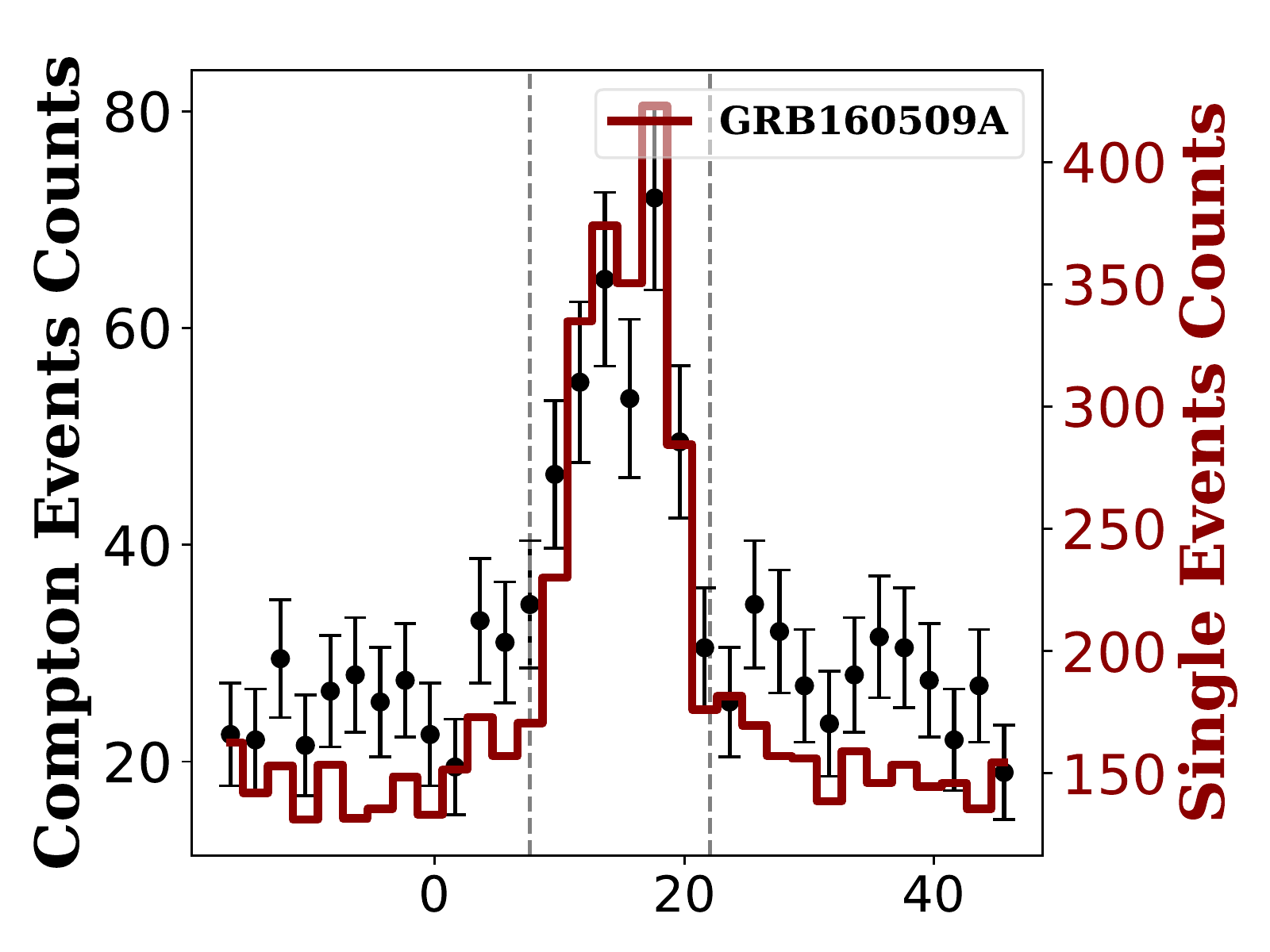}
\includegraphics[width=.49\linewidth]{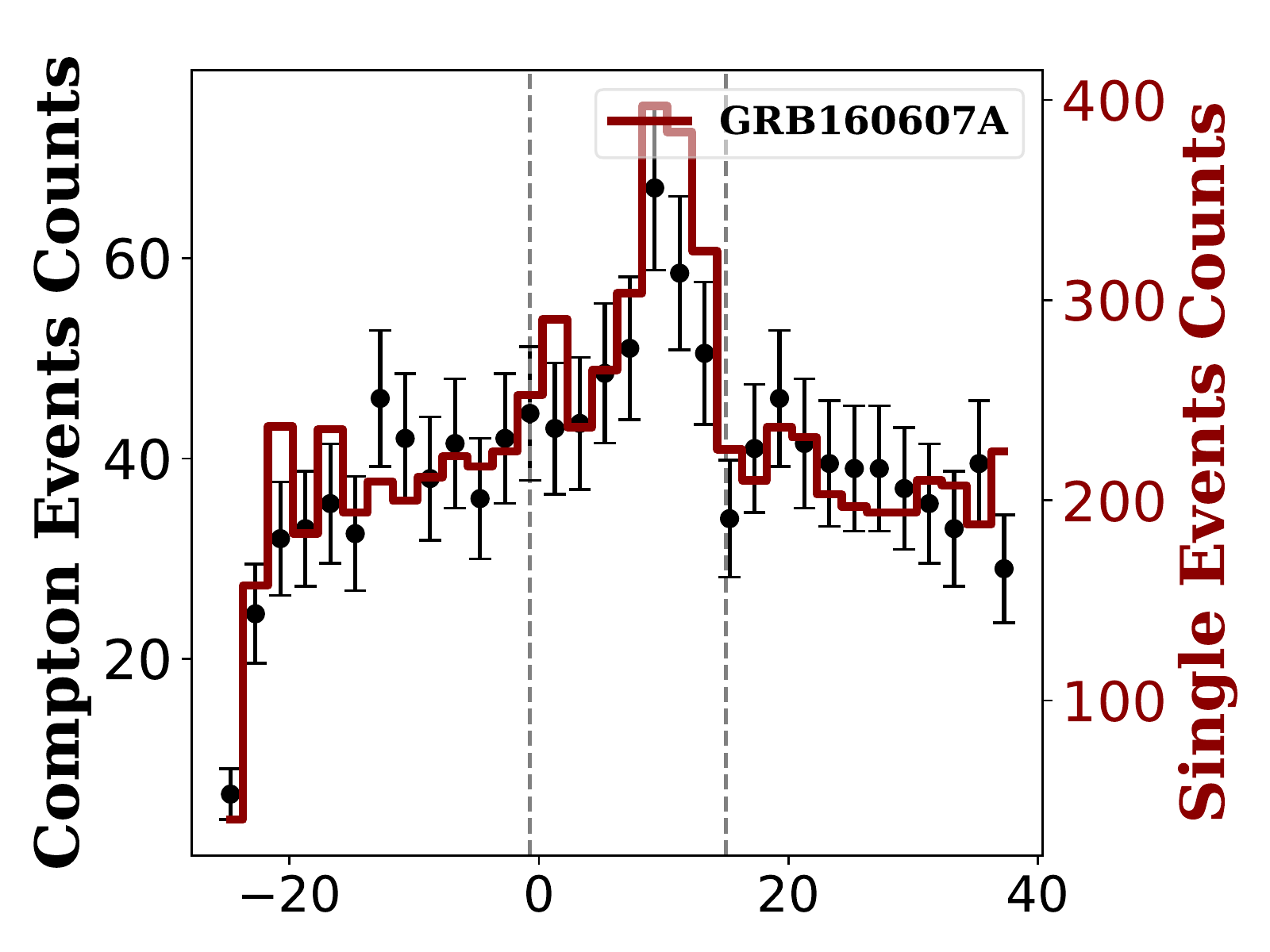}
\includegraphics[width=.49\linewidth]{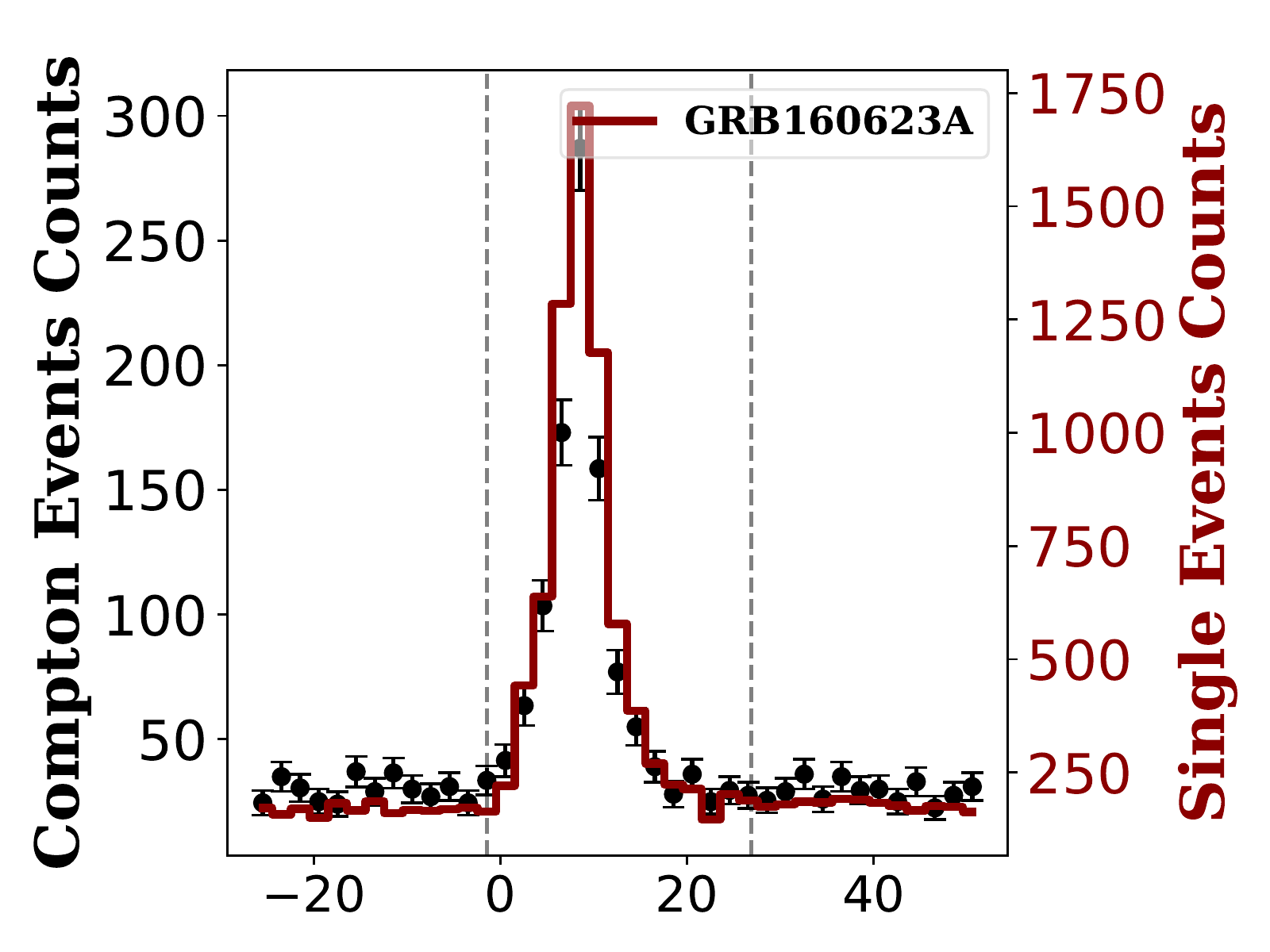}
\includegraphics[width=.49\linewidth]{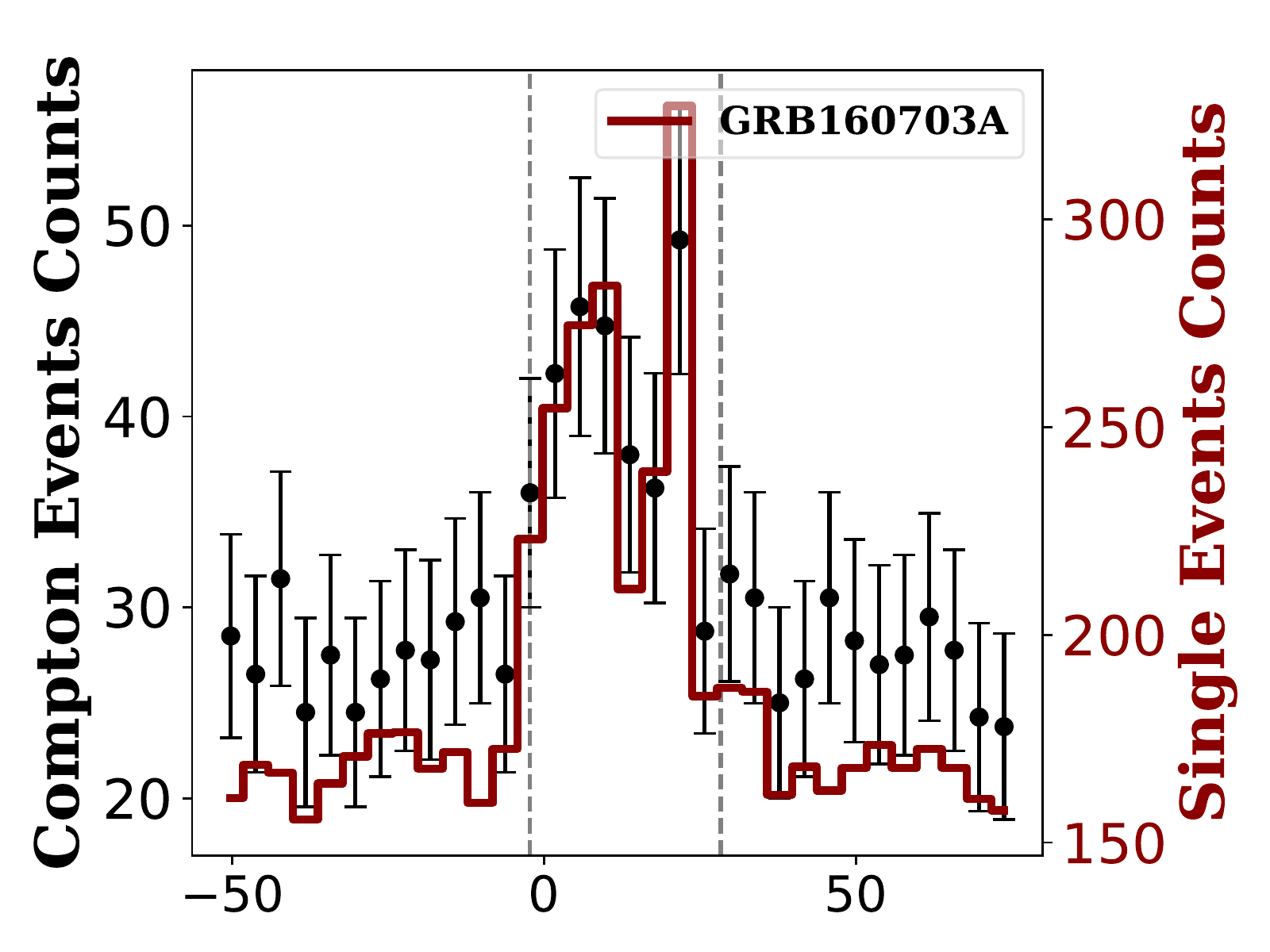}
\includegraphics[width=.49\linewidth]{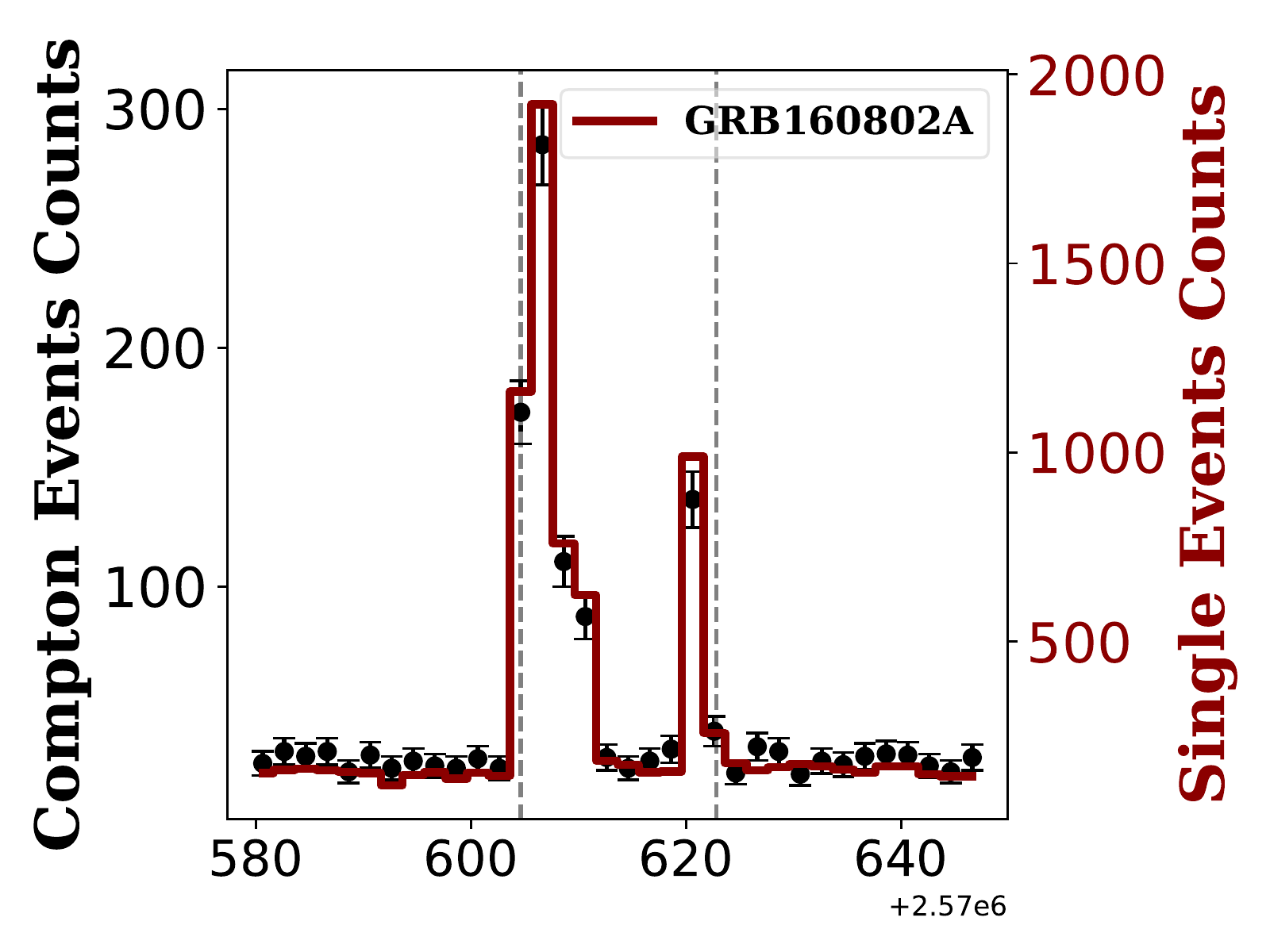}
\includegraphics[width=.49\linewidth]{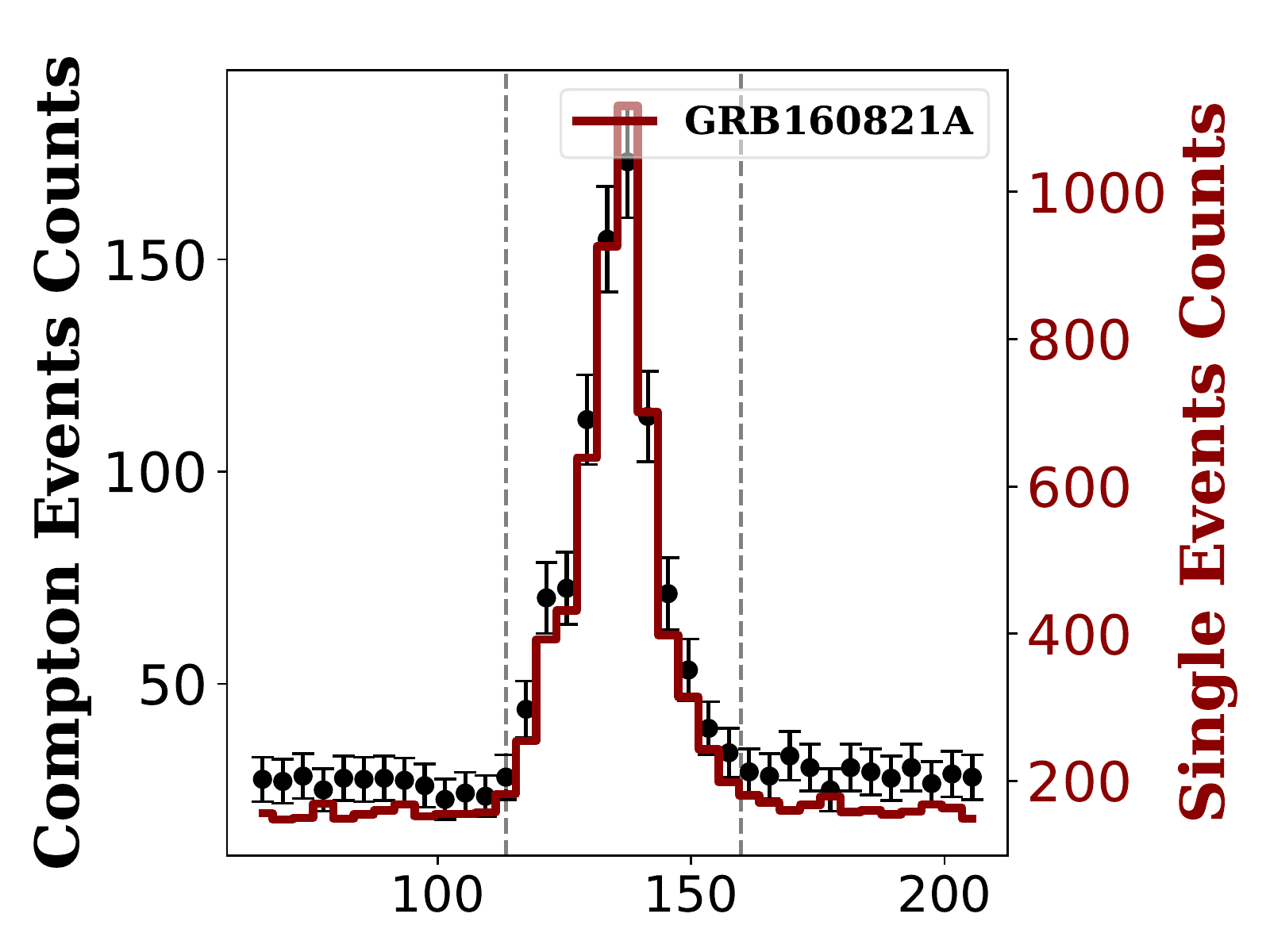}
\includegraphics[width=.49\linewidth]{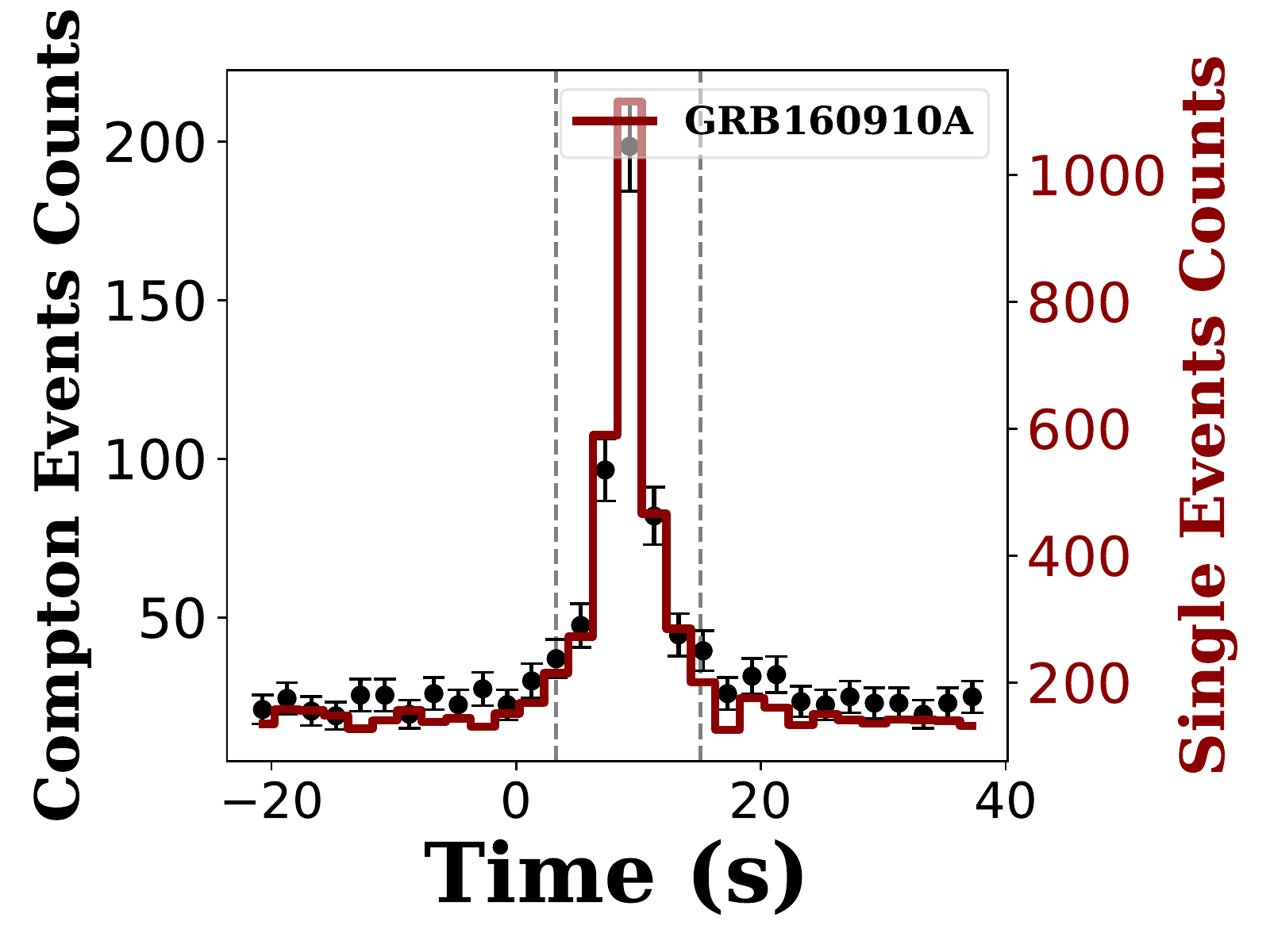}
\caption{GRB light curves from CZTI 1-pixel (red) and 2-pixel events (black points) for the 11 GRBs. The time intervals of the bursts are obtained from the Bayesian block analysis on the 1-pixel CZTI light curves as shown by the vertical dashed lines. The `zero' denoted in the time-axis stands for the trigger time reported by the {\em Fermi}-GBM.}
\label{BB_selection}
\end{figure}

In this work, the spectrum 
%and the polarisation 
analyses are conducted on the time integrated emission of the bursts. The time interval corresponding to the integrated emission is chosen by employing the Bayesian block algorithm \citep{scargle1998studies,Scargle2013, Burgess2014} of time binning on the single pixel event data of the bursts. The block with the minimum probability density value corresponding to the background region is taken as the guide to decide the start and stop times of the integrated emission. The onset time of the first block with the probability density greater than that of the background which is closer to the onset time of the burst and the end time of the last block after which the background continues are considered as the start and stop times of the time interval of integrated emission respectively (Figure \ref{Bayesian_selection} in Appendix).         

%%%%%%%%%%%%%%%%%%%%%

In the next section, we describe the methodology of spectroscopy using 1-pixel and 2-pixel CZTI events and CsI-veto detected events followed by broadband spectroscopy results for the eleven bright GRBs detected in 2015 $-$ 2016. For CZTI events, we utilize both the standard good pixels and the newly calibrated low-gain pixels to extend the energy spectra to sub-MeV region. 

\section{Methodology for Spectroscopy}
\label{spec}

\subsection{Spectroscopic response}
{The 2D spectral responses for CZTI 1-pixel, 2-pixel and CsI/veto spectroscopy are generated using GEANT4 simulation. Here we outline the basic steps of response generation. 
Response is computed using GEANT4 mono-energy simulations of the full {\em AstroSat} mass model at specific $\theta$ and $\phi$ viewing angles for each GRB ($\theta$ and $\phi$ for a given GRB are provided by either \emph{Swift} or \emph{Fermi}). The mono-energetic lines for simulations were selected between 100 keV and 2 MeV at every 20 keV till 1 MeV and at every 100 keV in 1 $-$ 2 MeV, totalling around 56 mono-energies. Simulation is done for a large number of photons (10$^9$ photons for each energy) in order to have a statistically significant energy distribution in CZTI for each mono-energetic line. The simulation file contains information of total seven interactions or steps for each incident photon (x, y, z-position of interactions in CZTI and deposited energy in each interaction, see \citet{chattopadhyay14}) in CZTI modules. We add up the energies from all the interactions happening within a pixel of 2.5 mm $\times$ 2.5 mm in Interactive Data Language (IDL \footnote{Research Systems, Inc. (1995). IDL user's guide : interactive data language version 4. Boulder, CO :Research Systems}) based routine outside the GEANT4. 
We apply the same CZTI pixel-level LLD (Lower Level Discriminator) values in the simulation data whereas the ULDs (Upper Level Discriminator) were computed from actual observational data for each module and is applied to simulation data accordingly. From this event list, the 1-pixel and 2-pixel events are separated and processed differently for final response generation.  
For 1-pixel events, the distribution of deposited energies is calculated at a bin size of 1 keV from 0 keV to 1000 keV (total 1000 bins) for each of the 56 mono-energies. 

It is to be noted that Geant4 simulation takes care of all types of interactions with appropriate probabilities including photoelectric, Compton, Rayleigh inside CZTI and photons scattered from the spacecraft or other surrounding payloads to CZTI. 
Because of these multiple interactions and scattered events from surrounding materials, the distribution of deposited energy in CZTI is broad and non-gaussian. However, the large number of photon simulation gives sufficient statistics to obtain the correct energy distribution in the full range of 100 bins for all 56 mono-energies.
The 2D matrix (56 $\times$ 1000) of deposited energy distributions for the mono-energetic lines 
is then convolved with a Gaussian function of appropriate width to generate the 1-pixel spectral response or the Redistribution Matrix File (RMF). On the other hand, we apply the Compton kinematics criteria on the 2-pixel events to select the valid Compton events. The energies of the two pixels are then added up to calculate the total deposited energy. The deposited energy distributions for the 56 mono-energies are then convolved with a Gaussian to obtain the 2-pixel response.   
The CsI (or veto) spectral responses are generated for each quadrant in the same fashion using the same {\em AstroSat} mass model simulation data where we only consider events and associated energies deposited in the CsI detectors to estimate the deposited energy distribution. 
It is to be noted that we use a $\mu\tau$ and charge diffusion based line profile model \citep{chattopadhyay16} for mask weighted response below 150 keV, whereas for this work, we use a simple Gaussian model for simplicity.
}
\subsection{Single Pixel (1-Pixel) Spectroscopy}
\label{1-pixel}
Because the CZTI surrounding structures and the collimators become increasingly transparent above 100 keV, the spectral analysis of the GRBs starts from 100 keV and extends to 900 keV after incorporating the low-gain pixels.  Detection efficiency of a 5 mm CZT drops below 10 \% above 1 MeV resulting in a low signal to noise ratio at those energies. 
The single pixel events are selected such that there are no other events reported in 100 $\mu$s time window on either side of the event. Energies deposited in all such events in the full burst region (interval obtained from Bayesian block analysis) are used to generate the spectrum with a 10 keV binning.

The 1-pixel light curves and the selected time intervals are shown in Figure \ref{BB_selection} (red solid lines).
The background spectrum is constituted by selecting at least 300 seconds of time window from the pre and post-burst regions.

\begin{figure*}
\centering
\includegraphics[width=0.45\linewidth]{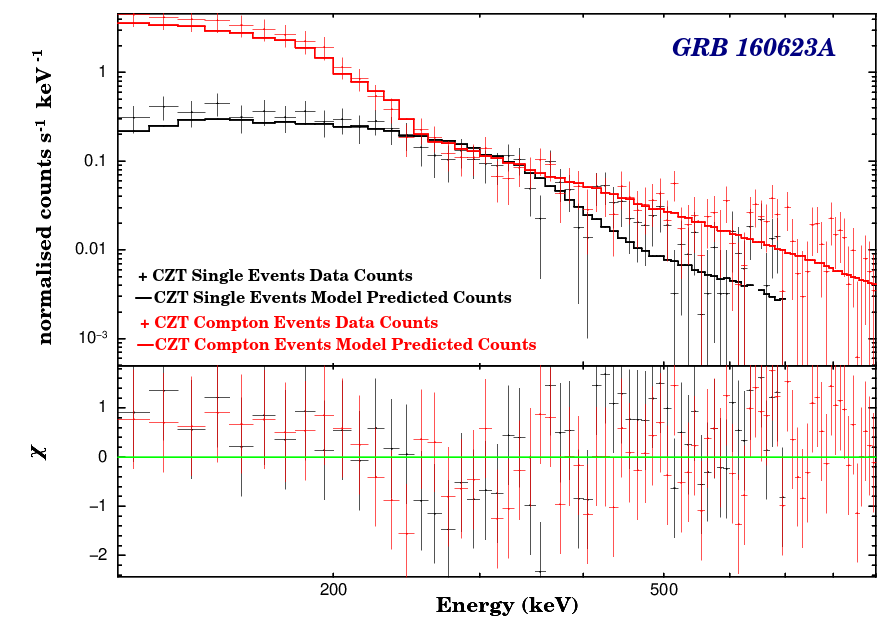}
\includegraphics[width=0.45\linewidth]{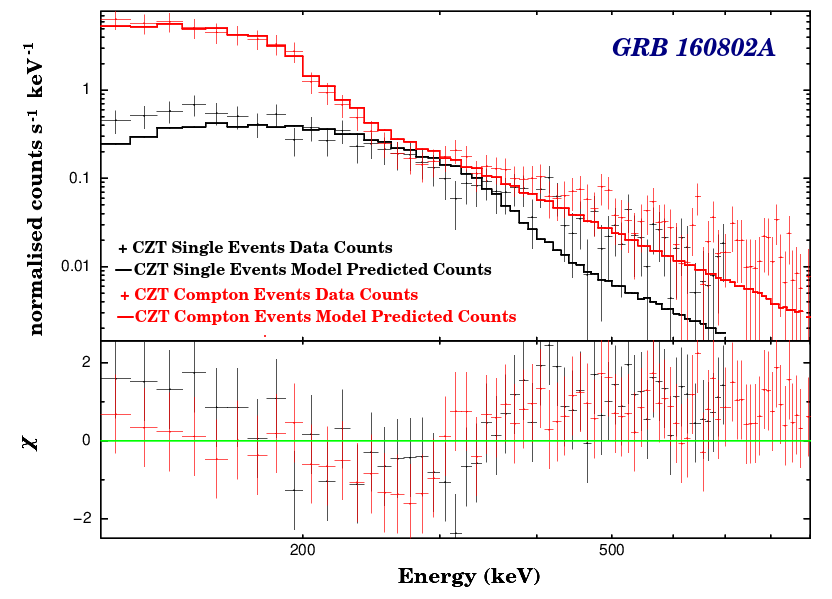}
\caption{Left: Example of count spectra (upper panel) and residuals (lower panel) obtained for the 1-pixel (red) and 2-pixel (black) CZTI events for one back Side GRB (GRB 160623A detected at angle $\sim$140$^\circ$, left panel) and one front side GRB (GRB 160802A, viewing angle $\sim$53$^\circ$, right panel). We fit the spectra with Band model keeping the spectral parameters frozen at the values reported in literature to check for the consistency in spectral shape with {\em Fermi} and {\em Konus-Wind}.} 
\label{1-pixel}
\end{figure*}

We quantify the systematics in the 1-pixel spectral data arising due to the uncertainties and inaccuracies in the AstroSat mass model and the CZTI detector via the analysis of the spectral data of GRBs detected at different incoming orientations.   
We use Band model \citep{band93} to fit the spectra while keeping the power law indices ($\alpha$ and $\beta$) and peak energy ($E_{peak}$) frozen at the values reported by either the \emph{Konus Wind} %(for GRB 160623A, detected only by \emph{Konus Wind} and {\em AstroSat}) 
or \emph{Fermi} spectral analysis, and the normalisation of the Band model is left free. 

For the different GRBs, listed in Table 1, detected at different orientations including most of the incoming angles around the spacecraft,  
%on either side of the spacecraft, 
once obtaining a ``best" fit,\footnote{{ The CZTI spectral data fit is considered to be reasonable when (a) the obtained residuals are roughly randomly distributed around zero, (b) the reduced chi-sq $\chi^2 < 2$ and the (c) normalization of the Band function is found to be consistent with what is obtained from {\it Konus-wind} and {\it Fermi} spectral analysis.}}  we find unresolvable discrepancies between calculations and the data, i.e., systematic errors.  Without knowing or assuming the origin of these features, we characterize the effect by adding systematic errors in an incremental fashion until we achieve a uniform residual with a reduced $\chi^2$ $<$~2.
%by adding a systematic of 10 $-$ 15 \% uniformly across the energy range $100-900$ keV}. We quantify the required systematic by adding systematic in small increments until we get a uniform residual with reduced $\chi^2$ $<$~2.
We, therefore, add 10 $-$ 15 \% systematic to the 1-pixel spectroscopic data for all the GRBs to take care of the inaccuracies in the {\em AstroSat} mass model. 

 For example, the spectral fits with the respective residuals obtained for the GRB 160623A (left) and GRB 160802A (right) that are detected on either side of the spacecraft are shown in the Figure \ref{1-pixel} %shows an example of 
where 1-pixel spectra in 100 $-$ 900 keV obtained from the full burst region are shown in red crosses.

\subsection{Compton (2-pixel) Spectroscopy}
The Compton spectroscopy is carried out in the energy range of 100 $-$ 700 keV since above 700 keV there is no sufficient Compton scattering efficiency of CZTI detectors. The 2-pixel Compton events are identified from adjacent pixel events within 20 $\mu$sec coincidence window with an additional Compton kinematics criteria of ratio of two deposited energies between 1 and 6 (also discussed in \citet{chattopadhyay14}). The energies from the two events recorded are added up to get the total energy and therefrom the spectrum with a bin size of 10 keV.

%2-pixel spectral response is generated in the same way using the mono-energetic simulation files from the {\em AstroSat} mass model. We apply the same Compton kinematic conditions to select the Compton events from simulation data while all other steps to generate the response is same as that for 1-pixel response. 

 The systematics involved in the Compton spectral data are assessed using the same methodology adopted for 1-pixel spectral data (section \ref{1-pixel}). 
The spectral fits and the residuals obtained for the Compton spectra of GRB 160623A and GRB 160802A in 100 $-$ 700 keV are shown in red data points in Figure \ref{1-pixel}. Similar to the 1-pixel spectra, we find reasonable fit to the data and agreement with {\em Fermi} norm %with an addition of 
by adding a systematic of 10 $-$ 15 \% uniformly throughout the energy range, 100-700 keV.  Therefore, this systematic is added to the 2-pixel spectral data of all the GRBs.  

\subsection{CsI (or Veto) spectroscopy}

There are four CsI(Tl) scintillator detectors (each 167 mm $\times$ 167 mm in size and 2 cm in thickness) below CZTI quadrants to veto the high energy particle induced background events reported in both CZTI and CsI detectors \citep{bhalerao16,rao16}. The veto detectors were initially not meant for spectroscopy. However since the detectors possess sufficient detection efficiency in the sub-MeV region, we explored the possibility of using them for spectroscopy to enhance the overall spectroscopic sensitivity.   
The existing CZTI pipeline provides the veto spectrum at every second. We employ the available data to generate spectrum for each Veto detector in a similar way that is used for CZTI single pixel events. However, we do not use the poorly calibrated 4$^{th}$ Veto quadrant for spectroscopy. 
It is to be noted that the Veto spectrum consists of all interactions in the CsI detectors and different from the Veto tagged events where the both CZTI and Veto are triggered due to simultaneous events recorded in those detectors. 
%The spectral responses are generated for each quadrant in the same fashion as CZTI 1-pixel response using the {\em AstroSat} mass model simulations. 

\begin{figure*}
\centering
\includegraphics[width=0.45\linewidth]{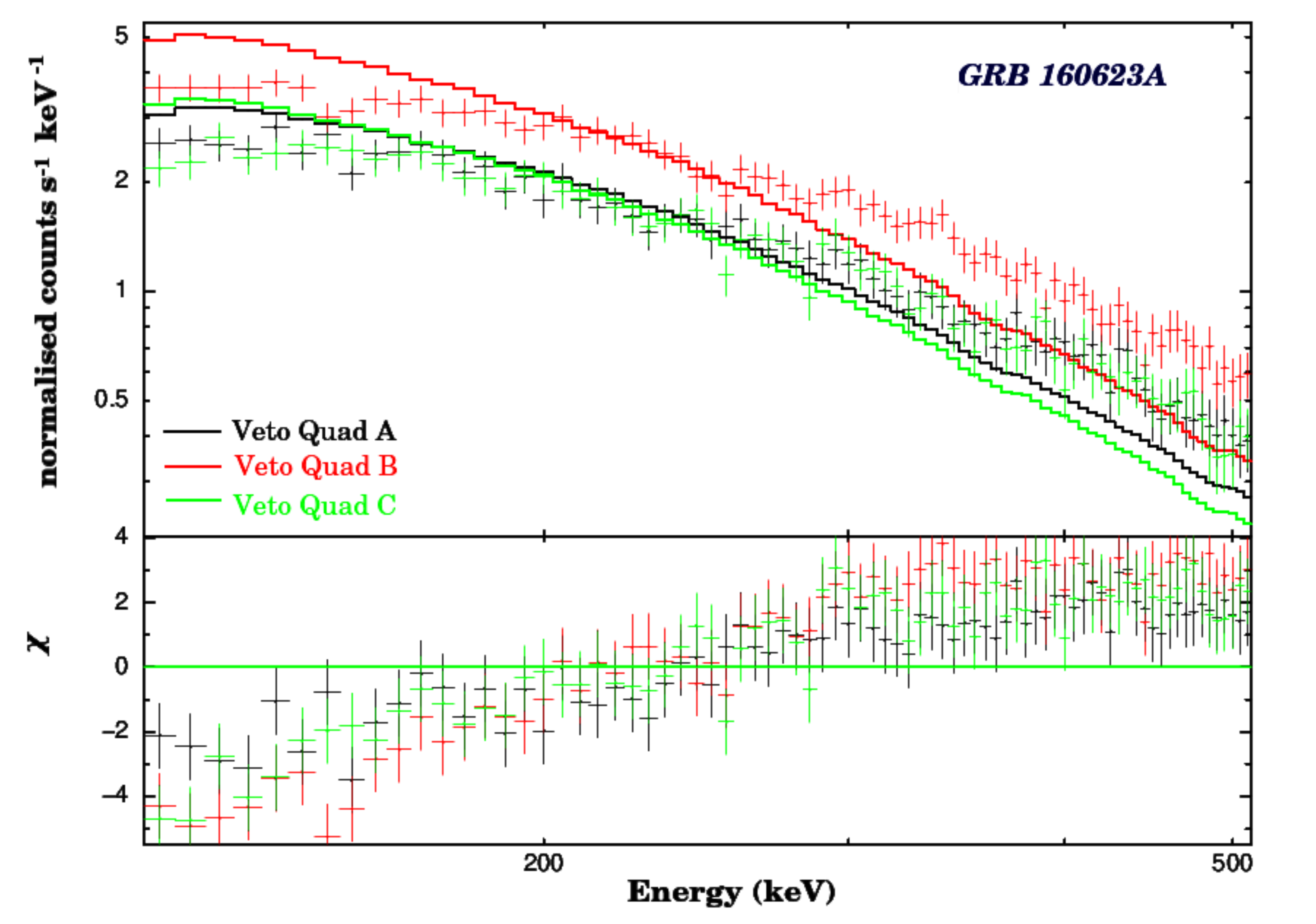}
\includegraphics[width=0.45\linewidth]{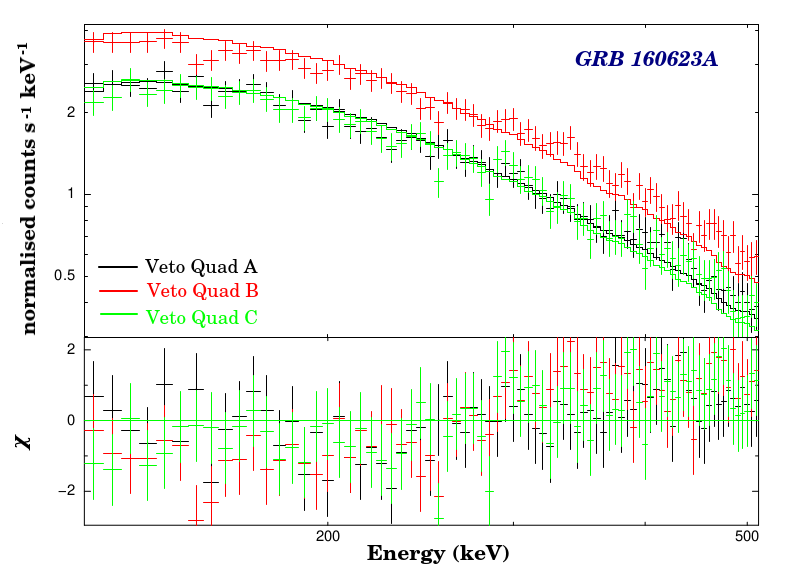}
\includegraphics[width=0.45\linewidth]{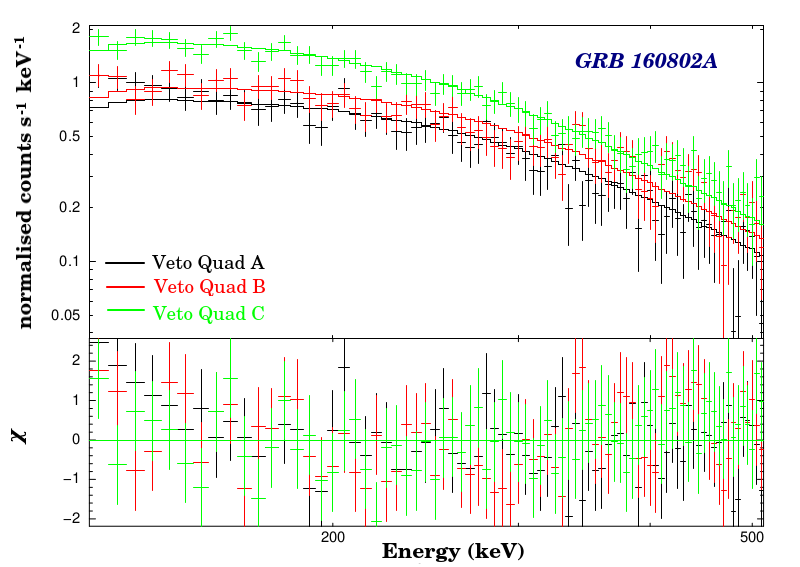}
\caption{Top left: the count spectra (upper panel) and their respective residuals (lower panel) obtained for the three quadrants of the Veto detectors (black: quadrant A, red: quadrant B and green: quadrant C) for GRB 160623A (detected from the back side of CZTI). We see a systematic trend in the residuals possibly due to lower detection probability by the scintillators around 100 keV which improves at higher energies; top right: Same as the left figure but after implementing an energy dependent correction ($1-e^{-Energy/E_{0}}$) where $E_{0}^{-1}~=~0.0045$ keV$^{-1}$ (see text for more details); bottom: same as the top figure but for GRB 160802A (detected from the front side) after implementing the energy dependent correction with a higher value of $E_{0}^{-1}~=~0.01$ keV$^{-1}$.}
\label{veto}
\end{figure*}

For all the GRBs detected from the rear side of the spacecraft, we find the observed spectra to be flatter than the response folded model. An example is shown in the top-left spectral plot of Figure \ref{veto} for GRB 160623A which is detected at $\theta$ of $\sim$140$^\circ$. We find an identical systematic trend in all the back side GRBs. However, we do not attribute the systematic to the mass model as CZTI 1-pixel and 2-pixel spectral fits for back side GRBs do not show such systematic trend in the residuals. On the other hand, the trend is significantly lower in Veto detectors for the front side GRBs. Therefore we believe that this systematic is originated in the CsI detectors but primarily for detections from back side. 
%A similar trend ($1-{e^{-energy}}$) is expected from scintillator detectors due to relatively lower detection probability at lower energies which improves at higher energies reaching to unity. 
CsI detectors are scintillator detectors where the scintillation light is collected by the PMTs (2 PMTs for each of the 4 CsI detectors). At lower energies ($\sim$100 keV), the number of scintillation photons generated is lower than that at higher energies. Given the fact that there are only two PMTs to collect the scintillation photons, the detection probability of the GRB photons at lower energies is expected to be relatively low.
We also note that the detectors were initially not meant for spectroscopy and therefore the number of readout photo-multiplier tubes and optical coupling between the crystal and the photo-multiplier tubes (PMTs) were not optimized to enhance the detection probability. The light collection efficiency might be significantly compromised for events happening in the back side of CsI because of the absence of optical reflecting coating on the back surface and a relatively higher level of cover shielding on the back side near the PMTs (light collecting area is relatively lower on the back side). 

To take care of this, we multiply the photon detection probability (represented by an empirical term, $1-e^{-Energy/E_{0}}$) to the model (same as multiplying to the CsI detector response) to mimic for an energy dependent systematic where the value of $E_{0}^{-1}$ depends on the location of transient observed with respect to CZTI. For the front side GRBs (example shown in the bottom panel of Figure \ref{veto} for GRB 160802A) i.e theta $<$ 60$^\circ$ the value of $E_{0}^{-1}$ is found to be around 0.01 keV$^{-1}$ which gives 90 \% detection probability at $\sim$200 keV, 
%implying an insignificant contribution of the exponential component, 
whereas for the orthogonal GRBs i.e  90$^\circ$ $<$ theta $<$ 110$^\circ$, the value of $E_{0}^{-1}$ comes out to be around 0.008 keV$^{-1}$. For the back side GRBs, value of $E_{0}^{-1}$ is found to be around 0.0045 keV$^{-1}$ signifying poor detection probability (90 \% detection probability at $\sim$600 keV).

Since we get similar values of $E_{0}^{-1}$ for front, back and orthogonal GRBs, we plan to incorporate the exponential feature observed in the Veto detectors in the response itself. We also include an additional 5 \% systematic in the data in case of back side GRBs.
%We compute $E_{0}^{-1} = 0.0045$ for the front side GRBs which gives 90 \% detection probability at ~200 keV) whereas for the back side GRBs, we find $E_{0}^{-1} = 0.01$ that gives 90 \% detection probability at $\sim$600 keV. 

%%%%%%%%%%%%%%%%%%%%%%%

\section{Results: Broadband Joint Spectroscopy of GRBs}
\label{result_spec}

{ With a fair assessment of the systematics present in the CZTI and Veto spectral data, we now conduct the broadband joint spectral analyses involving the spectral data from {\it Fermi}, Niel Gehrels {\it Swift} BAT, along with CZTI data including the single, Compton and Veto for the time integrated emission of different GRBs. 
We analyse the time integrated spectrum of 10 GRBs that were detected by CZTI in the first year of its operation (2015 $-$ 2016). The time interval of the integrated emission is selected using Bayesian Block binning technique and described in section \ref{interval}.}

The {\it Fermi} spectral data includes two bright sodium iodide (NaI) 
detectors with source angle less than ($<60^{\circ}$) and the brightest 
bismuth germanate (BGO) detector \citep{gruber2014fermi}. In case of 
GRB151006A, GRB160509A and GRB160821A, the low energy Large Area 
Telescope (LLE) data are also used. The {\it Fermi} spectral files are 
extracted using Fermi Burst Analysis GUI v. 02-03-00p33 
(gtburst\footnote{https://fermi.gsfc.nasa.gov/ssc/data/analysis/scitools
/gtburst.html}). The Swift BAT spectral files are prepared by the 
standard methodology\footnote{https://swift.gsfc.nasa.gov/analysis/threa
ds/bat\_threads.html}. 

The spectral analyses are performed using the X-Ray Spectral Fitting Package ({\tt XSPEC}, \citealt{arnaud1996xspec}) version: 12.11.0 and have followed chi-square statistics. Both BAT and CZTI spectral files are compatible with Gaussian statistics, however, the GBM and LLE files are consistent with Pgstat wherein the background and signal are assumed to be Gaussian and Poissonian respectively. Therefore, using Heasoft Ftool {\tt GRPPHA\footnote{https://heasarc.gsfc.nasa.gov/ftools/caldb/help/grppha.txt}}, we rebinned both GBM and LLE spectral files such that each energy channel contains a minimum of 20 photons.  

\begin{figure*}
\centering
\includegraphics[width=.45\linewidth]{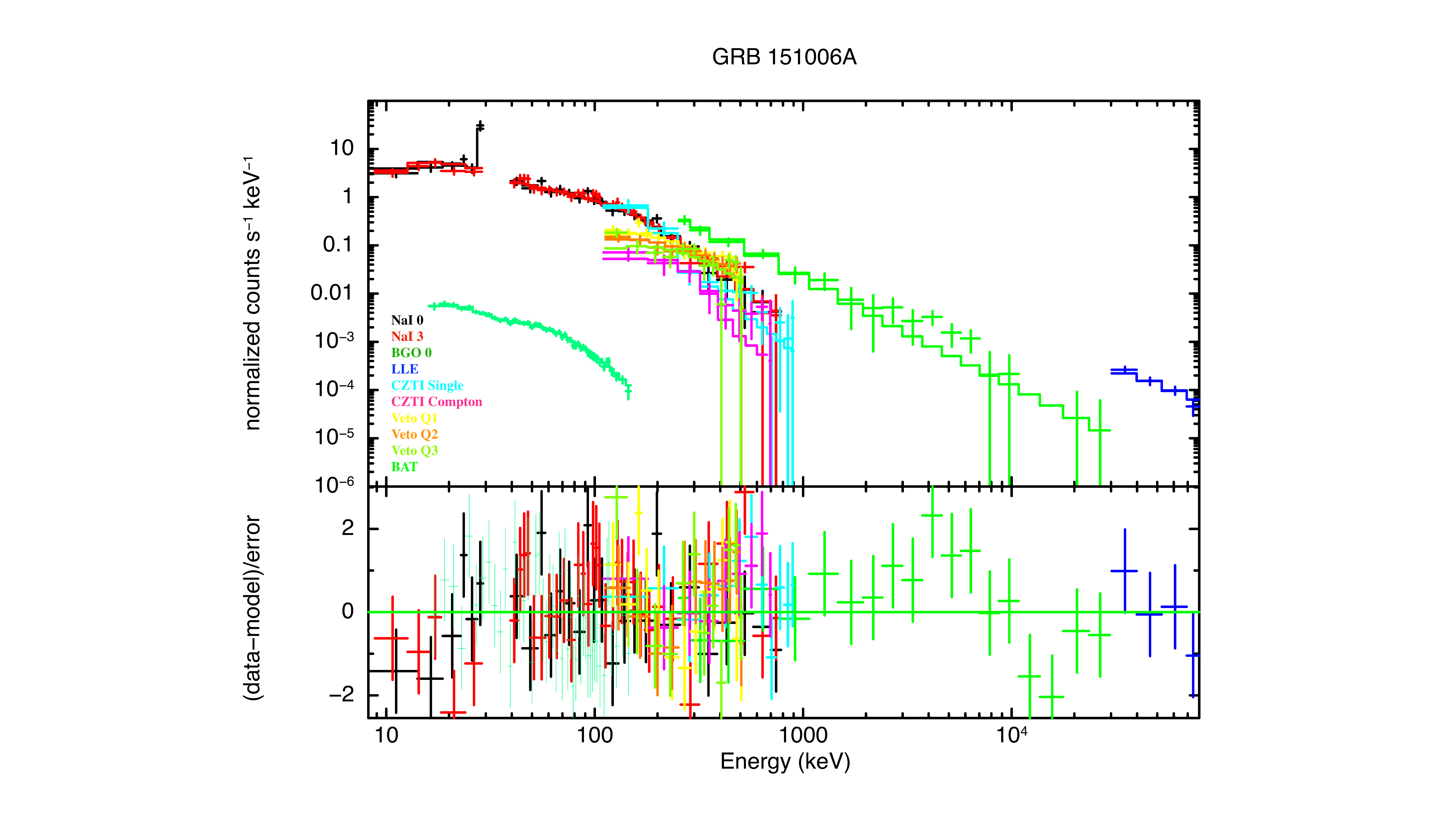}
\includegraphics[width=.45\linewidth]{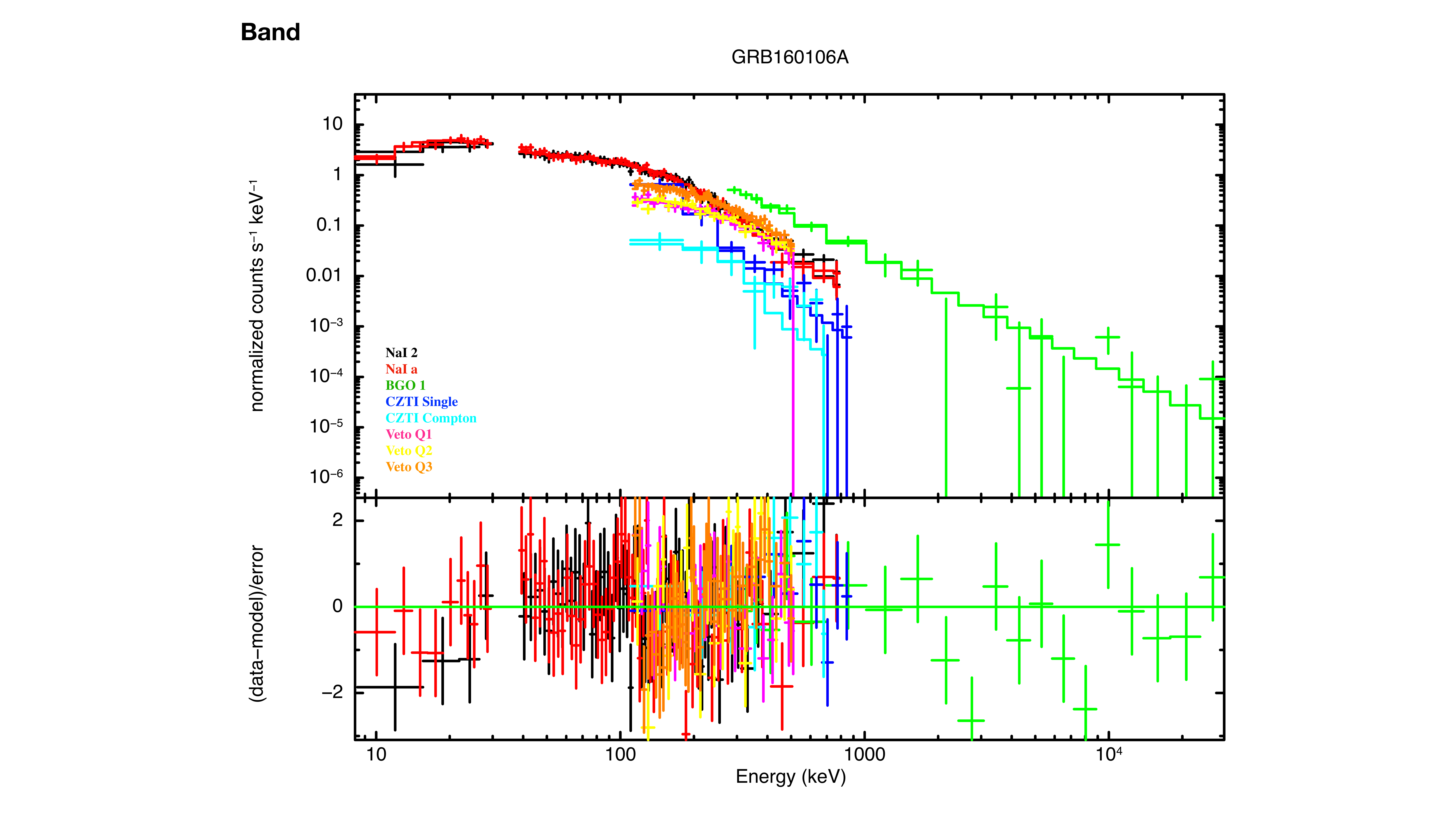}
\includegraphics[width=.45\linewidth]{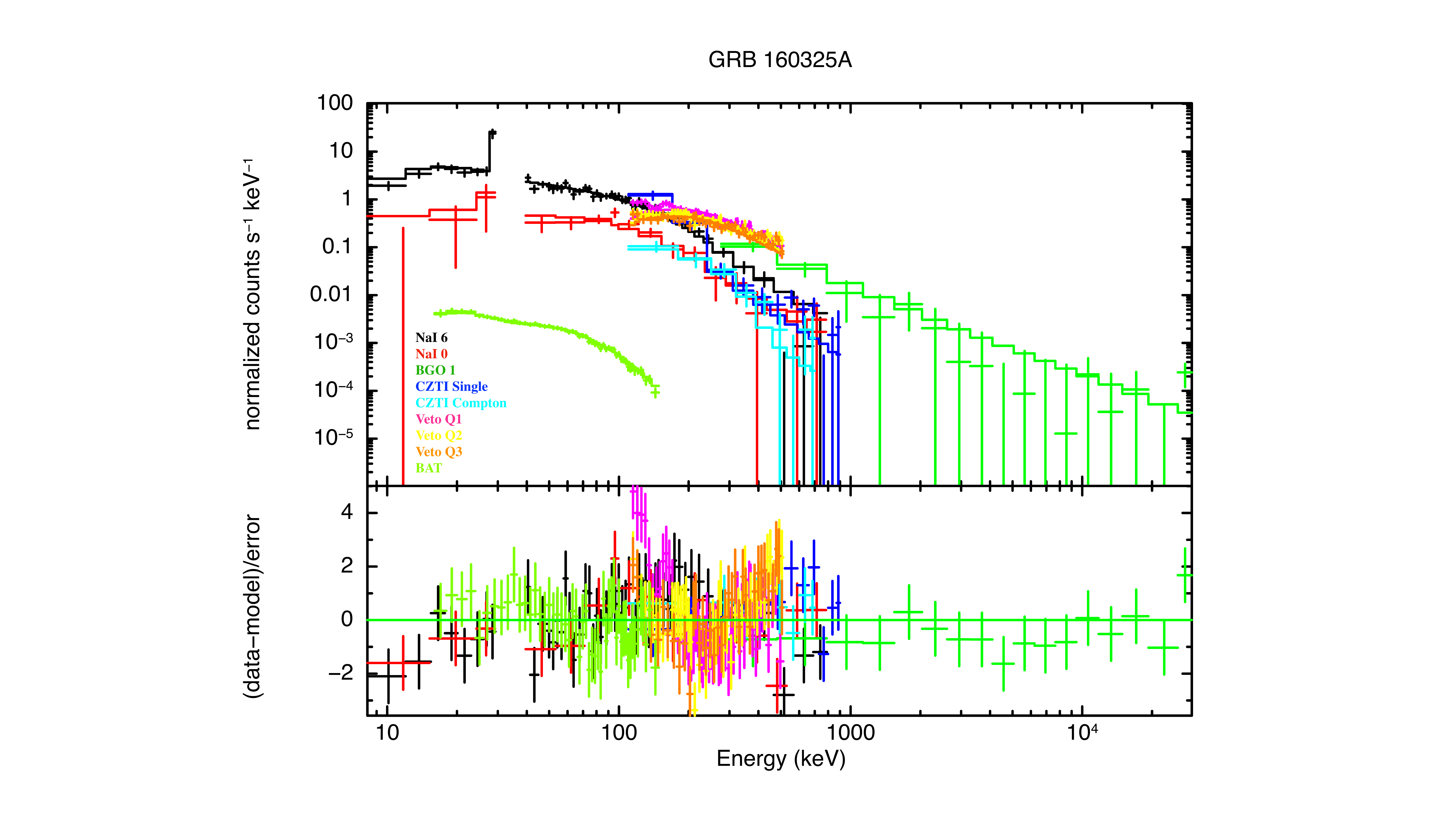}
\includegraphics[width=.45\linewidth]{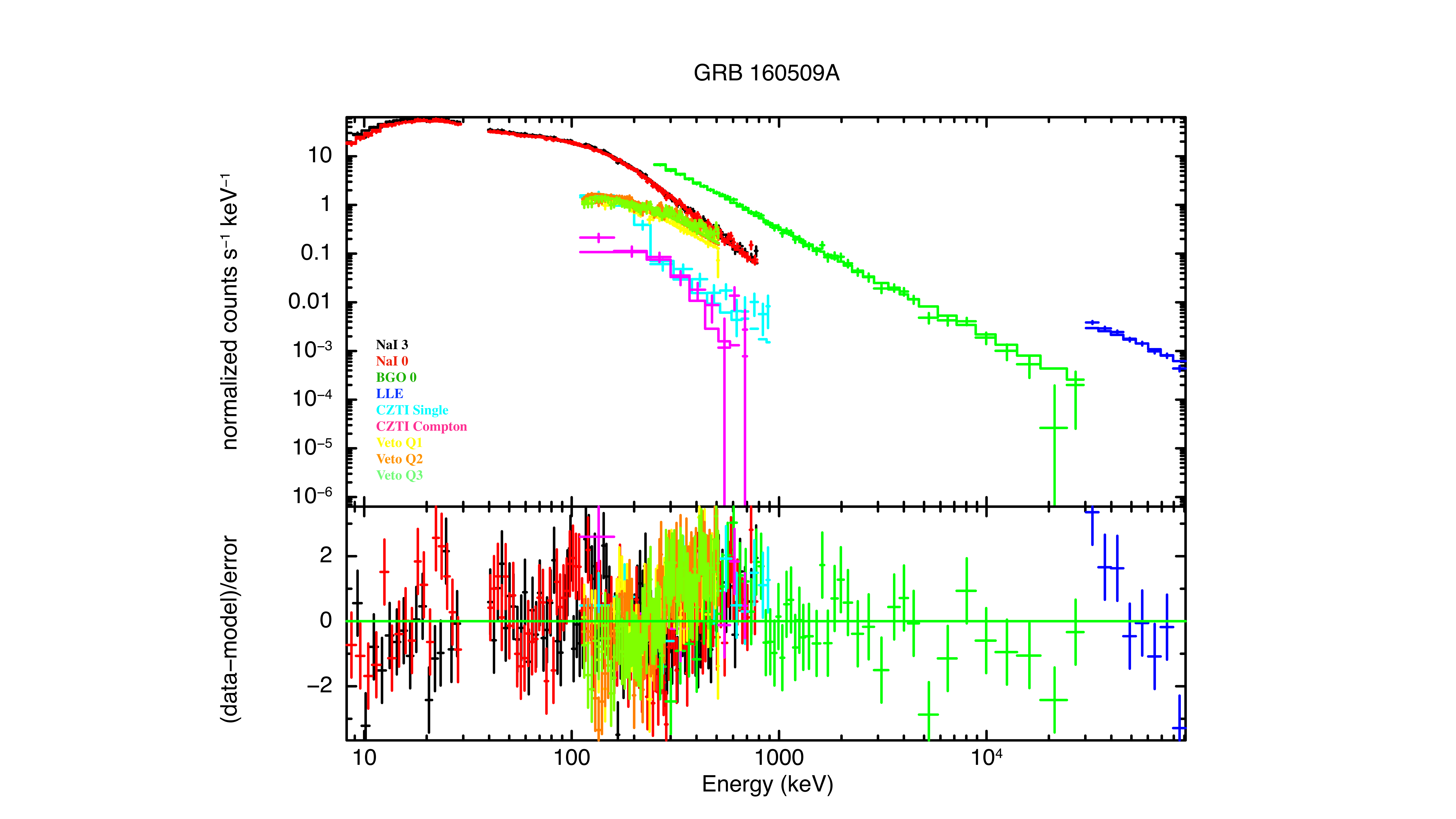}
\includegraphics[width=.43\linewidth]{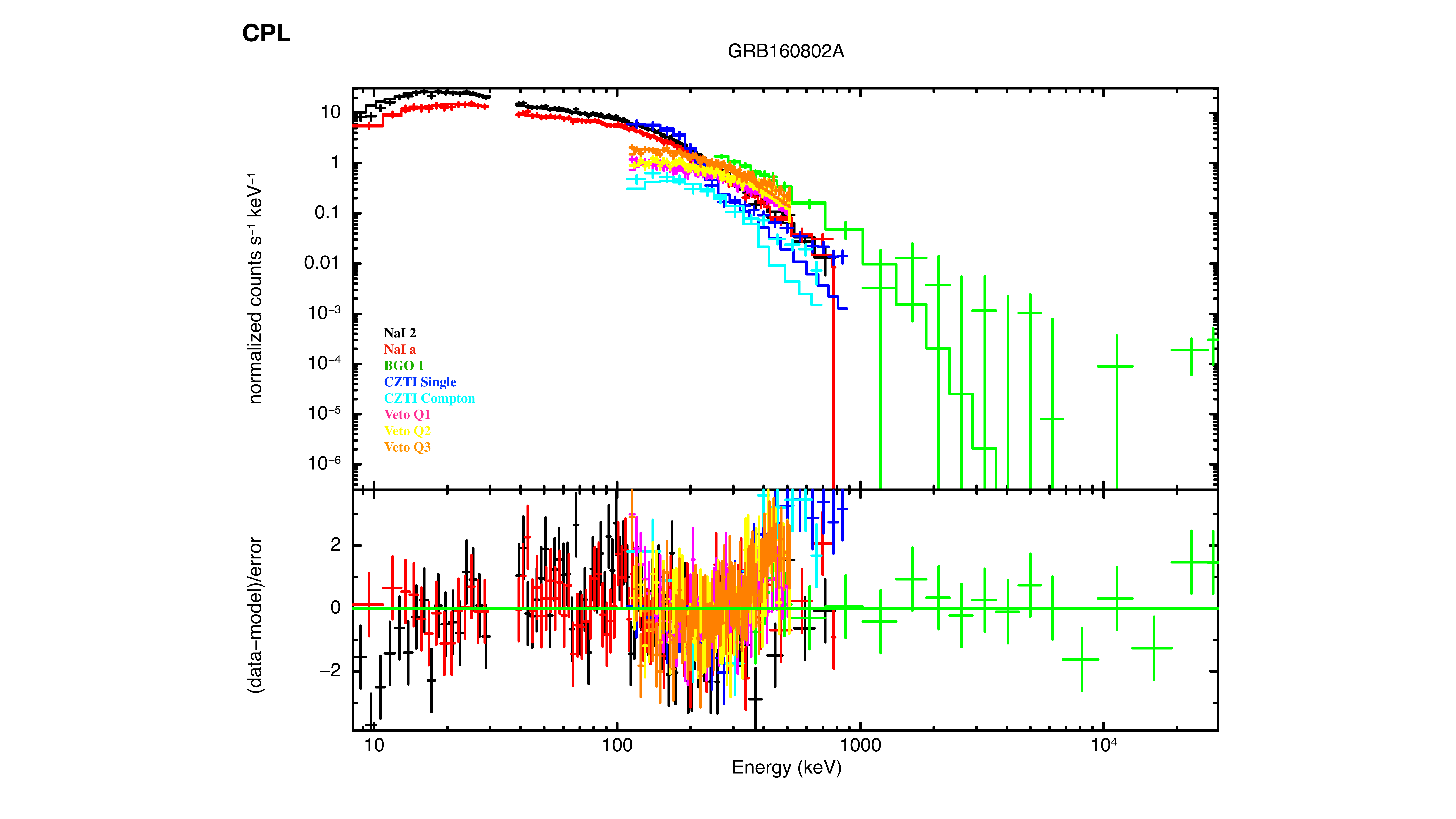}
\includegraphics[width=.45\linewidth]{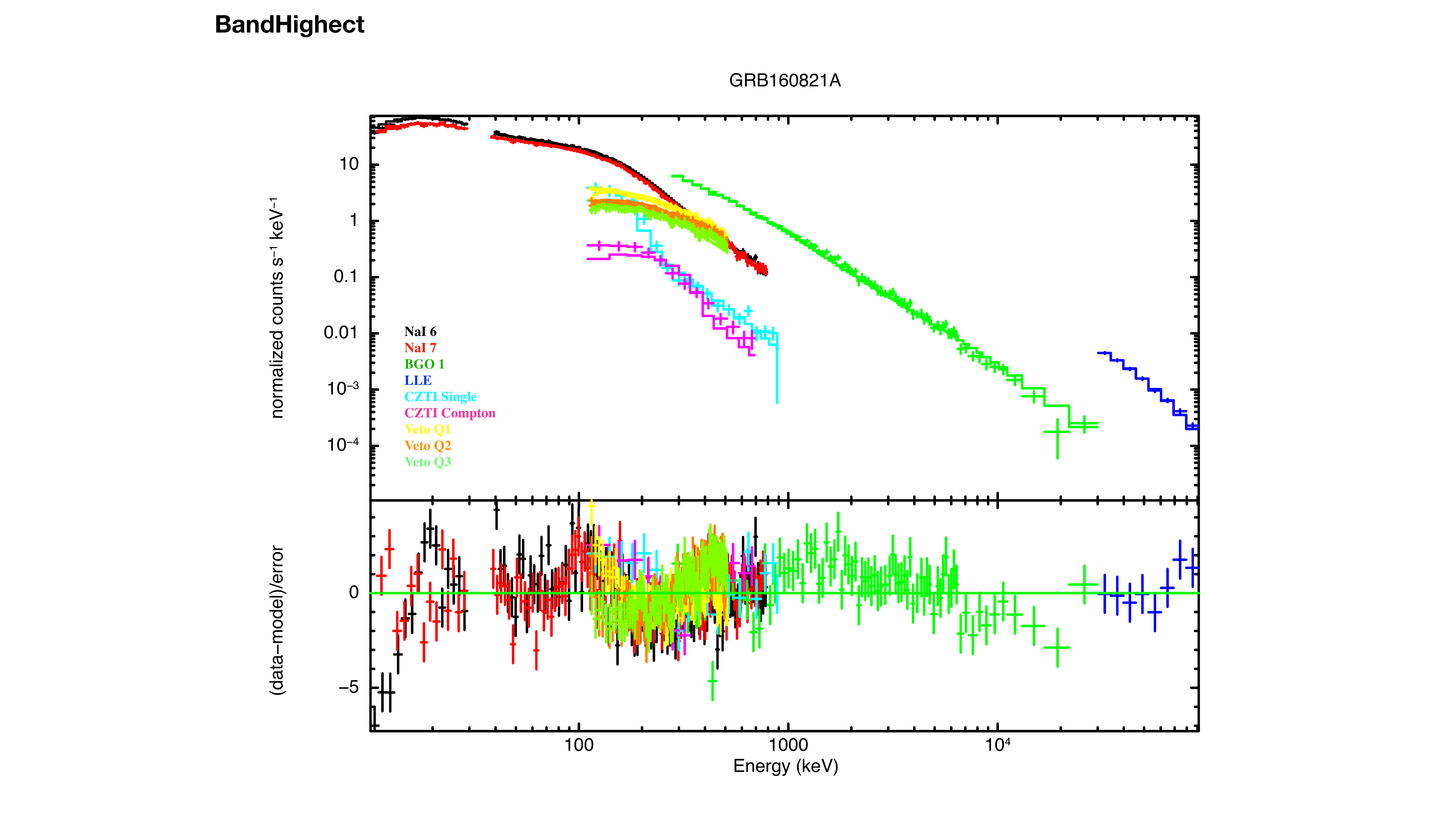}
\includegraphics[width=.45\linewidth]{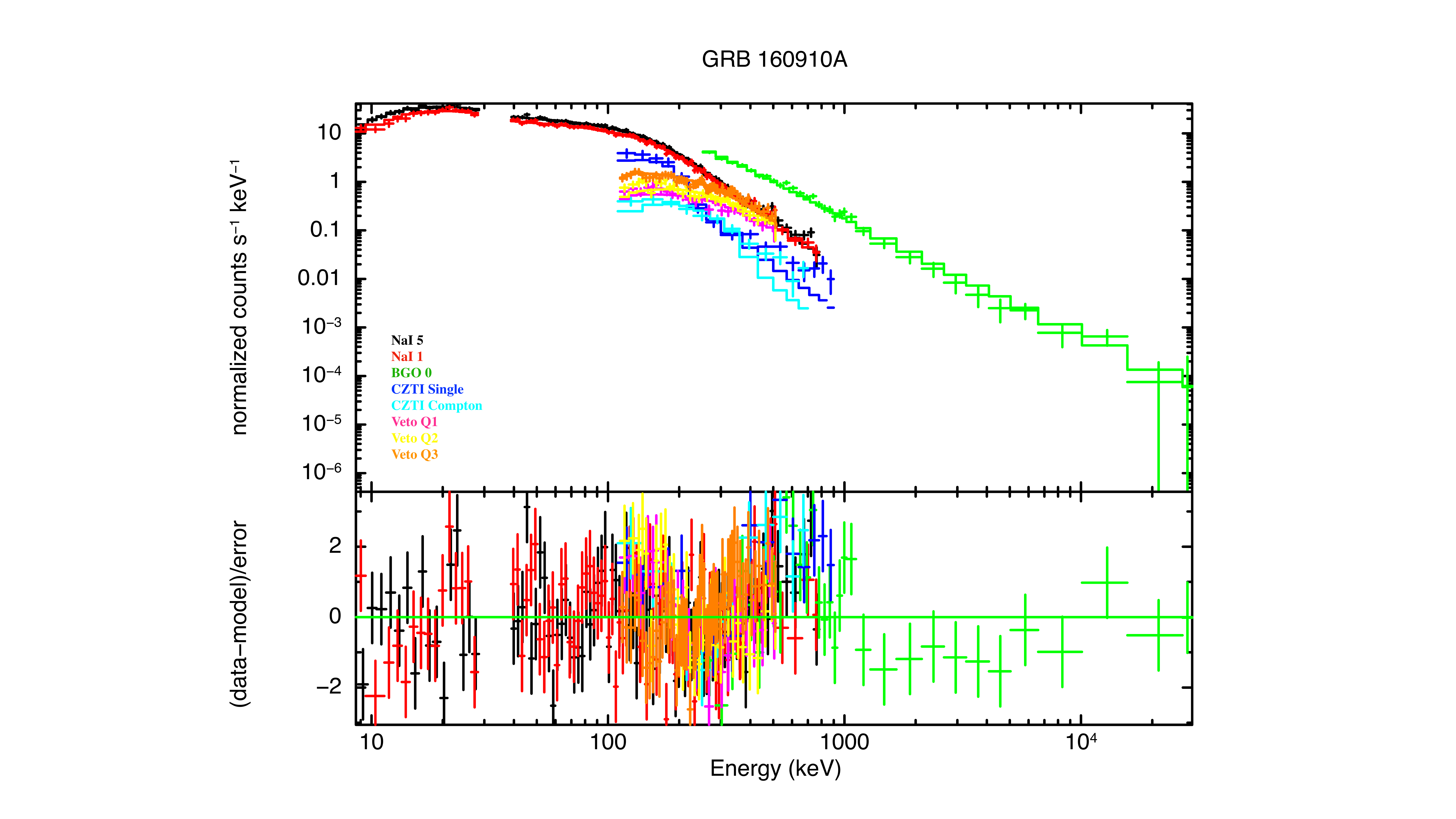}

\caption{The count spectra (upper panel) and the respective residuals (lower panel) obtained for the broadband joint spectral analysis consisting of {\it Fermi} + CZTI data (+ BAT data in cases where it is available) for GRB 151006A, GRB 160106A, GRB 160509A, GRB 160325A, GRB 160802A, GRB 160821A and GRB 160910A are shown.}
\label{fermi_czti_bat_spec}
\end{figure*}

\begin{figure*}
\centering
\includegraphics[width=.42\linewidth]{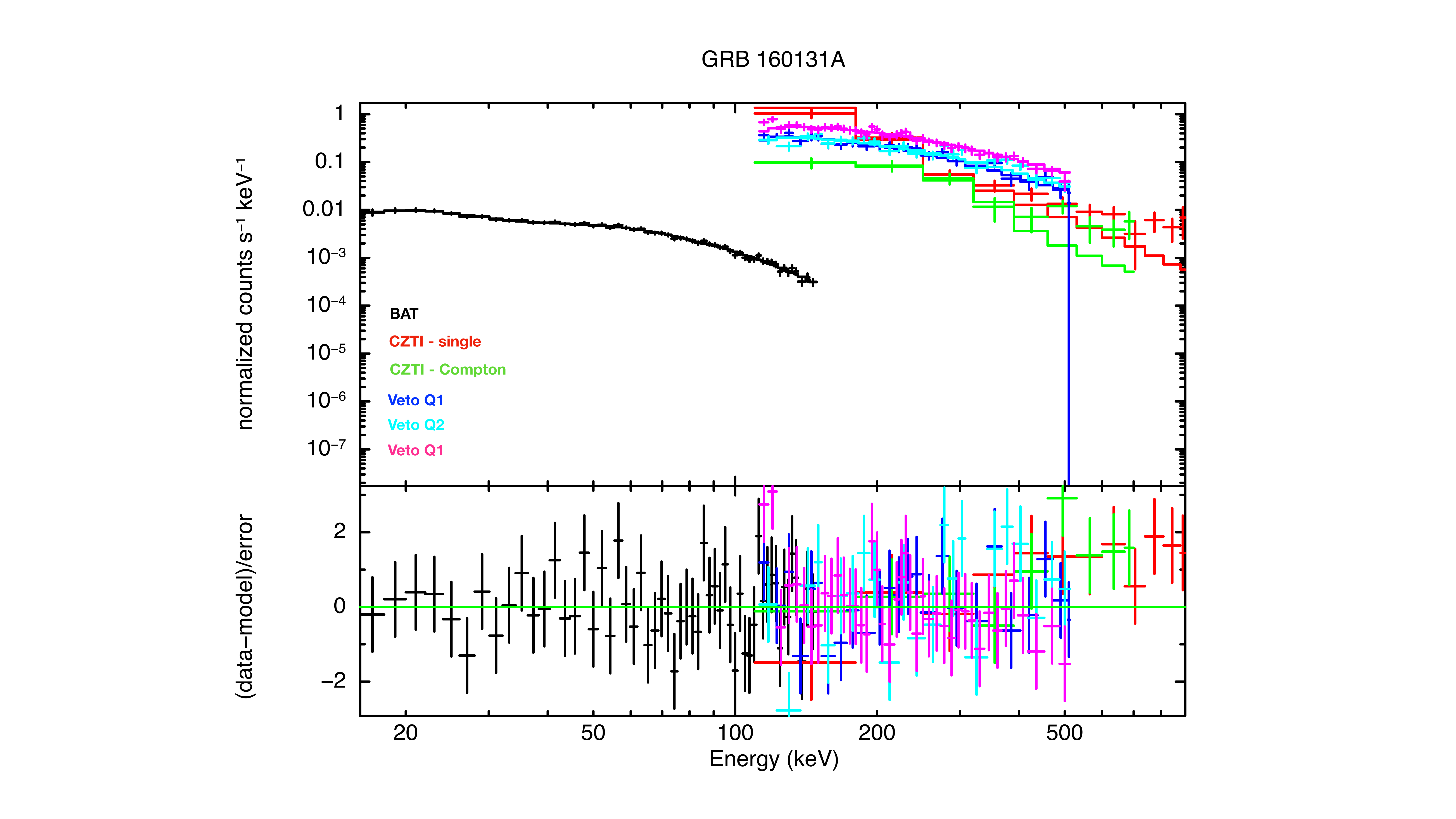}
\includegraphics[width=.42\linewidth]{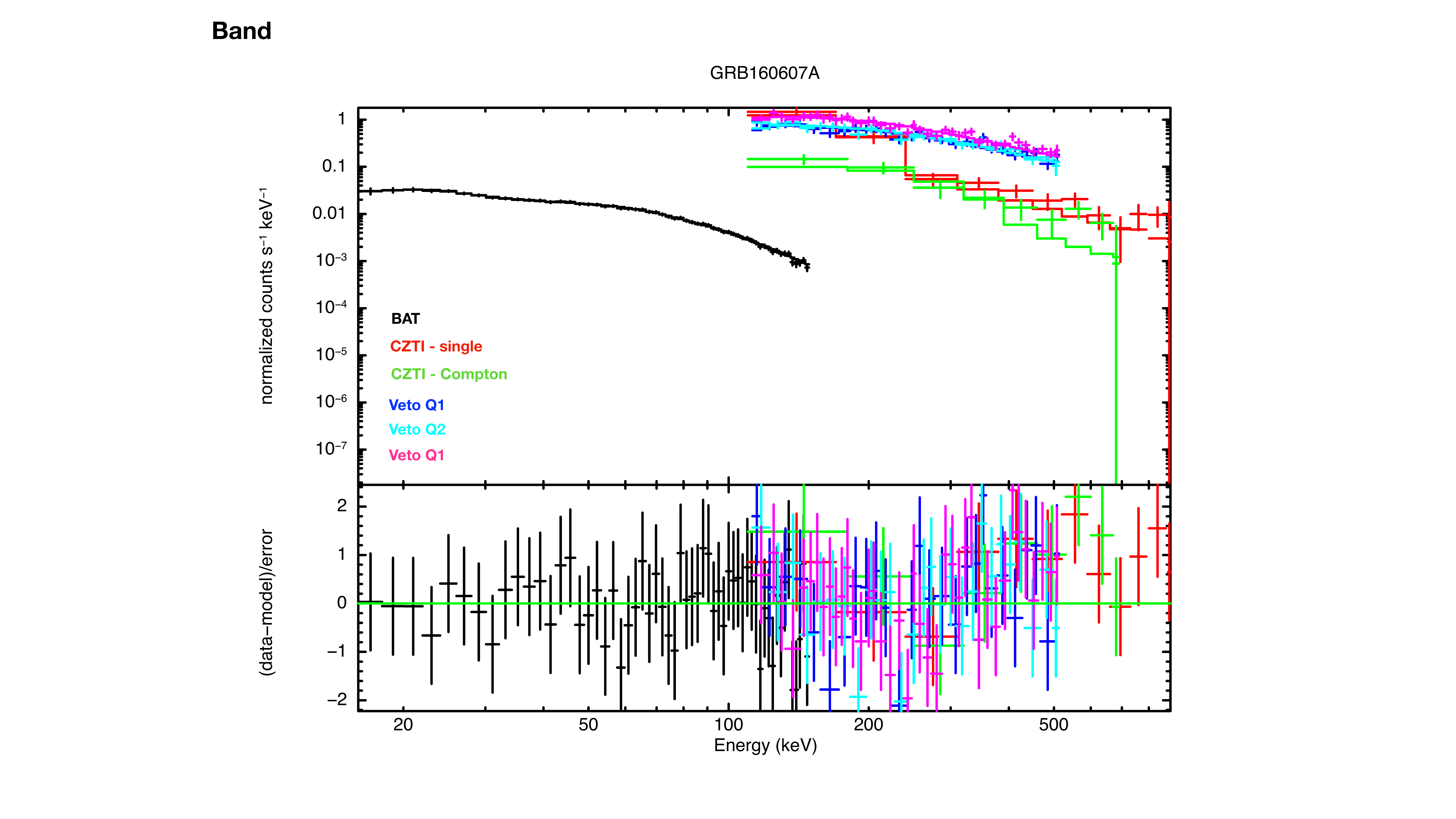}
\includegraphics[width=.42\linewidth]{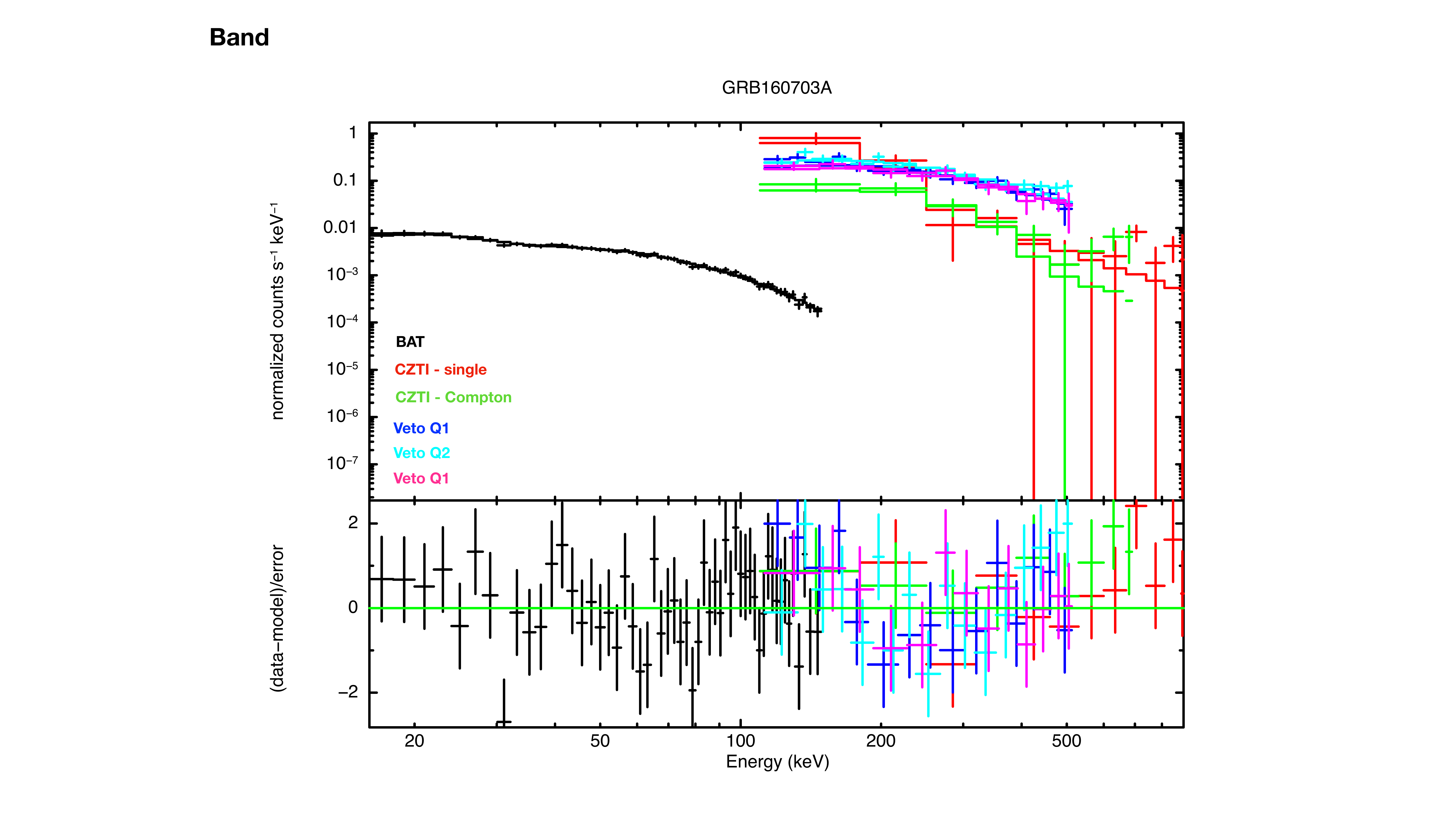}
\caption{The count spectra (upper panel) and the respective residuals (lower panel) obtained for the joint spectral analysis consisting of Niel Gehrels {\it Swift} BAT + CZTI data using the spectral model Band function for the bursts GRB 160131A (top left), GRB 160607A (top right) and GRB 160703A (bottom middle) are shown. Here we demonstrate that for bursts without {\it Fermi} detections, the usage of CZTI data extending until $900 \, \rm keV$ along with BAT, enables us to constrain the $E_{peak}$ of the GRB spectrum.}
\label{bat_czti_spec}
\end{figure*}

\begin{figure*}
\centering
\includegraphics[width=.8\linewidth]{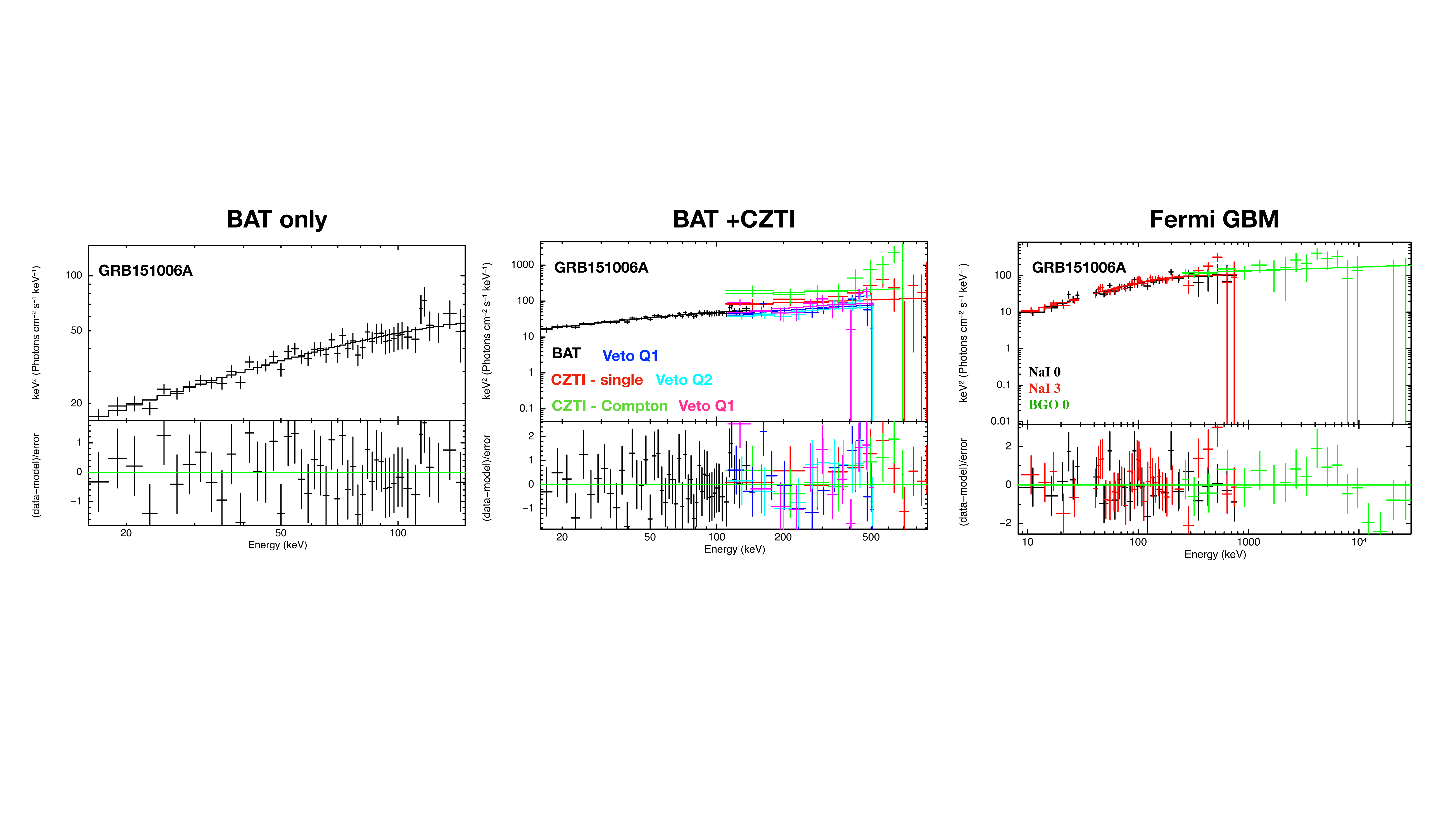}
\includegraphics[width=.8\linewidth]{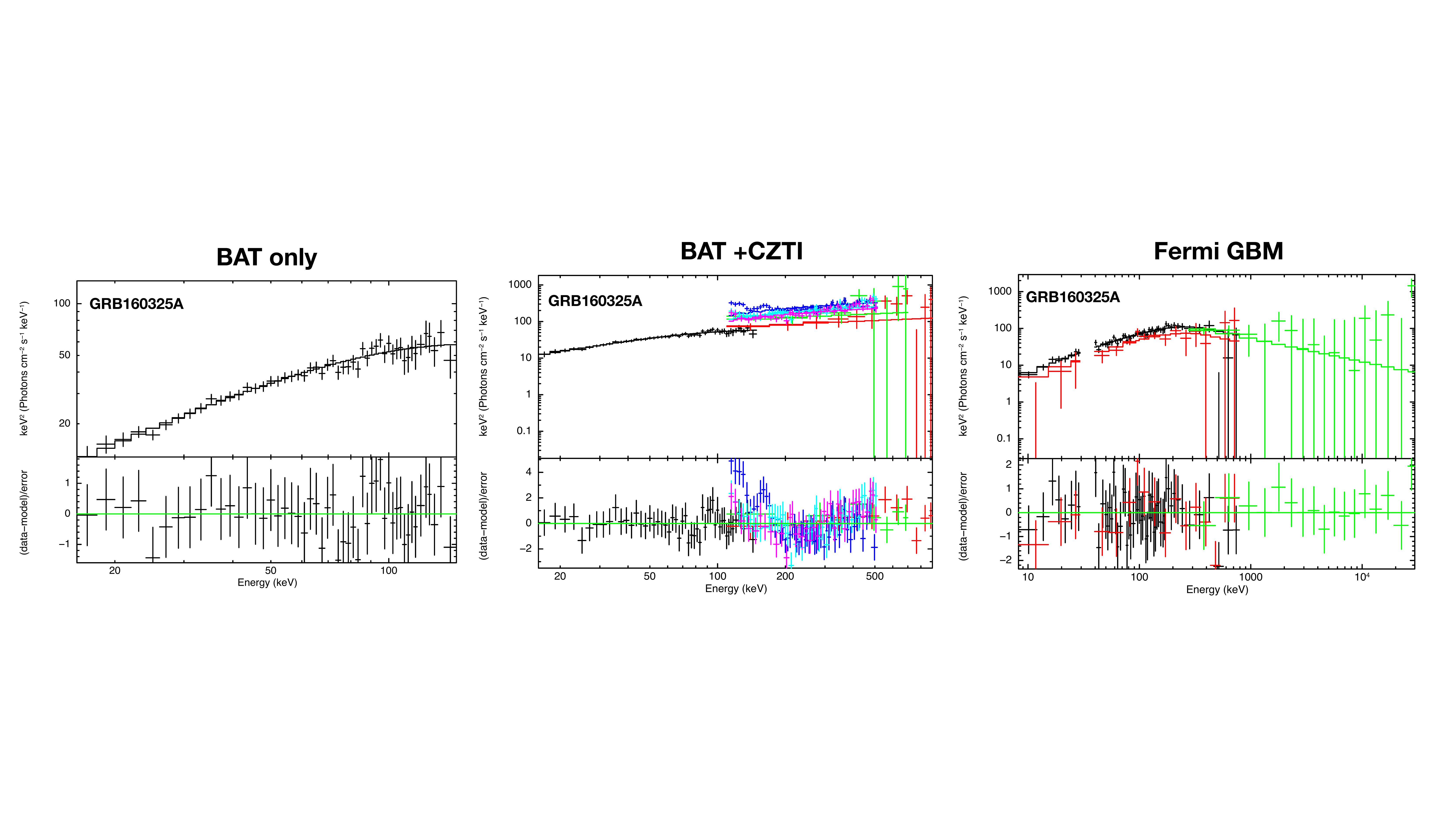}
\caption{For GRB 151006A (top panel) and GRB 160325A (bottom panel), we demonstrate that the analysis of the BAT data alone (top left and bottom left) does not allow us to ascertain the spectral peak and $\beta$ of the Band function fit to the data. However, when using CZTI data along with BAT (middle plot of top and below panel), we find that we can constrain the $E_{peak}$ of the spectrum which is reasonably consistent with that determined in solo {\it Fermi} GBM analysis (top right and bottom right). The fit parameter values are given in Table \ref{bat_czti_results}.}
\label{czti_comp}
\end{figure*}

The spectral fit results and the respective residuals obtained for the best fit empirical functions like Band function ({\tt Band}) and cutoff power law ({\tt CPL}) are reported in Table \ref{spec_fit_results} and shown in Figure \ref{fermi_czti_bat_spec}, \ref{bat_czti_spec} and \ref{czti_comp} respectively.  We find that residual obtained for CZTI spectral data are consistent with those obtained for ${\it Fermi}$. The residuals are found within $3\sigma$ for CZTI data. 

\begin{figure}
\centering
\includegraphics[width=.9\linewidth]{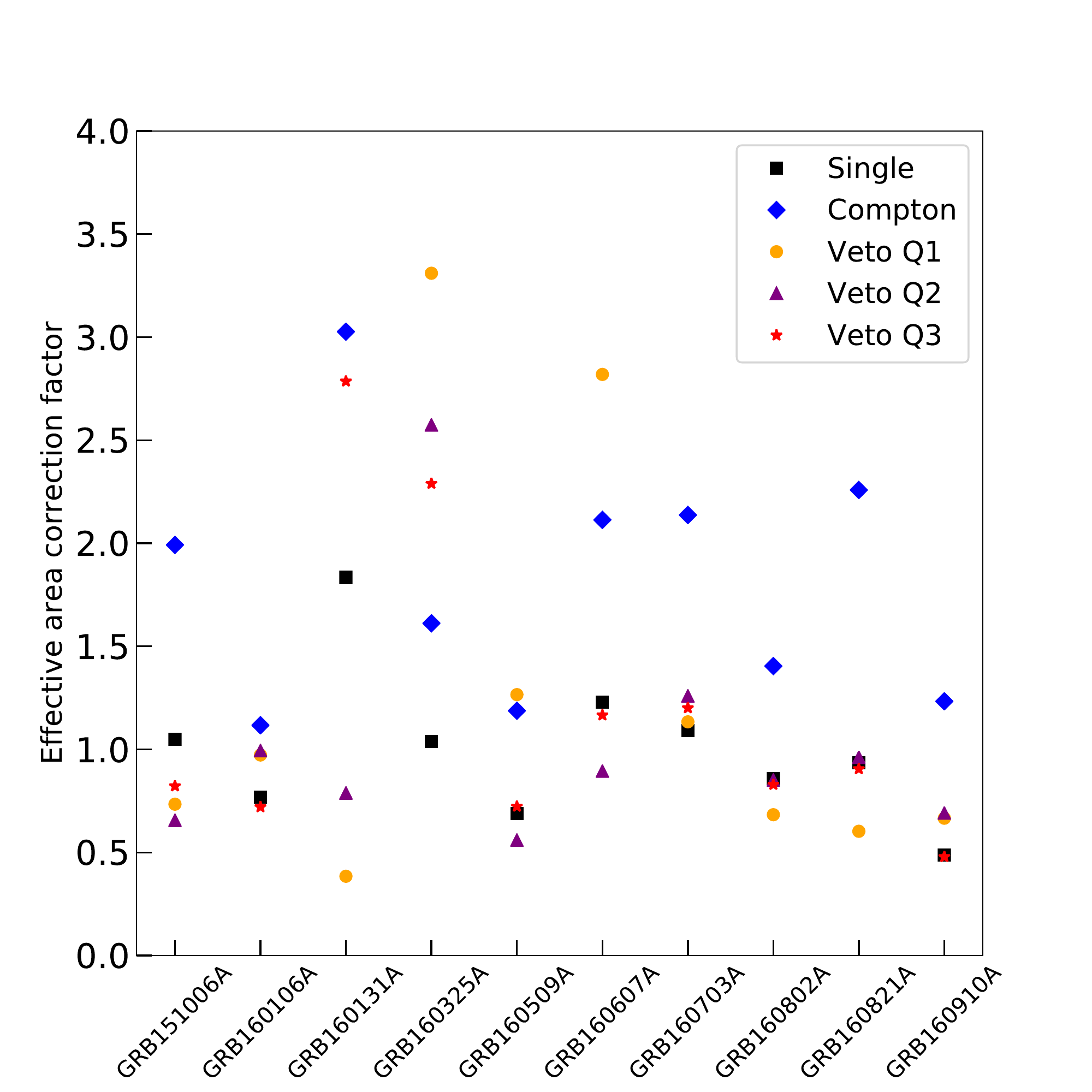}
\caption{The effective area correction factors obtained for the different CZTI datasets: single (black square), Compton (blue diamond), Veto Q1 (yellow circle), Veto Q2 (purple triangle) and Veto Q2 (red star) with respect to the brightest NaI detector of {\it Fermi} for different GRBs in the sample except for GRB 160131A, GRB 160607A and GRB 160703A where the values are obtained with respect to Swift BAT detector are shown.}
%\ref{eff_area}
\label{eff_area}
\end{figure}

The small energy window of {\it Swift} BAT ($15 - 150 \, \rm keV$) generally does not allow us to constrain the $E_{peak}$ of the spectrum in cases where there is only BAT detection. In case of GRB 160607A, GRB 160703A and GRB 160131A, where {\it Fermi} detections were not available, we conducted the spectral analysis using {\it Swift} BAT along with CZTI data. We demonstrate that the usage of CZTI data extending until $900 \, \rm keV$ allows us to well constrain the $E_{peak}$ of the spectrum. The effective area correction factor obtained between BAT and CZTI are shown in Figure \ref{eff_area}, where constant for BAT is frozen to unity.
The energy flux estimated in the range of $10 - 1000 \, \rm keV$ for the bursts are reported in the Table \ref{spec_fit_results}. 

{ During the joint spectral analysis using different detectors, we have tied the spectral parameters of all the detectors including the normalisation of the spectral model. The difference in count rates in different detectors are taken care of by including the effective area correction factor along with the spectral model that is used to analyse the data.}
To estimate the effective area correction factor between the different
detectors, we multiply an energy independent constant factor to the spectral model during the fitting process. 
The effective area correction factor obtained between {\it Fermi} and the different datasets of CZTI except in GRB 160131A, GRB 160607A and GRB 160703A where the values are obtained with respect to the {\it Swift} BAT
are shown in Figure \ref{eff_area}. 
On average, the normalization estimates of the empirical function fits done to the single, Compton, Veto 1, 2, and 3 data are found to vary around $20\%$, $55\%$, $55\%$, $40\%$ and $40\%$ of the normalization estimate of the brightest {\it Fermi} NaI detector respectively. While with respect to the BAT detector, the normalization estimates for the single, Compton, Veto 1, 2, and 3 events, vary around $40\%$, $140\%$, $85\%$, $20\%$ and $70\%$ respectively. In certain GRBs, we observe low normalizations for Compton and Veto data which results in an effective area correction factor $> 2$. The cause of such cases are being studied.   

For GRB 151006A and GRB 160325A, both {\it Fermi} and BAT observations are available. So, in these GRBs, we conduct a joint spectral analysis of BAT and CZTI data and then compare the spectral fit results with that obtained using {\it Fermi} GBM data alone. We are able to ascertain the $\alpha$, $E_{peak}$ and normalization 
values which are reasonably consistent with {\it Fermi} GBM results, within $90\%$ error limits (Table \ref{bat_czti_results} and Figure \ref{czti_comp}). This further endorse the capability of CZTI as a sub-MeV spectrometer along with BAT to determine the GRB spectrum. 

\begin{figure}
\centering
\includegraphics[width=.9\linewidth]{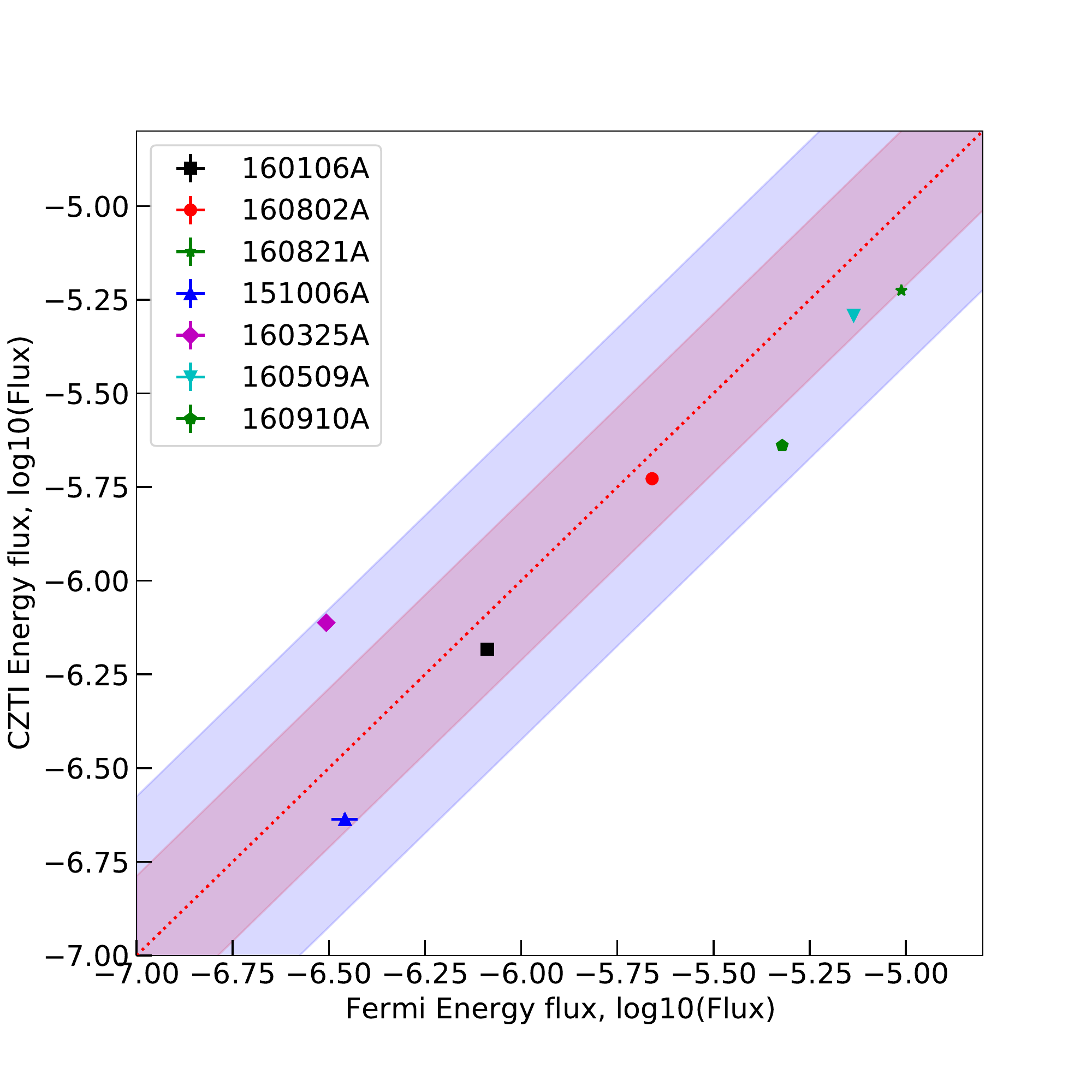}
\caption{Above the CZTI flux estimates (Y) done for the spectral fits done to CZTI data alone versus the energy flux estimates done for {\it Fermi} (X) alone spectral fits are shown. The red (blue) shaded region marks the $1 \sigma$ ($2 \sigma$) scatter of the distribution of points around the Y=X line shown in dotted red line.}
%\ref{eff_area}
\label{flux_comp}
\end{figure}

We note here that being opaque below $100$ keV, CZTI spectrum alone cannot measure the GRB spectral parameters fully. On the other hand, if we assume canonical values for the power law indices ($\alpha = -1$ and $\beta = -2.5$) of the Band function, we can constrain the $E_{peak}$ and normalisation of the spectrum.
In certain cases, the $E_{peak}$ estimates are found to lie close to the edge or outside the 
energy window of CZTI (e.g GRB160509A and GRB160821A). 
{ In Figure 
\ref{flux_comp}, the energy fluxes estimated in the energy range $100 \, \rm keV -1000 \, \rm keV$ for spectral fits of CZTI data alone (where the power law indices are frozen to the canonical values), are plotted against the respective energy flux estimated from the {\it Fermi} data only spectral fits (where all the fit parameters are left free) of the different bursts.} We find the CZTI flux estimates are consistent within $2 \sigma$ scatter\footnote{The scatter is the standard deviation of the Gaussian fit to the distribution of the displacement of the CZTI measured flux from the {\it Fermi} flux and is found to be $\sigma =0.21$.} around the line 
denoting CZTI energy flux is equivalent to {\it Fermi} flux.

\begin{table*}
	%\centering
	 \resizebox{0.7\textwidth}{!}{\begin{minipage}{\textwidth}
		 \caption{Spectral fit results of the analysis of the time integrated emission of the bursts in the sample using the CZTI, {\it Fermi} and  Niel Gehrel {\it Swift} BAT data.}
	\label{spec_fit_results}
	\begin{tabular}{*{10}{l}}
	
	%\begin{tabular}{l l{0.1\textwidth}l{0.1\textwidth}l{0.1\textwidth}l{0.1\textwidth}l l{0.1\textwidth}l{0.1\textwidth}} % four columns, alignment for each
		\hline
		GRB name & T$_{\rm start}$ & T$_{\rm stop}$ & $\alpha$ & $\beta$ & $\rm E_{peak}/E_{cut}$ &  log$_{10}$(Flux) & Chi-square/& Model Fit & Other \\
		& (s)& (s) & & &(keV) & (erg/cm$^2$/s)&DOF & & Instruments \\[0.1cm]
		\hline
		GRB151006A & -1.11 &30.21  & $-1.31_{-0.02}^{+0.02}$ & $-2.19_{-0.04}^{+0.03}$ & $1000_{-139}^{+1008}$ & $-6.37 \pm 0.005$ & $664.70/772$ & Band & GBM + LAT + BAT \\[0.2cm]
		GRB160106A & 0.41 & 47.26 &$-0.66 \pm 0.05$ &$-2.33_{-0.18}^{+0.13}$ &$246_{-24}^{+26}$ &$ -6.07 \pm 0.005$ & $520.04/697$  &Band & GBM \\[0.2cm]
		GRB160131A & 11.95 & 49.63 & $1.02 \pm 0.03$ & -- & $407^{+52}_{-39}$ & $-6.2^{+0.004}_{-0.005}$ & $297.04/436$ & Cutoffpl & BAT \\[0.2cm]
		GRB160325A & -1.31 & 46.41 & $-0.85^{+0.07}_{-0.05}$ & $-1.82^{+0.03}_{-0.04}$ & $156^{+20}_{-21}$ & $-6.37 \pm 0.004$ & 717.06/768 & Band & GBM + BAT \\[0.2cm]
		GRB160509A & 7.63 & 22.06 & $-0.75 \pm 0.01$ & $-2.16 \pm0.01$ & $279^{+7}_{-6}$ & $-5.09 \pm -0.001$ & $896.15/709$ & Band & GBM + LAT  \\[0.2cm]
		GRB160607A &-0.72 & 14.96  &$0.80_{-0.12}^{+0.18}$ &$-1.57 \pm 0.05$ &$ 108_{-32}^{+35}$ &$-5.55 \pm 0.003$& 250.34/420 & Band & BAT \\[0.2cm]
		%GRB160623A &-1.44 &26.86  & & & & & & -\\
		GRB160703A &-2.24 &28.31  &$-0.94_{-0.06}^{+0.07}$ &$-2.17_{-0.17}^{+0.15}$ &$195_{-34}^{+45}$ &$ -6.28 \pm 0.004$ & 264.04/423 &Band & BAT \\[0.2cm]
		GRB160802A &-0.36 & 17.84 &$0.82 \pm 0.02$ & - &$263_{-9}^{+10}$ &$ -5.61 \pm 0.002$ & $743.78/690$ &Cutoffpl & GBM \\[0.2cm]
		GRB160821A &113.47 &159.87  & $-0.98\pm 0.005$ &$-2.13\pm 0.02$ &$860_{-23}^{+22}$ &$-5.01 \pm 0.0008$ & $1178.55/699$ &Band $\times$ Highecut & GBM + LAT \\[0.2cm]
		GRB160910A & 3.27 & 15.04 & $-0.61_{-0.02}^{+0.02}$ & $-2.37_{-0.04}^{+0.04}$ & $222_{-8}^{+9}$ & $-5.34_{-0.002}^{+0.002}$ & 778.32/680 & Band & GBM \\[0.2cm]
 		
		\hline
		
\hline
\end{tabular}\\

%\tablenotetext{}
	  {The errors are reported for $68\%$ confidence interval.
The references for the burst detection in different instruments are provided. GRB151006A - GBM \citep{151006A_GBM}, LAT \citep{151006A_Lat_detection}, BAT \citep{151006A_BAT}; GRB160131A - Fermi-GBM trigger number 473813134; GRB160325A - GBM \citep{roberts16_160325A}, LAT \citep{axelsson16_160325A}, BAT \citep{160325A_bat}; GRB160509A - GBM \citep{160509A_gbm}, LLE \citep{160509A_LAT}; GRB160607A - BAT \citep{lien16_160607A}; GRB160703A - BAT \citep{lien16_160703A}; GRB160802A - GBM \citep{160802A_gbm}; GRB160821A - GBM \citep{160821A_gbm}, LAT \citep{160821A_lat}; GRB160910A - GBM \citep{160910A_gbm}} 
\end{minipage}}
\end{table*}

\begin{table*}
	\centering
	 \resizebox{0.9\textwidth}{!}{\begin{minipage}{\textwidth}
	\caption{The Band model fit comparison between BAT + CZTI and {\em Fermi} alone analysis of the bursts GRB 151006A and GRB 160325A. The errors are reported for $90\%$ confidence interval.}
	\label{bat_czti_results}
	\begin{tabular}{*{5}{l}}
		\hline
			GRB name & Band Parameters & BAT & BAT + CZTI & Fermi  \\
			\hline
			 GRB151006A & $\alpha$ & $-1.25^{+0.07}_{-0.14}$ & $-1.23^{+0.15}_{-0.12}$ & $-1.08^{+0.12}_{-0.13}$ \\[0.2cm]
			 & $\beta$  & $-9.37^{+19}_{-0.0}$ & $-1.79^{+0.18}_{-0.17}$ & $-1.89^{+0.11}_{-0.20}$ \\[0.2cm]
			 & $E_{peak} \, \rm (keV)$ & $288^{+257}_{-117}$ & $262^{+44}_{-24}$ & $350^{+400}_{-126}$ \\[0.2cm]
			 & $Norm$ & $0.007^{+0.001}_{-0.0009}$ & $0.007^{+0.002}_{-0.001}$ & $0.008^{+0.002}_{-0.001}$ \\[0.2cm]
			 & $\chi_{red}^2$ & 0.68 & 0.69 & 1.02 \\[0.2cm]
			 \hline 
			 GRB160325A & $\alpha$ & $-0.87^{+0.13}_{-0.12}$ & $-0.82^{+0.08}_{-0.16}$ & $-0.77^{+0.10}_{-0.09}$ \\[0.2cm]
			 & $\beta$  & $-10^{+1e-15}_{-0.0}$ & $-1.74^{+0.06}_{-0.09}$ & $-2.63^{+0.42}_{-2.36}$ \\[0.2cm]
			 & $E_{peak} \, \rm (keV)$ & $137^{+54}_{-27}$ & $124^{+44}_{-24}$ & $214^{+53}_{-43}$ \\[0.2cm]
			 & $Norm$ & $0.02^{+0.003}_{-0.002}$ & $0.01^{+0.002}_{-0.003}$ & $0.01^{+0.002}_{-0.001}$ \\[0.2cm]
			 & $\chi_{red}^2$ & 0.55 & 0.91 & 0.81 \\[0.2cm]

\hline
\end{tabular}\\
\end{minipage}}
\end{table*}

\section{Summary and future plan}

CZT-Imager on board {\em AstroSat} has been a prolific GRB monitor with around detection of nearly 83 GRBs per year. I %While spectro-polarimetric performance of CZTI has been demonstrated earlier for multiple GRBs (see \citet{rao16,basak17,chand18a,chand18b,chattopadhyay19,vidushi19,Sharma20}), 
In this article, we explored the spectroscopic sensitivity of CZTI in the sub-MeV region by attempting spectroscopic 
%and polarization 
analysis for some of the bright GRBs detected in the first year (October 2015 $-$ Spetember 2016) of {\em AstroSat} operation. 
The improvement in the spectroscopic sensitivity has been possible because of (1) inclusion of the low-gain CZTI pixels after a thorough calibration which consists of around 20 \% of the CZTI detection area, (2) identification and removal of 2-pixel noisy events. Both the methods improve the S/N of the bursts significantly and in particular the low-gain pixels enable the spectroscopy all the way up to 900 keV (1-pixel Compton spectroscopy: 100 $-$ 700 keV, 1-pixel spectroscopy: 100 $-$ 900 keV) .
%and GRB polarimetry up to 600 keV. 
We also utilize the CsI (or Veto) detectors for spectroscopy in 100 $-$ 500 keV to enhance the overall sensitivity.       

In Section \ref{result_spec}, we performed joint {\it Fermi} and {\em AstroSat} (and BAT wherever available) spectral analysis for 10 out of the eleven first year GRBs (except GRB 160623A where a concurrent observation with {\it Fermi} was not available) in the full burst region. We are able to obtain
spectral fit parameter values that are in close agreement with those obtained in solo {\it Fermi} analysis. This provides an independent validation of the {\em AstroSat} mass model, thereby boosting the confidence in the spectral 
%and polarization 
analysis of the CZTI GRBs. Spectral validation of the mass model and availability of CZTI spectra up to 900 keV also allows to explore spectral study of the GRBs detected only by Swift-BAT and CZTI but not by {\it Fermi}. This aspect has been particularly demonstrated in the case of GRB151006A and GRB160325A where we find reasonably consistent spectral fit values for BAT + CZTI in comparison to solo {\it Fermi} data analysis of these bursts. Thus, the satisfactory spectral fits obtained in 15 $-$ 900 keV (15 $-$ 150 keV from BAT, 100 $-$ 900 keV from CZTI) for GRB 160607A, GRB 160131A and GRB 160703A demonstrates the importance of CZTI sub-MeV spectroscopic capability particularly to characterize the GRBs that are not detected by {\it Fermi}. 
We also identify possible systematics involved in the mass model and attempt to quantify them ($<$15 \%) in the front and rear sides of the spacecraft.                   
%In section \ref{result_pol}, we report the polarization results for six of the eleven first year GRBs in the full burst region in 100 $-$ 600 keV after the implementation of the new techniques. We ignore the GRBs detected in 60 $-$ 90$^\circ$ angle with respect to the CZTI normal for all the future CZTI GRBs unless the detected GRB is extremely bright. The GRBs are found to be unpolarized when the full burst regions are considered for analysis. However, it has been shown in previous works such as \citet{chattopadhyay19} (and \citet{vidushi19} for GRB 160821A) that some intervals within the burst duration show hints of polarization in certain energy ranges. Thus, the unpolarized nature of the GRBs for the entire prompt emission is likely to be produced by the result of varying polarisation angle within the burst duration. We find our results to be consistent with the recent polarimetry findings reported by POLAR \citep{zhang19}.     

This paper primarily describes the new methods of sub-MeV spectroscopy with CZTI. We are continuing to refine these methods further, and will extensively test them against a much larger sample of bright GRBs detected by CZTI in the last five years. 
We also plan to explore the feasibility of using the CZT detectors and the CsI detectors in Compton camera configuration to enhance the spectroscopic sensitivity of the instrument. From a preliminary analysis, we could successfully detect the GRBs in the veto-tagged events (Compton scattered photons from CZT detectors which are absorbed by the CsI detectors) after applying Compton scattering kinematic conditions. 
\begin{figure}
\centering
\includegraphics[width=\hsize]{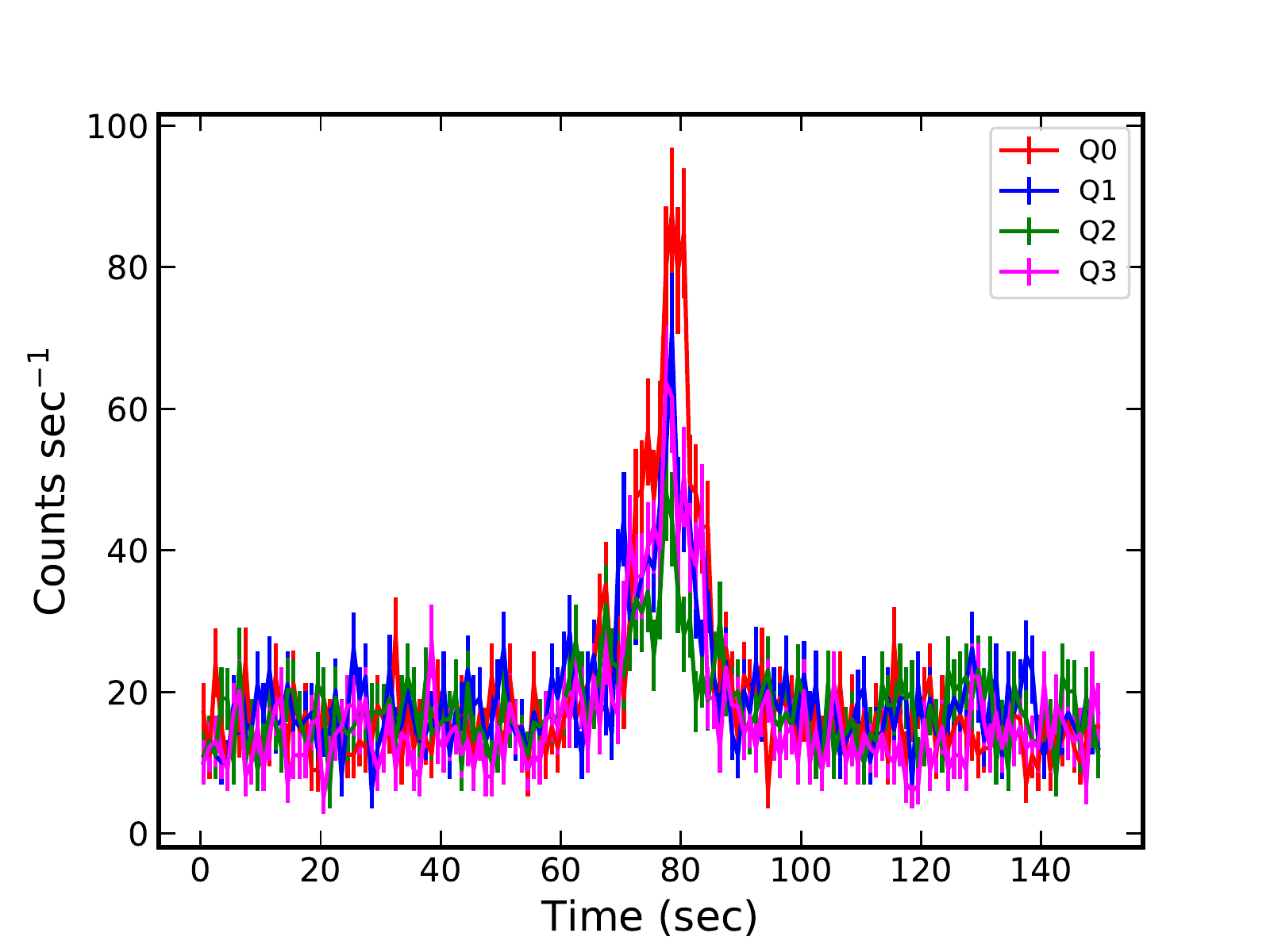}
\caption{Detection of GRB 160821A in the veto-tagged events. Different colours stand for different CZTI quadrants. The time-axis is plotted from {\em AstroSat} time  209507728 seconds (marked as zero) onward.}
\label{grb160821A_vetotagged}
\end{figure}
We plan to use the {\em AstroSat} mass model to generate response matrix for the veto-tagged events. 
%On the other hand, in order to further improve the polarimetric sensitivity of CZTI,we plan to explore the possibility of using the Compton scattering events from next neighboring pixels in the analysis. In the current analysis pipeline, we extract the Compton events from the neighboring pixels to get the 8-bin azimuthal angle distribution. The simulation results predict around 20 \% of additional Compton events from the next neighboring pixels. Identification of these events will not only enhance the signal strength but also offer additional azimuthal bins to obtain more comprehensive modulation curves for the polarized GRBs.     

\section*{Appendix A}

\begin{figure*}[h]
\centering
\includegraphics[width=.45\linewidth]{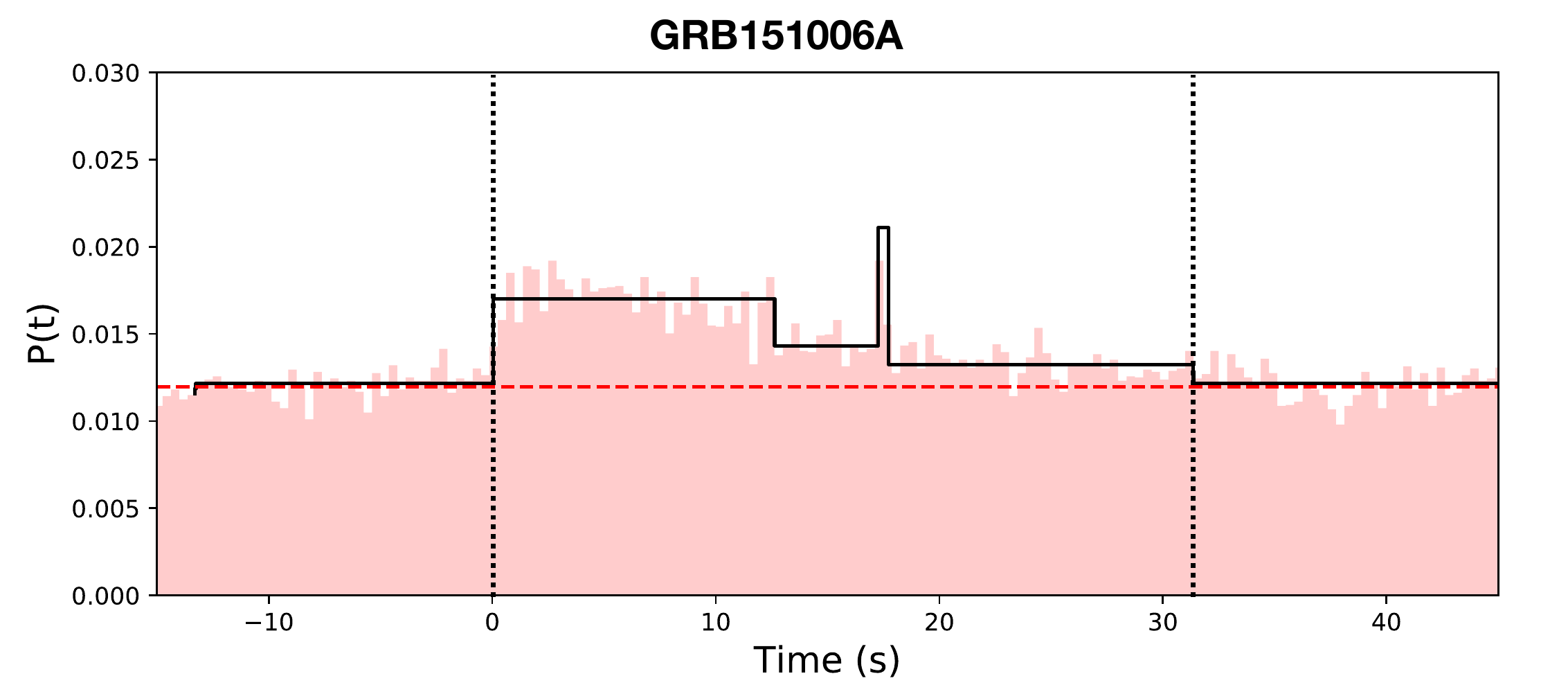}
\includegraphics[width=.45\linewidth]{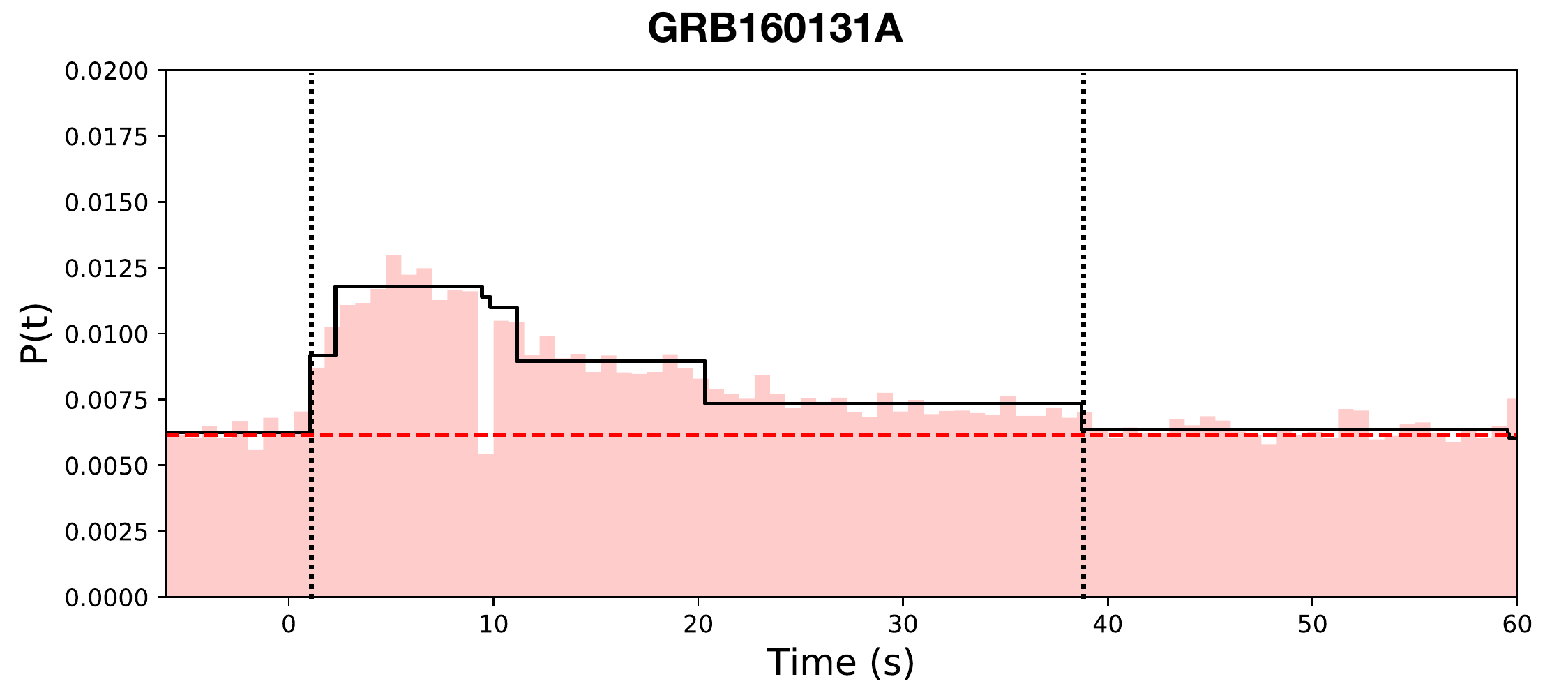}
\includegraphics[width=.45\linewidth]{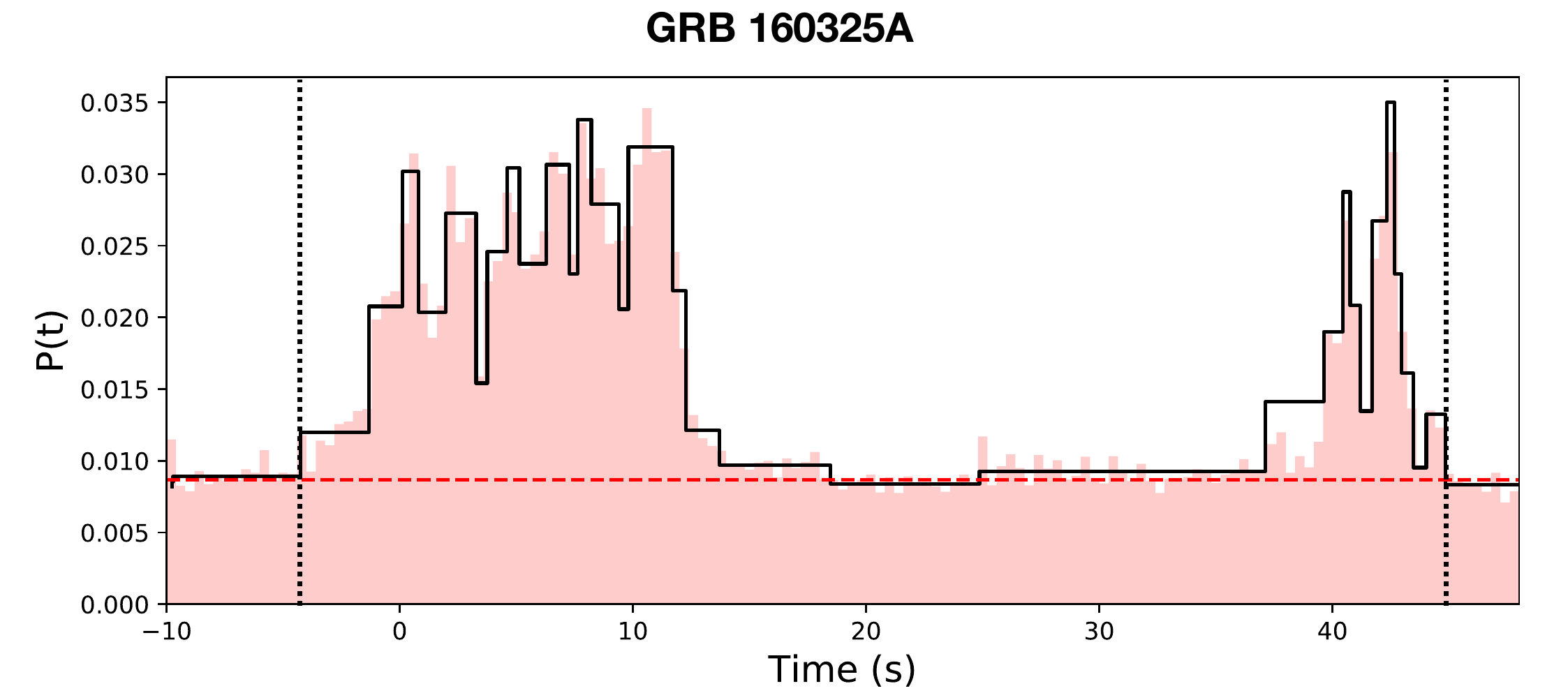}
\includegraphics[width=.45\linewidth]{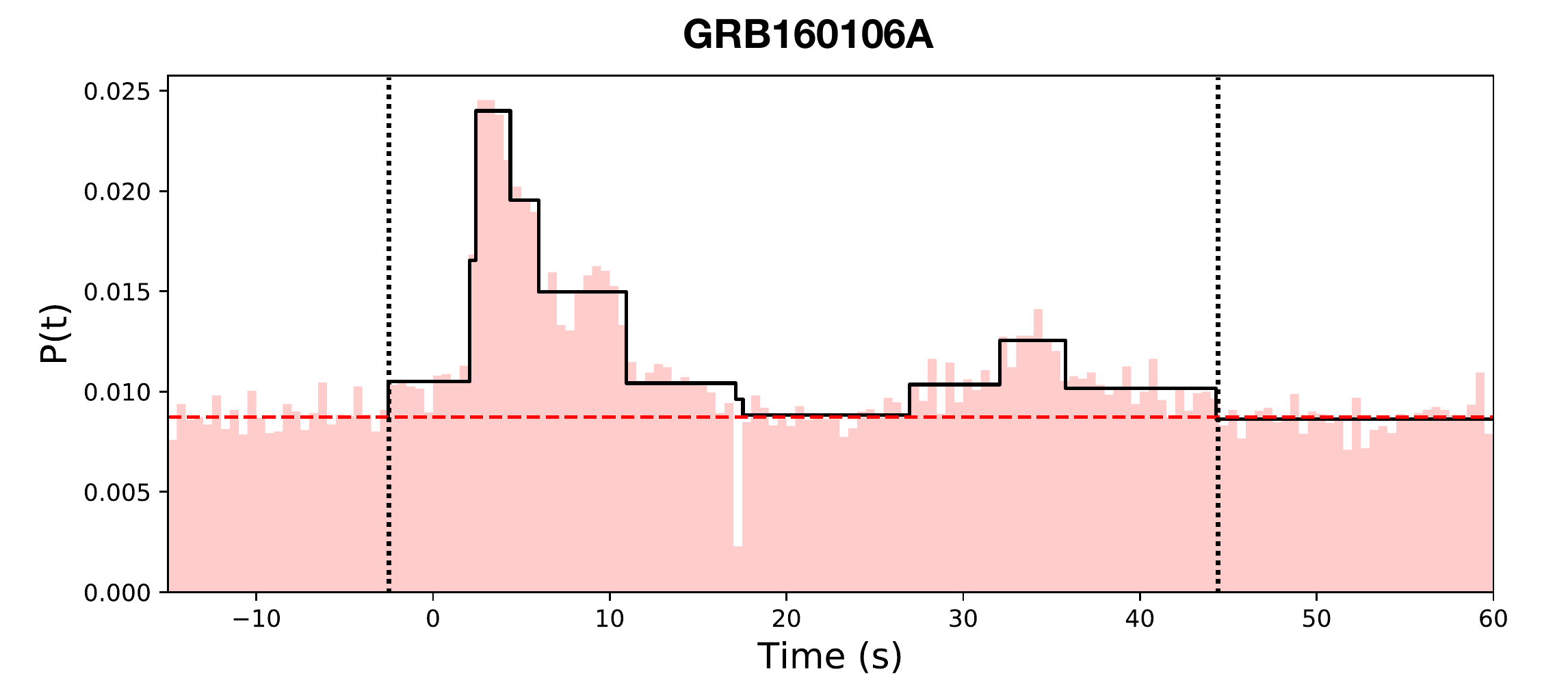}
\includegraphics[width=.45\linewidth]{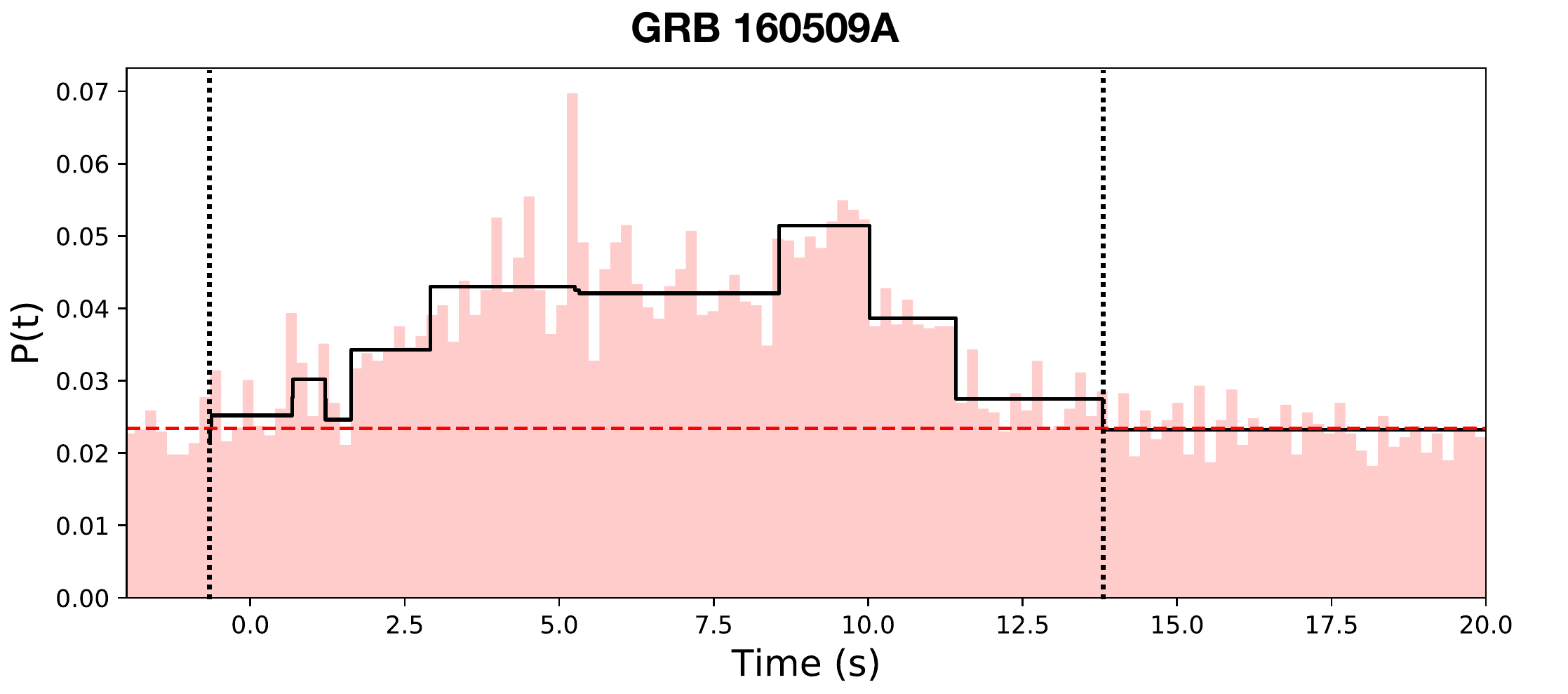}
\includegraphics[width=.45\linewidth]{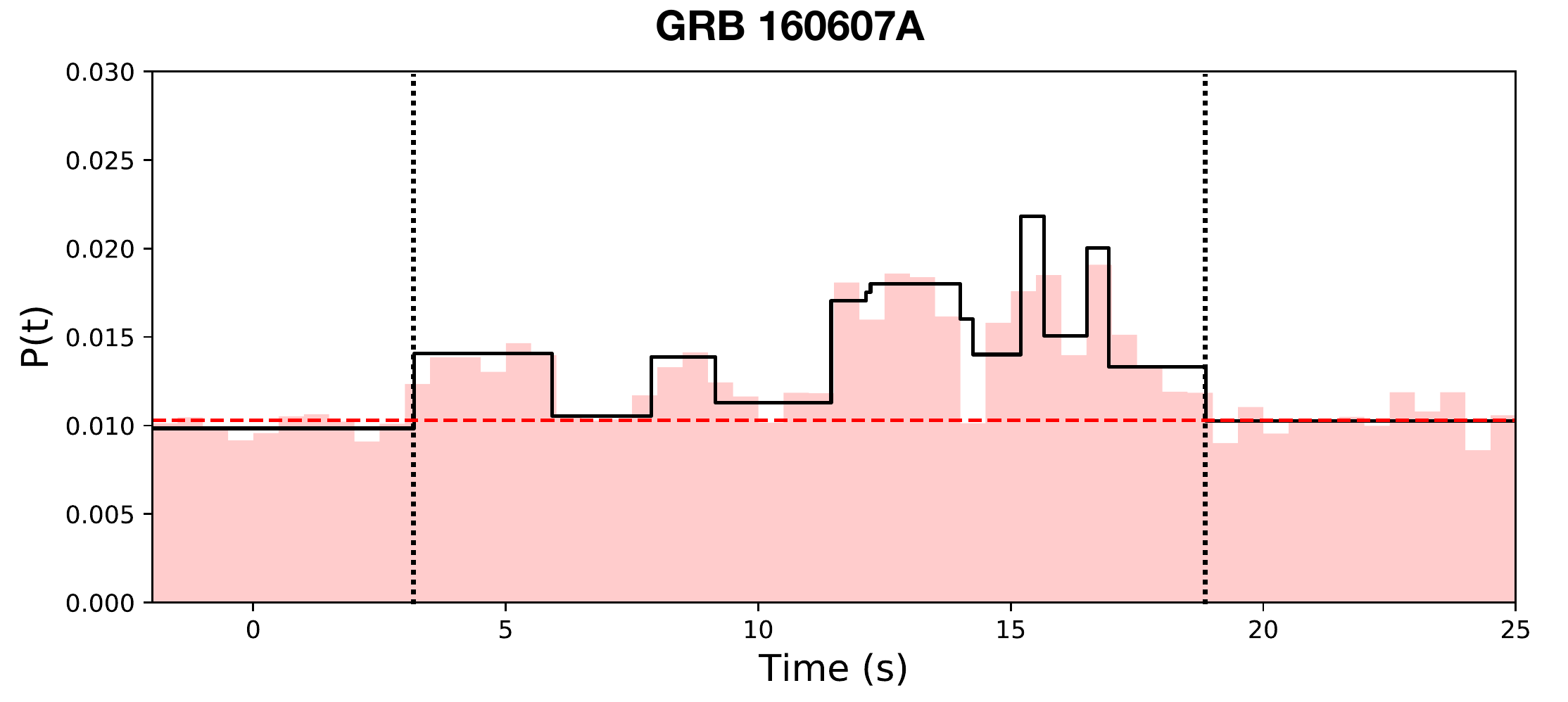}
\includegraphics[width=.45\linewidth]{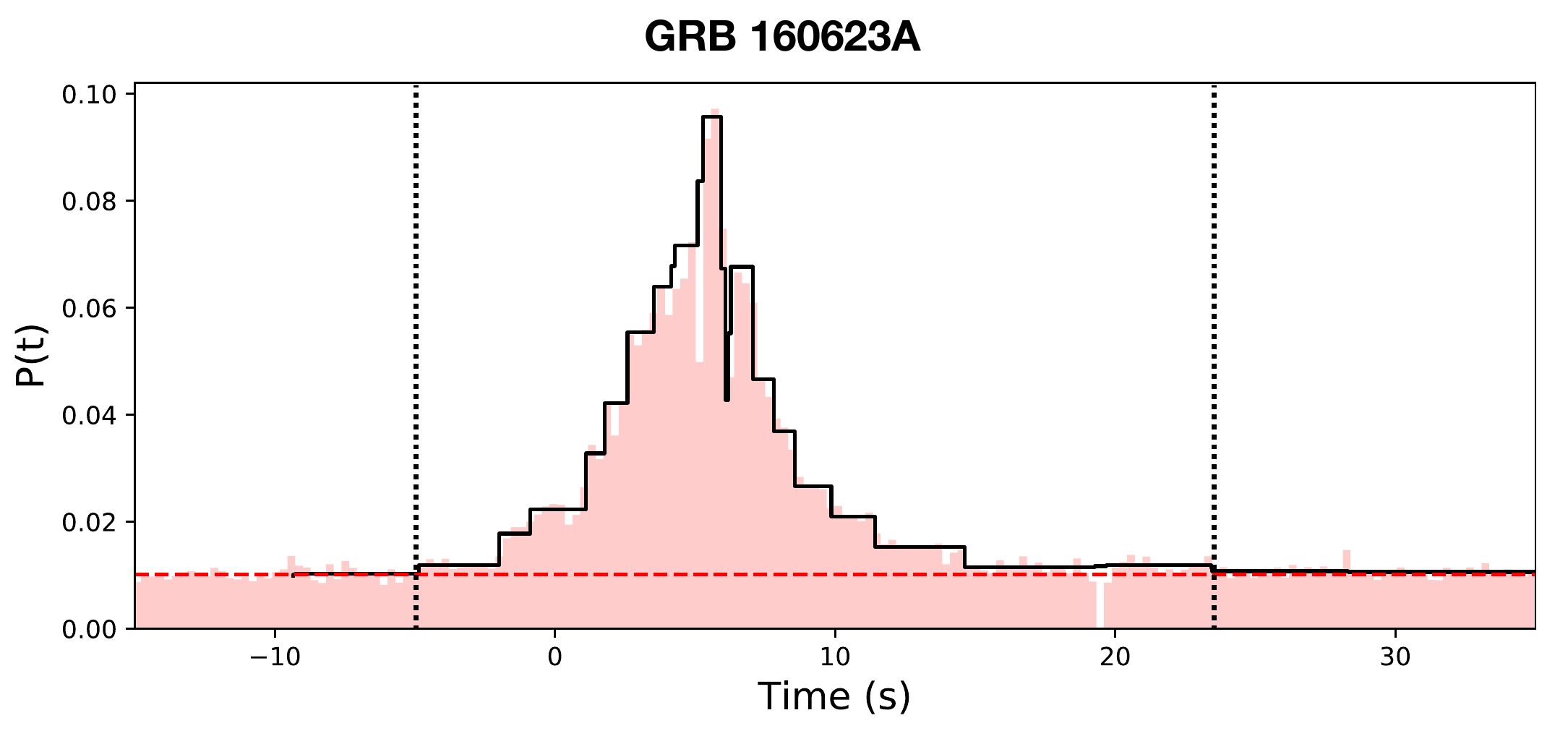}
\includegraphics[width=.45\linewidth]{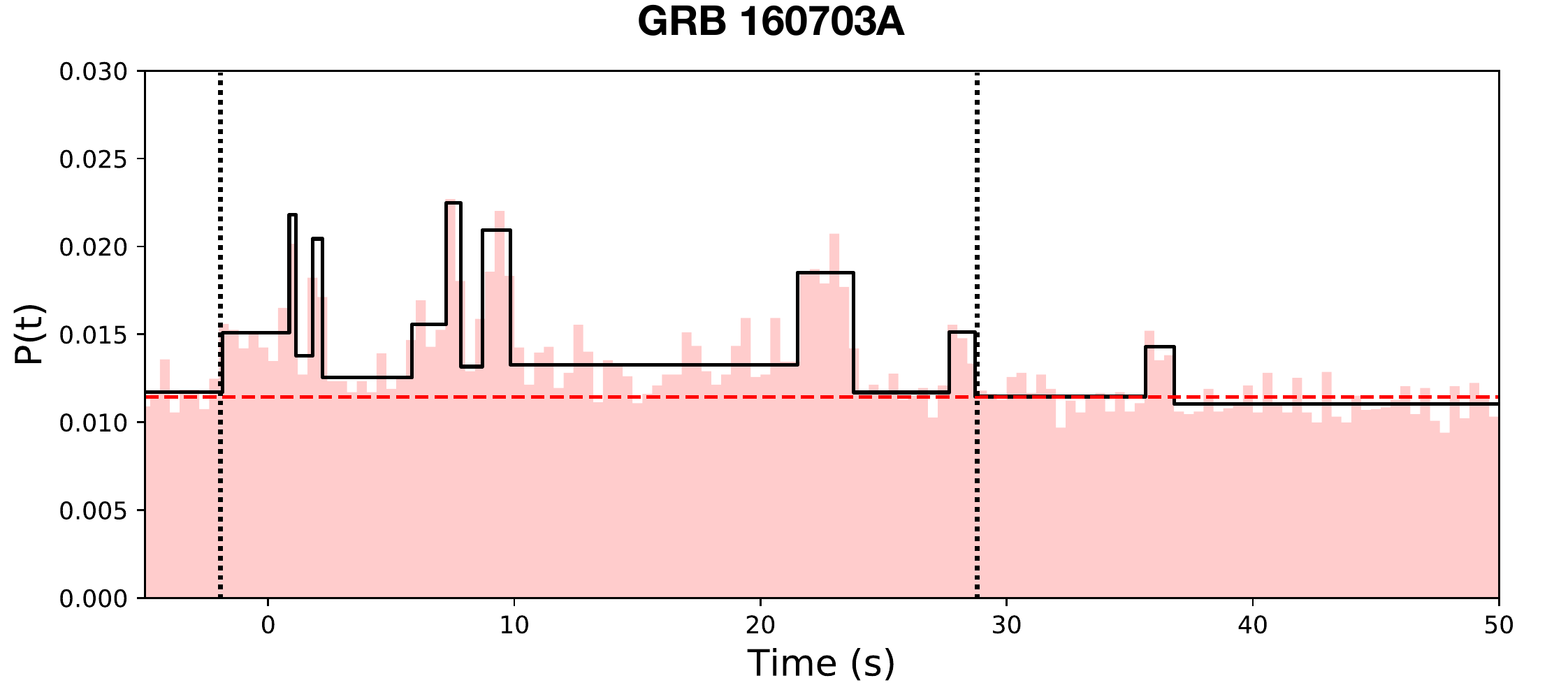}
\includegraphics[width=.45\linewidth]{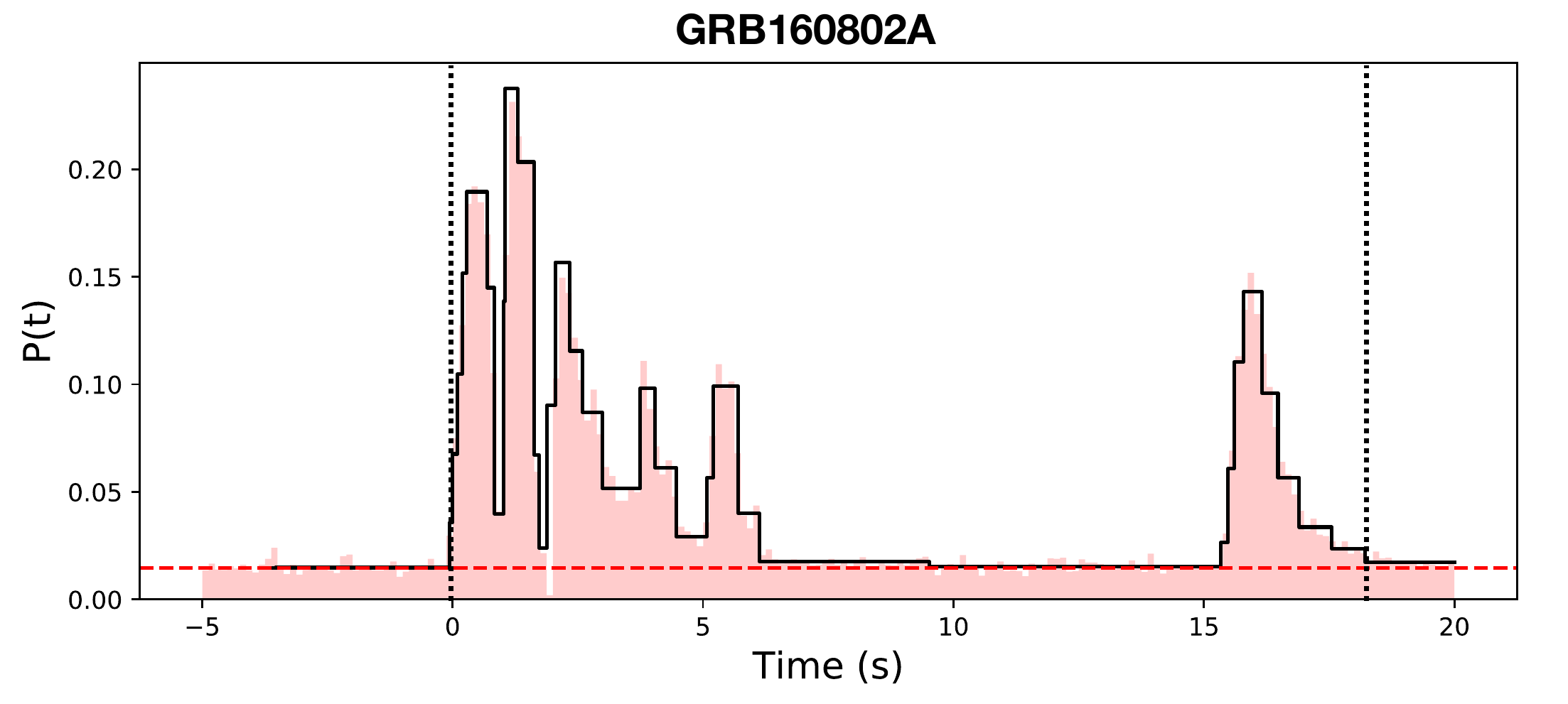}
\includegraphics[width=.45\linewidth]{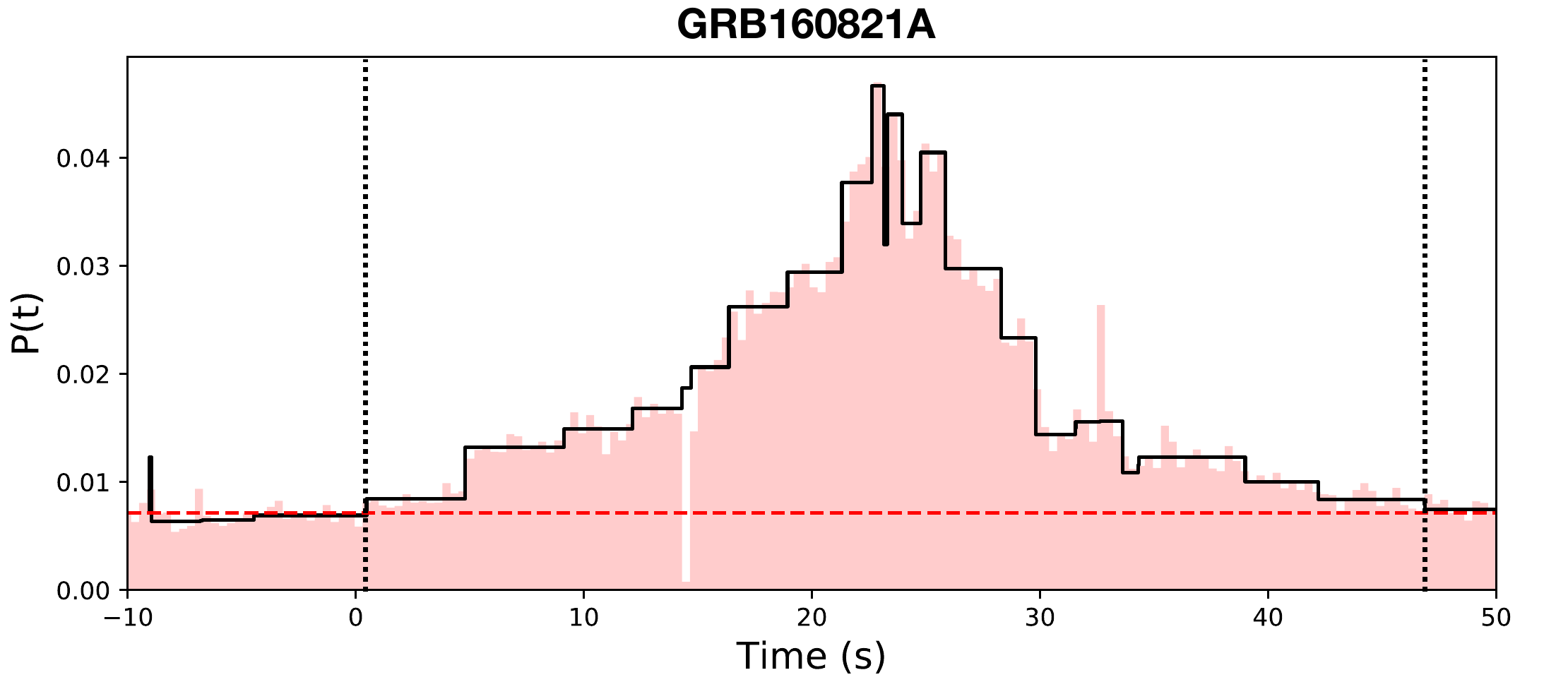}
\includegraphics[width=.45\linewidth]{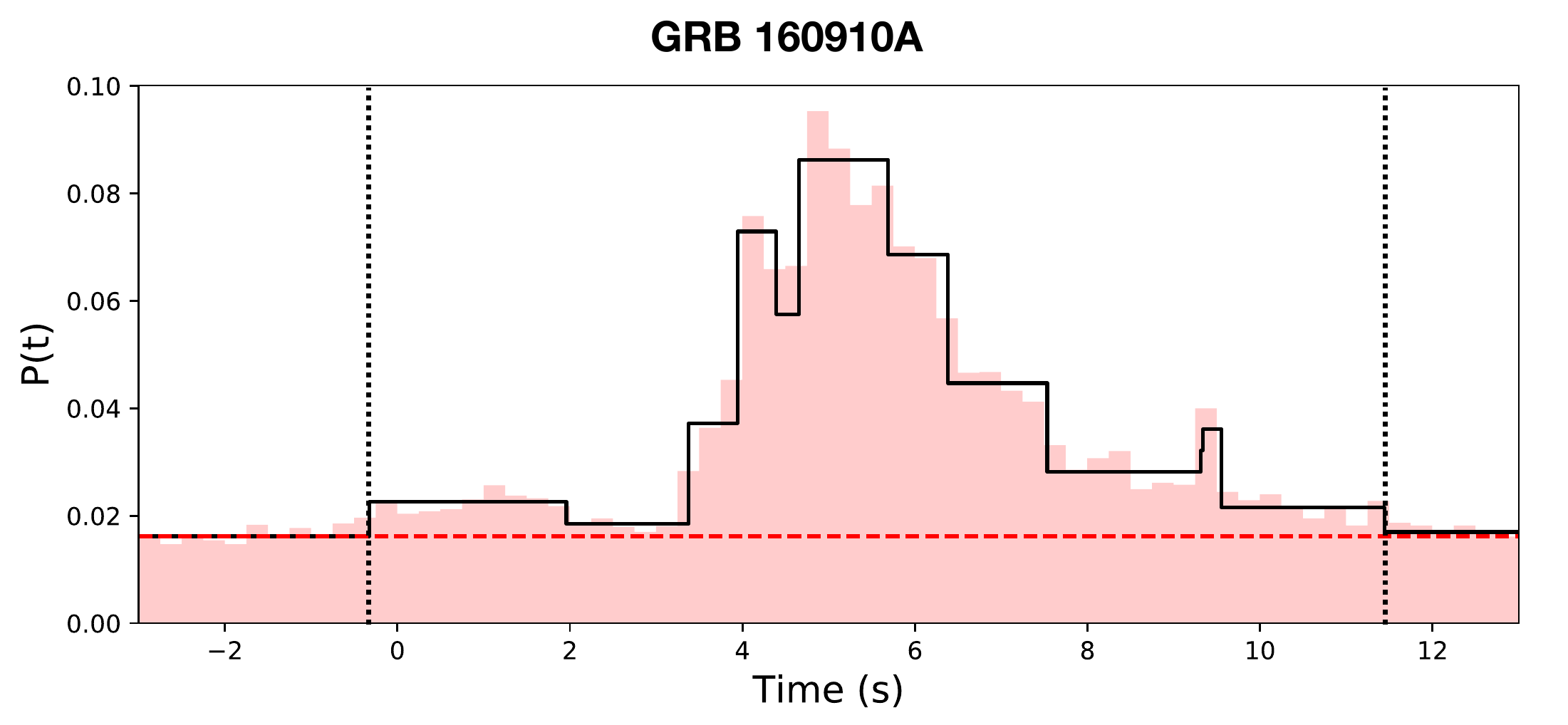}
\caption{The Bayesian block binning of the single event CZTI light curve of the bursts are shown above in black solid lines. The time interval of the integrated emission of each burst is marked by the vertical dotted lines on the respective plots. The red dashed horizontal line marks the background level. The basic light curve is plotted in the background in pink colour. We note that here the $0$ marks the start of the $T_{90}$ region of the burst.}
\label{Bayesian_selection}
\end{figure*}

Plots for the Bayesian block analysis conducted on single event data of the GRBs are shown here.

%%Use section* for acknowledgements
\section*{Acknowledgements}
This publication uses data from the {\em AstroSat} mission of the
Indian Space Research Organization (ISRO), archived at the
Indian Space Science Data Centre (ISSDC). CZT-Imager is built
by a consortium of Institutes across India including Tata Institute
of Fundamental Research, Mumbai, Vikram Sarabhai Space
Centre, Thiruvananthapuram, ISRO Satellite Centre, Bengaluru,
Inter University Centre for Astronomy and Astrophysics, Pune,
Physical Research Laboratory, Ahmedabad, Space Application
Centre, Ahmedabad: contributions from the vast technical team
from all these institutes are gratefully acknowledged. 

We acknowledge the use of Vikram-100 HPC at the Physical Research Laboratory (PRL), Ahmedabad and Pegasus HPC at the Inter University Centre for Astronomy and Astrophysics (IUCAA), Pune.

This research has also made use of data obtained through the High Energy Astrophysics Science Archive Research Center Online Service, provided by the NASA/Goddard Space Flight Center.

\vspace{-1em}

%%use \balance somewhere in the left column of the last page to balance the two columns in the end page

%%References section

%\bibliographystyle{apj}
%  \bibliography{reference}

\begin{thebibliography}{}
\expandafter\ifx\csname natexlab\endcsname\relax\def\natexlab#1{#1}\fi

\bibitem[{Arnaud(1996)}]{arnaud1996xspec}
Arnaud, K. 1996, in Astronomical Data Analysis Software and Systems V, Vol.
  101, 17

\bibitem[{{Axelsson} {$et~al$.}(2016){Axelsson}, {Bissaldi}, {Desiante}, \&
  {Longo}}]{axelsson16_160325A}
{Axelsson}, M., {Bissaldi}, E., {Desiante}, R., \& {Longo}, F. 2016, GRB
  Coordinates Network, 19227

\bibitem[{{Band} {$et~al$.}(1993){Band}, {Matteson}, {Ford}, {Schaefer},
  {Palmer}, {Teegarden}, {Cline}, {Briggs}, {Paciesas}, {Pendleton}, {Fishman},
  {Kouveliotou}, {Meegan}, {Wilson}, \& {Lestrade}}]{band93}
{Band}, D., {Matteson}, J., {Ford}, L., {$et~al$.} 1993, \apj, 413, 281

\bibitem[{{Bhalerao} {$et~al$.}(2017){Bhalerao}, {Bhattacharya}, {Vibhute},
  {Pawar}, {Rao}, {Hingar}, {Khanna}, {Kutty}, {Malkar}, {Patil}, {Arora},
  {Sinha}, {Priya}, {Samuel}, {Sreekumar}, {Vinod}, {Mithun}, {Vadawale},
  {Vagshette}, {Navalgund}, {Sarma}, {Pandiyan}, {Seetha}, \&
  {Subbarao}}]{bhalerao16}
{Bhalerao}, V., {Bhattacharya}, D., {Vibhute}, A., {$et~al$.} 2017, Journal of
  Astrophysics and Astronomy, 38, 31

\bibitem[{{Bissaldi}(2016)}]{160802A_gbm}
{Bissaldi}, E. 2016, GRB Coordinates Network, 19754, 1

\bibitem[{{Burgess}(2014)}]{Burgess2014}
{Burgess}, J.~M. 2014, \mnras, 445, 2589

\bibitem[{{Chand} {$et~al$.}(2019){Chand}, {Chattopadhyay}, {Oganesyan}, {Rao},
  {Vadawale}, {Bhattacharya}, {Bhalerao}, \& {Misra}}]{chand18b}
{Chand}, V., {Chattopadhyay}, T., {Oganesyan}, G., {$et~al$.} 2019, \apj, 874,
  70

\bibitem[{{Chand} {$et~al$.}(2018){Chand}, {Chattopadhyay}, {Iyyani}, {Basak},
  {Aarthy}, {Rao}, {Vadawale}, {Bhattacharya}, \& {Bhalerao}}]{chand18a}
{Chand}, V., {Chattopadhyay}, T., {Iyyani}, S., {$et~al$.} 2018, \apj, 862, 154

\bibitem[{{Chattopadhyay} {$et~al$.}(2016){Chattopadhyay}, {Vadawale}, {Rao},
  {Bhattacharya}, {Mithun}, \& {Bhalerao}}]{chattopadhyay16}
{Chattopadhyay}, T., {Vadawale}, S.~V., {Rao}, A.~R., {$et~al$.} 2016, in
  \procspie, Vol. 9905, Society of Photo-Optical Instrumentation Engineers
  (SPIE) Conference Series, 99054D

\bibitem[{{Chattopadhyay} {$et~al$.}(2014){Chattopadhyay}, {Vadawale}, {Rao},
  {Sreekumar}, \& {Bhattacharya}}]{chattopadhyay14}
{Chattopadhyay}, T., {Vadawale}, S.~V., {Rao}, A.~R., {Sreekumar}, S., \&
  {Bhattacharya}, D. 2014, Experimental Astronomy, 37, 555

\bibitem[{{Chattopadhyay} {$et~al$.}(2019){Chattopadhyay}, {Vadawale},
  {Aarthy}, {Mithun}, {Chand}, {Ratheesh}, {Basak}, {Rao}, {Bhalerao}, {Mate},
  {Arvind}, {Sharma}, \& {Bhattacharya}}]{chattopadhyay19}
{Chattopadhyay}, T., {Vadawale}, S.~V., {Aarthy}, E., {$et~al$.} 2019, \apj,
  884, 123

\bibitem[{{Cummings} {$et~al$.}(2015){Cummings}, {Barthelmy}, {Gehrels},
  {Kocevski}, {Krimm}, {Lien}, {Markwardt}, {Palmer}, {Sakamoto}, {Stamatikos},
  \& {Ukwatta}}]{151006A_BAT}
{Cummings}, J.~R., {Barthelmy}, S.~D., {Gehrels}, N., {$et~al$.} 2015, GRB
  Coordinates Network, 18410, 1

\bibitem[{Gruber {$et~al$.}(2014)Gruber, Goldstein, von Ahlefeld, Bhat,
  Bissaldi, Briggs, Byrne, Cleveland, Connaughton, Diehl,
  {$et~al$.}}]{gruber2014fermi}
Gruber, D., Goldstein, A., von Ahlefeld, V.~W., {$et~al$.} 2014, The
  Astrophysical Journal Supplement Series, 211, 12

\bibitem[{{Kocevski} \& {Longo}(2016)}]{160509A_LAT}
{Kocevski}, D., \& {Longo}, F. 2016, in Eighth Huntsville Gamma-Ray Burst
  Symposium, Vol. 1962, 4092

\bibitem[{{Lien} {$et~al$.}(2016{\natexlab{a}}){Lien}, {Barthelmy}, {Cummings},
  {Gehrels}, {Krimm}, {Markwardt}, {Palmer}, {Sakamoto}, {Sonbas},
  {Stamatikos}, \& {Ukwatta}}]{160325A_bat}
{Lien}, A.~Y., {Barthelmy}, S.~D., {Cummings}, J.~R., {$et~al$.}
  2016{\natexlab{a}}, GRB Coordinates Network, 19234, 1

\bibitem[{{Lien} {$et~al$.}(2016{\natexlab{b}}){Lien}, {Barthelmy}, {Cummings},
  {Gehrels}, {Krimm}, {Markwardt}, {Palmer}, {Sakamoto}, {Stamatikos}, \&
  {Ukwatta}}]{lien16_160607A}
---. 2016{\natexlab{b}}, GRB Coordinates Network, 19506

\bibitem[{{Lien} {$et~al$.}(2016{\natexlab{c}}){Lien}, {Barthelmy}, {Cenko},
  {Cummings}, {Gehrels}, {Krimm}, {Markwardt}, {Palmer}, {Sakamoto},
  {Stamatikos}, \& {Ukwatta}}]{lien16_160703A}
{Lien}, A.~Y., {Barthelmy}, S.~D., {Cenko}, S.~B., {$et~al$.}
  2016{\natexlab{c}}, GRB Coordinates Network, 19648

\bibitem[{Lyne {$et~al$.}(1999)Lyne, Pritchard, \&
  Roberts}]{crab_Lyne1999JodrellBC}
Lyne, A., Pritchard, R., \& Roberts, M. 1999

\bibitem[{{McEnery} {$et~al$.}(2016){McEnery}, {Racusin}, \&
  {Longo}}]{160821A_lat}
{McEnery}, J., {Racusin}, J., \& {Longo}, F. 2016, GRB Coordinates Network,
  19831, 1

\bibitem[{{Odaka} {$et~al$.}(2018){Odaka}, {Asai}, {Hagino}, {Koi}, {Madejski},
  {Mizuno}, {Ohno}, {Saito}, {Sato}, {Wright}, {Enoto}, {Fukazawa}, {Hayashi},
  {Kataoka}, {Katsuta}, {Kawaharada}, {Kobayashi}, {Kokubun}, {Laurent},
  {Lebrun}, {Limousin}, {Maier}, {Makishima}, {Mimura}, {Miyake}, {Mori},
  {Murakami}, {Nakamori}, {Nakano}, {Nakazawa}, {Noda}, {Ohta}, {Ozaki},
  {Sato}, {Sato}, {Tajima}, {Takahashi}, {Takahashi}, {Takeda}, {Tanaka},
  {Tanaka}, {Terada}, {Uchiyama}, {Uchiyama}, {Watanabe}, {Yamaoka}, {Yasuda},
  {Yatsu}, {Yuasa}, \& {Zoglauer}}]{odaka18}
{Odaka}, H., {Asai}, M., {Hagino}, K., {$et~al$.} 2018, Nuclear Instruments and
  Methods in Physics Research A, 891, 92

\bibitem[{{Ohno} {$et~al$.}(2015){Ohno}, {Bissaldi}, {Vianello}, {Kocevski}, \&
  {Longo}}]{151006A_Lat_detection}
{Ohno}, M., {Bissaldi}, E., {Vianello}, G., {Kocevski}, D., \& {Longo}, F.
  2015, GRB Coordinates Network, 18406, 1

\bibitem[{{Paul}(2013)}]{paul13}
{Paul}, B. 2013, International Journal of Modern Physics D, 22, 41009

\bibitem[{{Rao} {$et~al$.}(2016){Rao}, {Chand}, {Hingar}, {Iyyani}, {Khanna},
  {Kutty}, {Malkar}, {Paul}, {Bhalerao}, {Bhattacharya}, {Dewangan}, {Pawar},
  {Vibhute}, {Chattopadhyay}, {Mithun}, {Vadawale}, {Vagshette}, {Basak},
  {Pradeep}, {Samuel}, {Sreekumar}, {Vinod}, {Navalgund}, {Pandiyan}, {Sarma},
  {Seetha}, \& {Subbarao}}]{rao16}
{Rao}, A.~R., {Chand}, V., {Hingar}, M.~K., {$et~al$.} 2016, \apj, 833, 86

\bibitem[{{Roberts}(2016)}]{roberts16_160325A}
{Roberts}, O.~J. 2016, GRB Coordinates Network, 19224

\bibitem[{{Roberts} {$et~al$.}(2016){Roberts}, {Fitzpatrick}, \&
  {Veres}}]{160509A_gbm}
{Roberts}, O.~J., {Fitzpatrick}, G., \& {Veres}, P. 2016, GRB Coordinates
  Network, 19411, 1

\bibitem[{{Roberts} \& {Meegan}(2015)}]{151006A_GBM}
{Roberts}, O.~J., \& {Meegan}, C. 2015, GRB Coordinates Network, 18404, 1

\bibitem[{Scargle(1998)}]{scargle1998studies}
Scargle, J.~D. 1998, The Astrophysical Journal, 504, 405

\bibitem[{{Scargle} {$et~al$.}(2013){Scargle}, {Norris}, {Jackson}, \&
  {Chiang}}]{Scargle2013}
{Scargle}, J.~D., {Norris}, J.~P., {Jackson}, B., \& {Chiang}, J. 2013, \apj,
  764, 167

\bibitem[{{Sharma} {$et~al$.}(2019){Sharma}, {Iyyani}, {Bhattacharya},
  {Chattopadhyay}, {Rao}, {Aarthy}, {Vadawale}, {Mithun}, {Bhalerao}, {Ryde},
  \& {Pe'er}}]{vidushi19}
{Sharma}, V., {Iyyani}, S., {Bhattacharya}, D., {$et~al$.} 2019, \apjl, 882,
  L10

\bibitem[{{Singh} {$et~al$.}(2014){Singh}, {Tandon}, {Agrawal}, {Antia},
  {Manchanda}, {Yadav}, {Seetha}, {Ramadevi}, {Rao}, {Bhattacharya}, {Paul},
  {Sreekumar}, {Bhattacharyya}, {Stewart}, {Hutchings}, {Annapurni}, {Ghosh},
  {Murthy}, {Pati}, {Rao}, {Stalin}, {Girish}, {Sankarasubramanian},
  {Vadawale}, {Bhalerao}, {Dewangan}, {Dedhia}, {Hingar}, {Katoch}, {Kothare},
  {Mirza}, {Mukerjee}, {Shah}, {Shah}, {Mohan}, {Sangal}, {Nagabhusana},
  {Sriram}, {Malkar}, {Sreekumar}, {Abbey}, {Hansford}, {Beardmore}, {Sharma},
  {Murthy}, {Kulkarni}, {Meena}, {Babu}, \& {Postma}}]{singh14}
{Singh}, K.~P., {Tandon}, S.~N., {Agrawal}, P.~C., {$et~al$.} 2014, in Society
  of Photo-Optical Instrumentation Engineers (SPIE) Conference Series, Vol.
  9144, Society of Photo-Optical Instrumentation Engineers (SPIE) Conference
  Series

\bibitem[{{Stanbro} \& {Meegan}(2016)}]{160821A_gbm}
{Stanbro}, M., \& {Meegan}, C. 2016, GRB Coordinates Network, 19835, 1

\bibitem[{{Vadawale} {$et~al$.}(2018){Vadawale}, {Chattopadhyay}, {Mithun},
  {Rao}, {Bhattacharya}, {Vibhute}, {Bhalerao}, {Dewangan}, {Misra}, {Paul},
  {Basu}, {Joshi}, {Sreekumar}, {Samuel}, {Priya}, {Vinod}, \&
  {Seetha}}]{vadawale17}
{Vadawale}, S.~V., {Chattopadhyay}, T., {Mithun}, N.~P.~S., {$et~al$.} 2018,
  Nature Astronomy, 2, 50

\bibitem[{{Veres} \& {Meegan}(2016)}]{160910A_gbm}
{Veres}, P., \& {Meegan}, C. 2016, GRB Coordinates Network, 19901, 1

\end{thebibliography}

\end{document}